\documentclass[10pt, conference, letterpaper]{IEEEtran}

\usepackage{subcaption}
\usepackage[table,xcdraw]{xcolor}
\usepackage{multirow}
\usepackage{url}
\usepackage{makecell}
 
\usepackage{verbatim}
 
\IEEEoverridecommandlockouts
\usepackage{cite}
\usepackage{amsmath,amssymb,amsfonts}
\usepackage{algorithmic}
\usepackage{graphicx}
\usepackage{textcomp}
\def\BibTeX{{\rm B\kern-.05em{\sc i\kern-.025em b}\kern-.08em
T\kern-.1667em\lower.7ex\hbox{E}\kern-.125emX}}

\definecolor{light-gray}{gray}{0.9}
\definecolor{dark-gray}{gray}{0.5}

\IEEEaftertitletext{\vspace{-11.5pt}} 

\begin{document}

\title{On the Distribution of Traffic Volumes in the Internet and its Implications}

\author{\IEEEauthorblockN{Mohammed Alasmar}
\IEEEauthorblockA{\textit{Department of Informatics} \\
\textit{University of Sussex}\\
Brighton, UK \\
m.alasmar@sussex.ac.uk}
\and
\IEEEauthorblockN{George Parisis}
\IEEEauthorblockA{\textit{Department of Informatics} \\
\textit{University of Sussex}\\
Brighton, UK  \\
g.parisis@sussex.ac.uk}
\and
\IEEEauthorblockN{Richard Clegg}
\IEEEauthorblockA{\textit{School of Computer Science } \\
\textit{Queen Mary University of London}\\
London, UK \\
r.clegg@qmul.ac.uk }
\and
\IEEEauthorblockN{Nickolay Zakhleniuk}
\IEEEauthorblockA{\textit{School of Computer Science} \\
\textit{University of Essex}\\
Colchester, UK\\
naz@essex.ac.uk}
}

\maketitle

\begin{abstract}

Getting good statistical models of traffic on network links is a well-known, often-studied problem.  A lot of attention has been given to correlation patterns and flow duration.  The distribution of the amount of traffic per unit time is an equally important but less studied problem.  We study a large number of traffic traces from many different networks including academic, commercial and residential networks using state-of-the-art statistical techniques.  We show that the log-normal distribution is a better fit than the Gaussian distribution commonly claimed in the literature. We also investigate a second heavy-tailed distribution (the Weibull) and show that its performance is better than Gaussian but worse than log-normal.  We examine anomalous traces which are a poor fit for all distributions tried and show that this is often due to traffic outages or links that hit maximum capacity.

We demonstrate the utility of the log-normal distribution in two contexts: predicting the proportion of time traffic will exceed a given level (for service level agreement or link capacity estimation) and predicting 95th percentile pricing.  We also show the log-normal distribution is a better predictor than Gaussian or Weibull distributions.

\end{abstract}

\begin{IEEEkeywords}
Traffic modelling, network planning, bandwidth provisioning, traffic billing 
\end{IEEEkeywords}

\section{Introduction}
\label{introduction}

Internet traffic characterisation is an important problem for network researchers and vendors. The subject has a long history. Early works~\cite{selfSimilarity95, self-sim97} discovered that the correlation structure of traffic exhibits self-similarity and that the durations of individual flows of packets exhibit heavy-tails~\cite{heavy-tailed-2010-trans}. These works were later challenged and refined (see Section~\ref{sec:related} for a summary). By comparison the distribution of the amount of traffic seen on a link in a given time period has seen comparatively less research interest. This is surprising as this quantity can be extremely useful in network planning. 

In this paper we use a rigorous statistical approach to fitting a statistical distribution to the amount of traffic within a given time period. Formally, we choose some timescale $T$ and let $X_i$ be the amount of traffic seen in the time period $[iT, (i+1)T)$. We investigate the distribution of the random variable $X$ over a wide range of values of $T$.  We show that the distribution of the variable has considerable implications for network planning; for assessing how often a link is over capacity and in particular for service level agreements (SLAs), and for traffic pricing, particularly using the 95th percentile scheme~\cite{95percentileIMC}.

Previous authors have claimed that $X$ has a normal (or Gaussian) distribution~\cite{Gaussian-everywhere,proveGaussianIFIP,Gaussian-revisited}. Others claim $X$ is Gaussian plus a tail associated with bursts~\cite{2014-ifip-conf,12-GLOBECOM2002}.  A variable $X$ has a log-normal distribution if its logarithm is normally distributed $\ln(X) \sim N(\mu,\sigma^2)$ where $\mu \in \mathbb{R}$ is the mean and $\sigma > 0$ is the standard deviation of the distribution.  We use a well-established statistical methodology~\cite{clauset} to show that a log-normal distribution is a better fit than Gaussian or Weibull\footnote{A variable $X$ has a Weibull distribution with parameters $k > 0$ (known as shape) and $\lambda > 0$ (known as scale) if its probability density function follows $f(x) = \frac{k}{\lambda}\left(\frac{x}{\lambda}\right)^{k-1} \exp(-(x/\lambda)^k)$ when $x \geq 0$ and is $0$ otherwise.} for the vast majority of traces.  This holds over a wide range of timescales $T$ (from $5$ msec to $5$ sec). This paper is the most comprehensive investigation of this phenomenon the authors know about. We study a large number of publicly available traces from a diverse set of locations (including commercial, academic and residential networks) with different link speeds and spanning the last 15 years. 


The structure of the paper is as follows. In Section~\ref{sec:dataset} we describe the  datasets used. In Section~\ref{sec:fitting} we describe our best-practice procedure for fitting traffic and demonstrate that log-normal is the best fit distribution for our traces under a variety of circumstances. We examine those few traces that do not follow this distribution and find it occurs when a link spends considerable time either having an outage or completely at maximum capacity. In Section~\ref{sec:provision} we demonstrate that the log-normal distribution is the most useful for estimating how often a link is over capacity. In Section~\ref{sec:pricing} we show that the log-normal distribution provides good estimates when looking at 95th percentile pricing. In Section~\ref{sec:related} we give related work. Finally, Section~\ref{sec:conclusion} gives our conclusions.

\section{Network Traffic Traces}
\label{sec:dataset}
 
A key contribution of our work stems from the spatial and temporal diversity of the studied traces. The dataset spans a period of 15 years and comprises $229$ traces.

\noindent\textbf{CAIDA traces.} We have used $27$ CAIDA traces captured at an Internet data collection monitor which is located at an Equinix data centre in Chicago~\cite{caidaRef}. The data centre is connected to a backbone link of a Tier 1 ISP. The  monitor records an hour-long traces four times a year, usually from $13{:}00$ to $14{:}00$ UTC. Each trace contains billions of IPv4 packets, the headers of which are anonymised. The average captured data rate is 2.5~Gbps. At the time of capturing, the monitored link had a capacity of 10~Gbps. Traces were captured between $2013$ and $2016$. 
\break 
\noindent\textbf{MAWI traces.} The MAWI archive~\cite{mawiRef} consists of a collection of Internet traffic traces, captured within the WIDE backbone network that connects Japanese universities and research institutions to the Internet. Each trace consists of IP level traffic observed daily from $14{:}00$ to $14{:}15$ at a vantage point within WIDE. Traces include anonymised IP and MAC headers, along with an \textit{ntpd} timestamp~\cite{mawiRef}. We have looked at $107$ traces (each one being $15$ minutes long). Traces were captured between $2014$ and $2018$. On average, each trace consists of 70 million packets; the average captured data rate is 422~Mbps. The monitored link had a capacity of 1~Gbps.
\break 
\noindent\textbf{Twente University traces.} We used $40$ traffic traces captured at five different locations ($8$ traces from each location). Traces are diverse in terms of the link rates, types of users and captured time~\cite{Meent2010Traces}. Each trace is $15$ minutes long. The first location is a residential network with a 300~Mbps link, which connects 2000 students (each one having a 100~Mbps access link); traces were captured in July 2002. The second location is a research institute network with a 1~Gbps link which connects 200 researchers (each one having a 100~Mbps access link); traces were captured between May and August 2003. The third location is at a large college with a 1~Gbps link which connects 1000 employees (each one having a 100~Mbps access link); traces were captured between February and July 2004. The fourth location is an ADSL access network with a 1~Gbps ADSL link used by hundreds of users (each one having a 256~Kbps to 8~Mbps access link); traces were captured between February and July 2004. The fifth location is an educational organisation with a 100~Mbps link connecting 135 students and employees (each one having a 100~Mbps access link);  traces were captured between May and June 2007.
\break 
\noindent\textbf{Waikato University VIII traces.} The Waikato dataset consists of traffic traces captured by the WAND group at the University of Waikato, New Zealand~\cite{waikatoRef}. The capture point is at the link interconnecting the University with the Internet. All of the traces were captured using software that was specifically developed for the Waikato capture point and a DAG 3 series hardware capture card. All IP addresses within the traces are anonymised. In our study, we have used $30$ traces captured between April 2011 and November 2011.
\break 
\noindent\textbf{Auckland University IX  traces.} The Auckland dataset consists of traffic traces captured by the WAND group at the University of Waikato~\cite{aucklandRef}. The traces were collected at the University of Auckland, New Zealand. The capture point is at the link interconnecting the University with the Internet. All IP addresses within the traces are anonymised. In our study, we have used $25$ traces captured in 2009.
\break

\section{Fitting a statistical distribution to Internet traffic data}
\label{sec:fitting}

In this section we present an extensive statistical analysis applied to the datasets described in the previous section. The aim is to discover which statistical distribution best fits the traces. In contrast to the existing research (see Section~\ref{sec:related}), we are basing our analysis on the framework proposed by Clauset et al.~\cite{clauset}, a comprehensive statistical framework developed specifically for testing power-law behaviour in empirical data\footnote{We have used the source code discussed in~\cite{Alstott}.}. The framework combines maximum-likelihood fitting methods with goodness-of-fit tests based on the Kolmogorov--Smirnov statistic and likelihood ratios. The method reliably tests whether the power-law distribution is the best model for a specific dataset, or, if not, whether an alternative statistical distribution (e.g., log-normal, exponential, Weibull) is. The framework performs the tests described above as follows: (1) the parameters of the power-law model are estimated for a given dataset; (2) the goodness-of-fit between the data and the power-law is calculated, under the hypothesis that the power-law is the best fit to the provided traffic samples. 
If the resulting $p$-value is greater than $0.1$ the hypothesis is accepted (i.e. the power law is a plausible fit to the given data), otherwise the hypothesis is rejected; (3) alternative distributions are tested against the power-law as a fit to the data by employing a likelihood ratio test.

For the vast majority of the traces examined, the hypothesis was rejected; i.e. the power-law distribution was not a good fit. Consequently, we investigate alternative distributions by performing the likelihood ratio (LLR) test (following Clauset's methodology), as follows:
\[
 \Re, p = \mathrm{fit.distributionCompare(powerlaw, alternative)}
\]
where $\Re$ is the normalised LLR\footnote{$\Re$ is calculated as $\mathcal{R} /(\sigma\sqrt{n}) $, where $\mathcal{R}$ is the log likelihood ratio~\cite{clauset}.} between the power-law and alternative distributions and $p$ is the significance value for this test. $\Re$ is positive if the power-law distribution is a better fit for the data, and negative if the alternative distribution is a better fit for the data. A  $p$-value less than $0.1$ means that the value of $\Re$ can be trusted to make a conclusion that one candidate distribution (power-law or alternative, depending on the sign of $\Re$) is a good fit for the data. In contrast, a $p$-value greater than $0.1$ means that there is nothing to be concluded from the likelihood ratio test. 

\begin{figure*}[t]
	\centering
	\subcaptionbox{CAIDA traces}[.32\linewidth][c]{%
		\includegraphics[width=.32\linewidth]{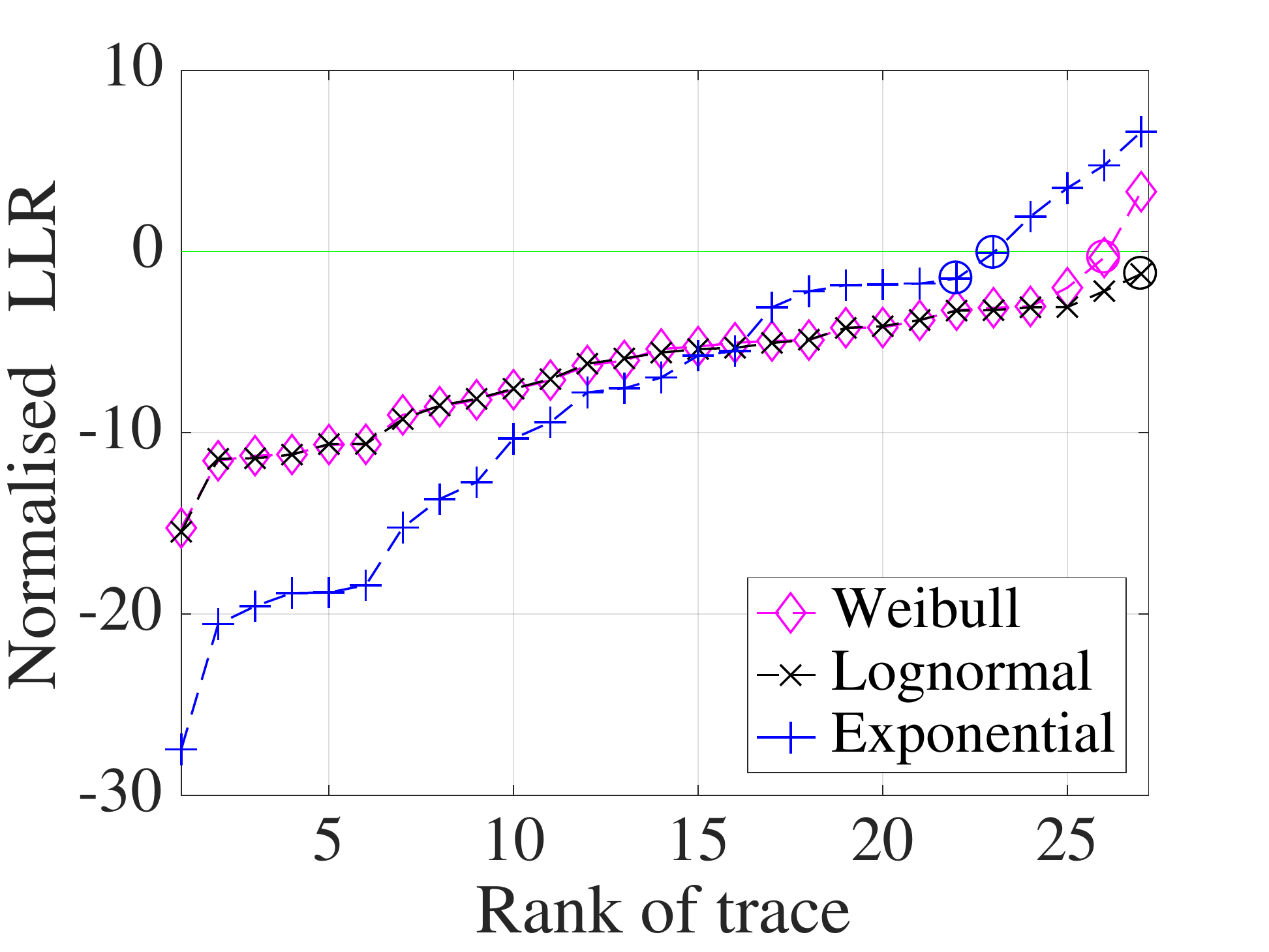}}\quad
	\subcaptionbox{Waikato traces}[.32\linewidth][c]{%
		\includegraphics[width=.32\linewidth]{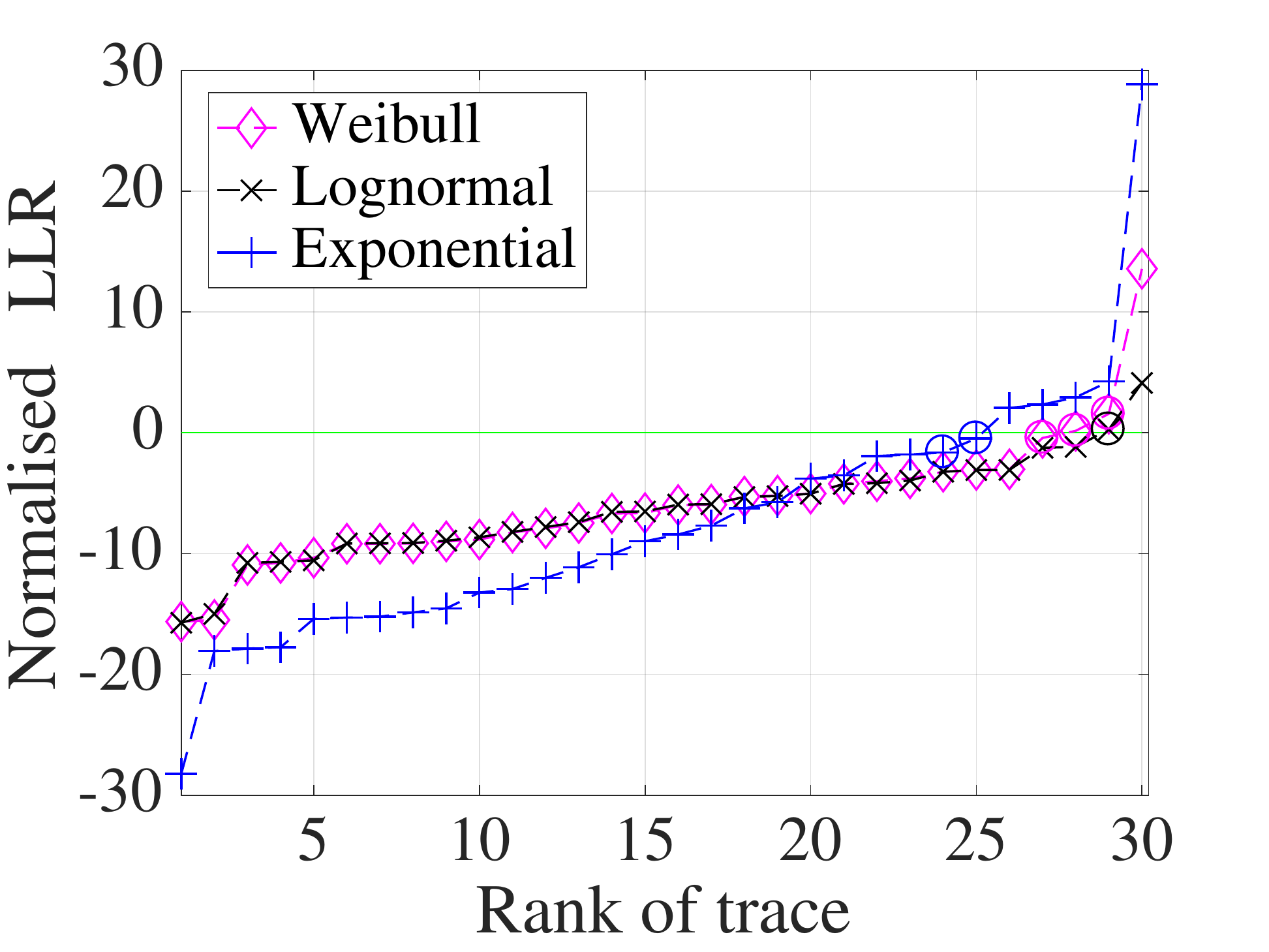}}\quad
	\subcaptionbox{Auckland traces}[.32\linewidth][c]{%
		\includegraphics[width=.32\linewidth]{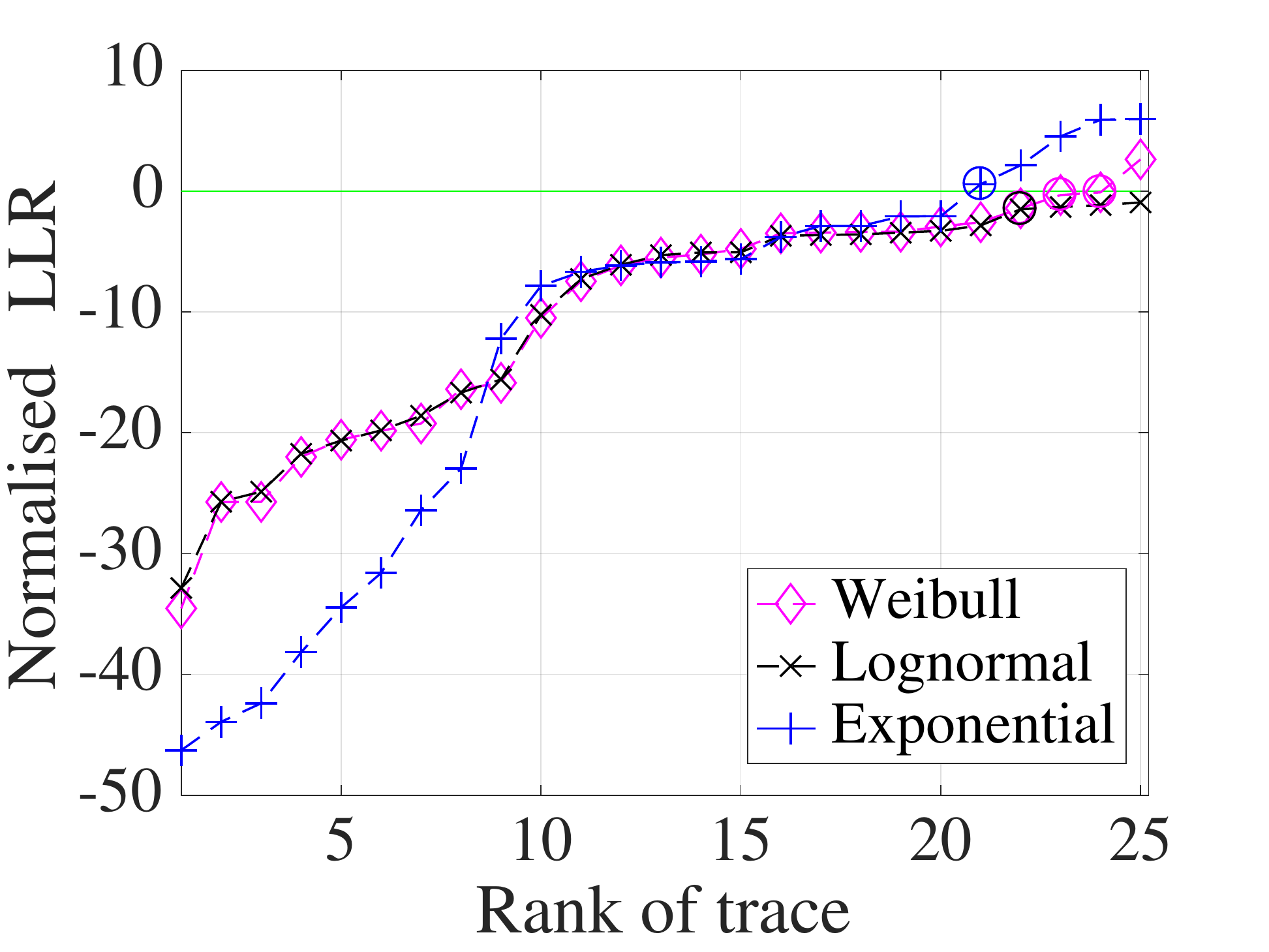}}
	\bigskip
	\subcaptionbox{Twente traces}[.32\linewidth][c]{%
		\includegraphics[width=.32\linewidth]{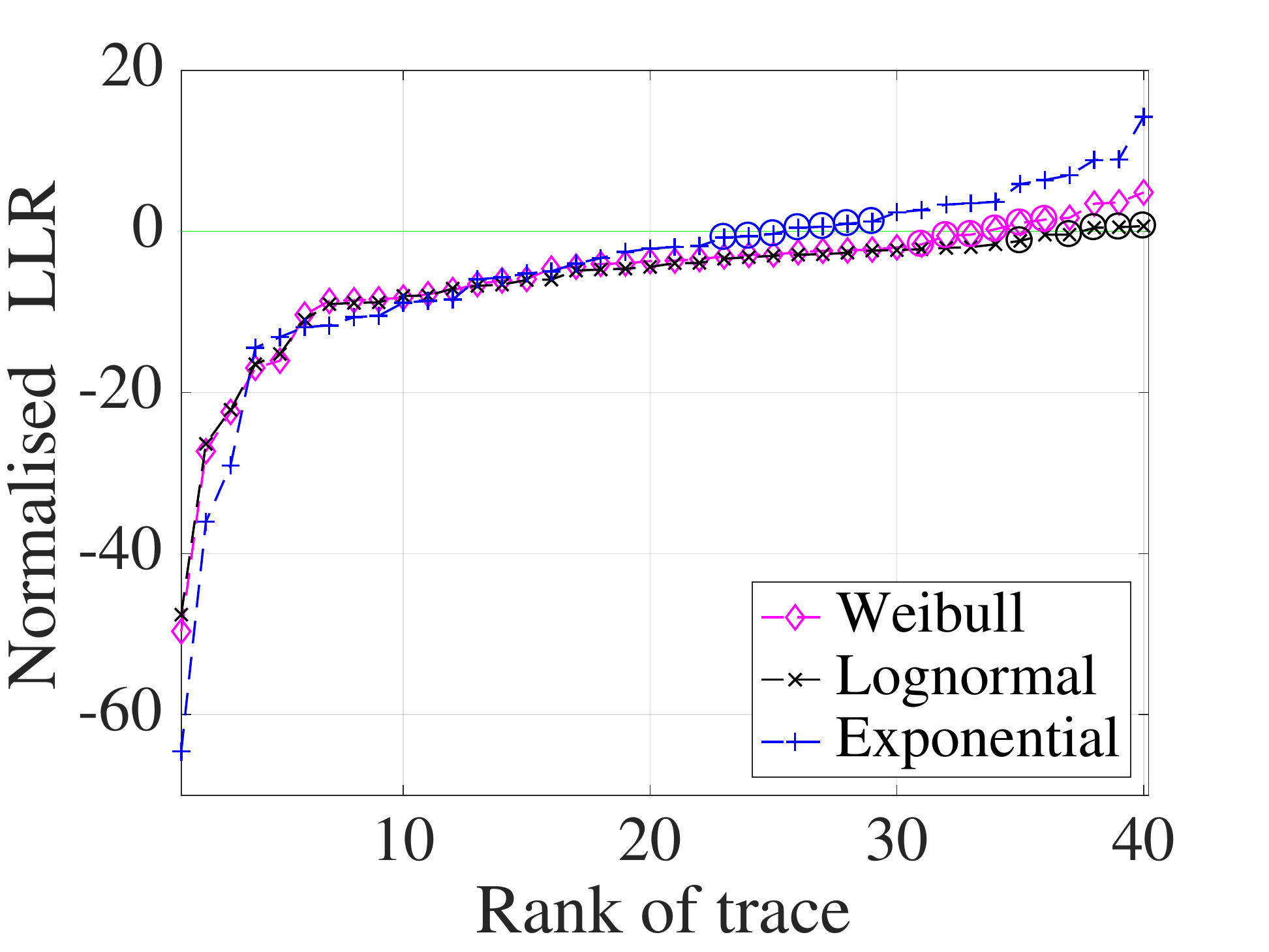}}\quad
	\subcaptionbox{MAWI traces}[.32\linewidth][c]{%
		\includegraphics[width=.32\linewidth]{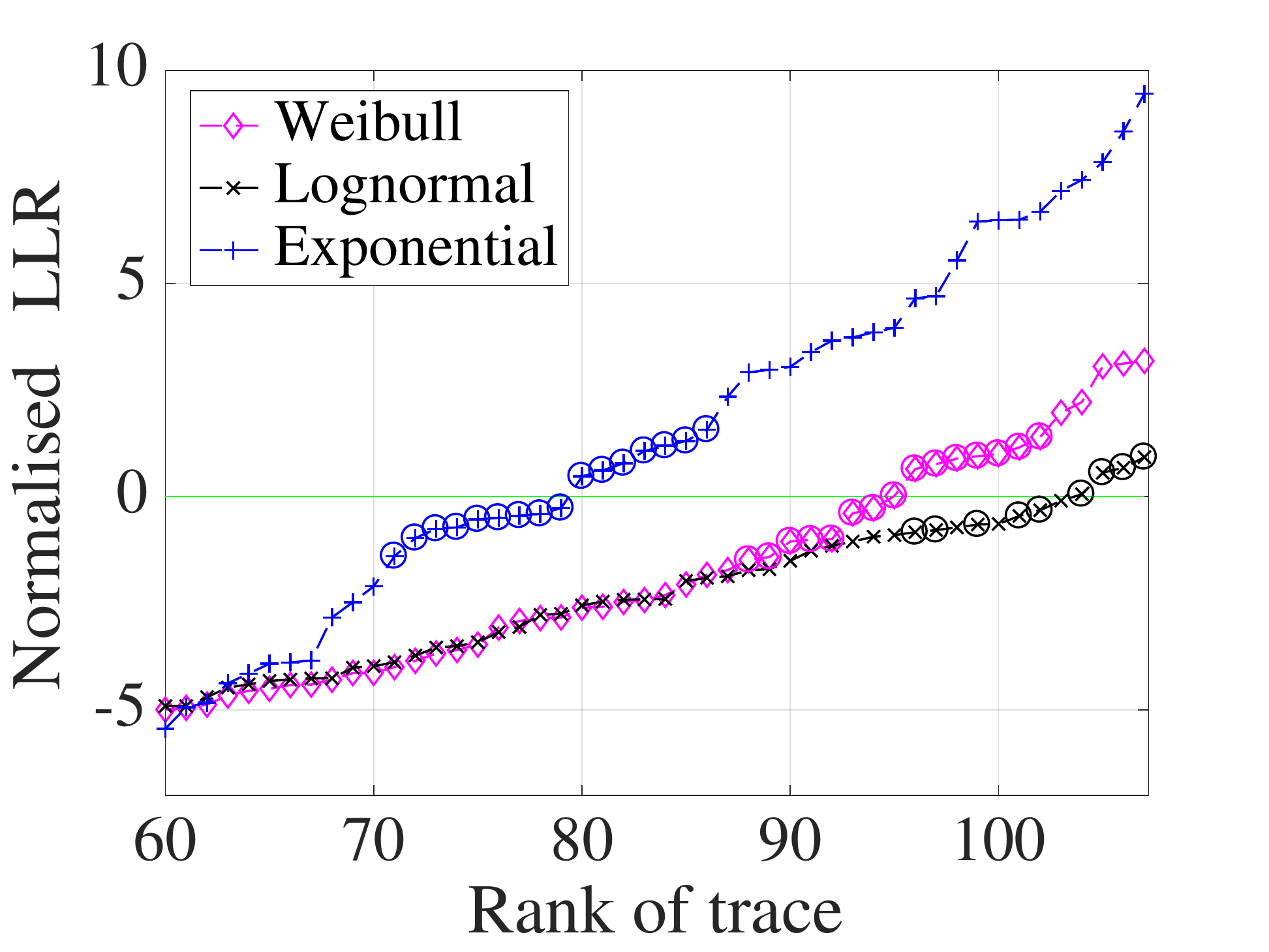}}\quad
	
	\caption{Normalised Log-Likelihood Ratio (LLR) test results for all studied traces and candidate distributions. Aggregation timescale $T$ is 100 msec. Circled points in the plot are the ones with $p$-value greater than $0.1$; i.e. likelihood test is inconclusive with respect to fitting any of the candidate distributions to the traffic data.}
	\label{LRtestResults} 	
\end{figure*}
\begin{figure*}[t]
	\centering
	\subcaptionbox{CAIDA traces}[.32\linewidth][c]{%
		\includegraphics[width=.32\linewidth]{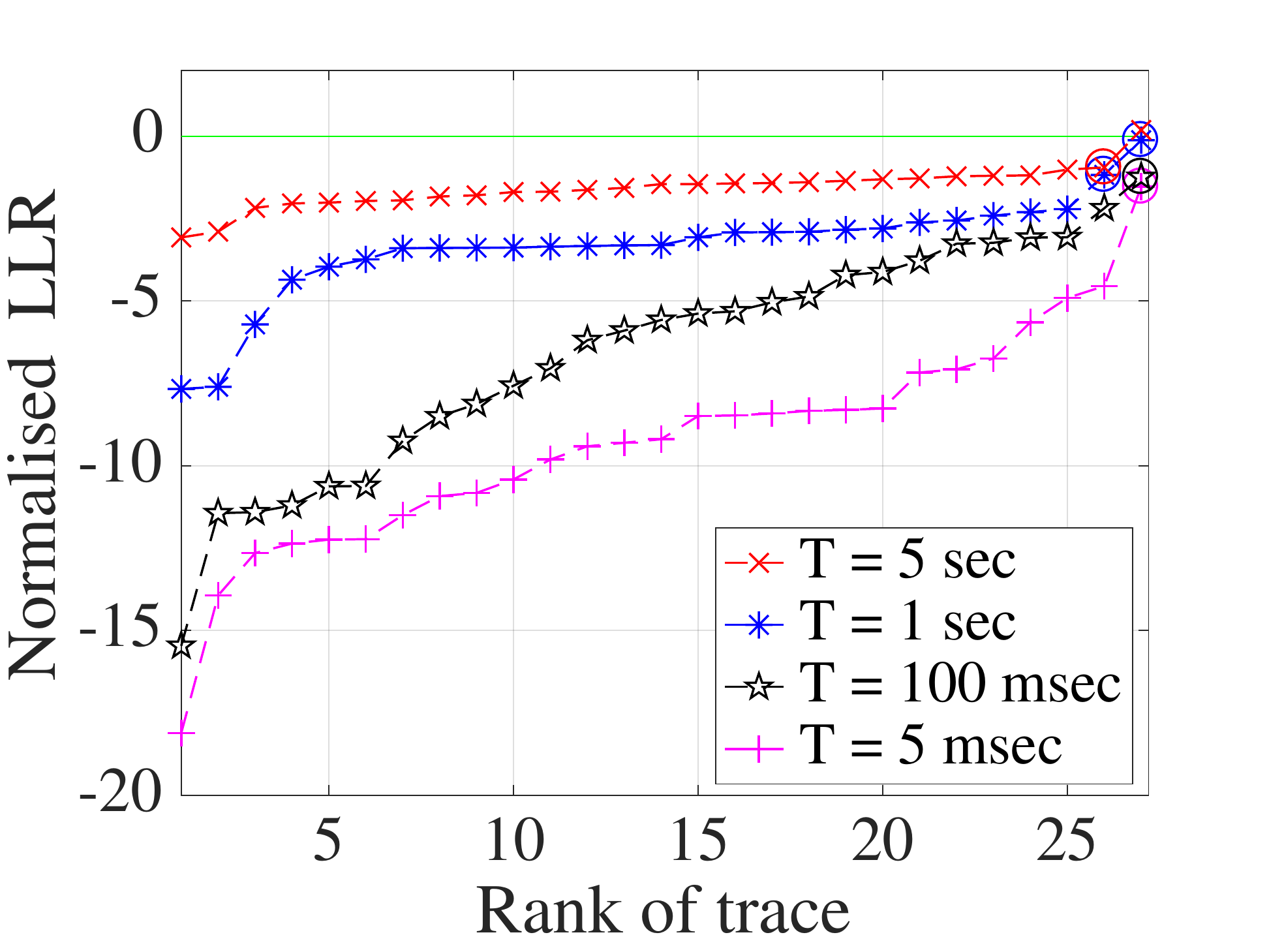}}\quad
	\subcaptionbox{Waikato traces}[.32\linewidth][c]{%
		\includegraphics[width=.32\linewidth]{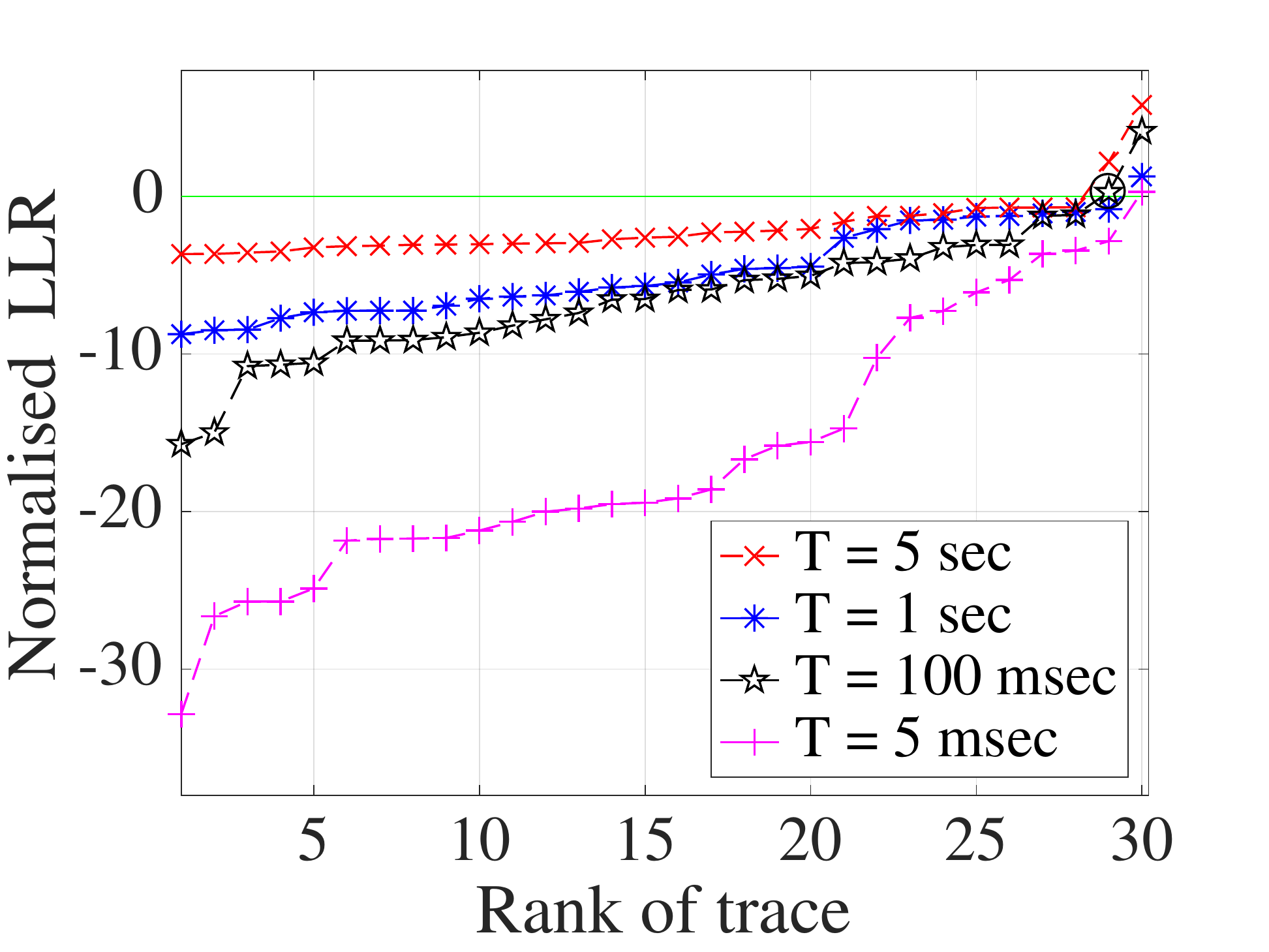}}\quad
	\subcaptionbox{Auckland traces}[.32\linewidth][c]{%
		\includegraphics[width=.32\linewidth]{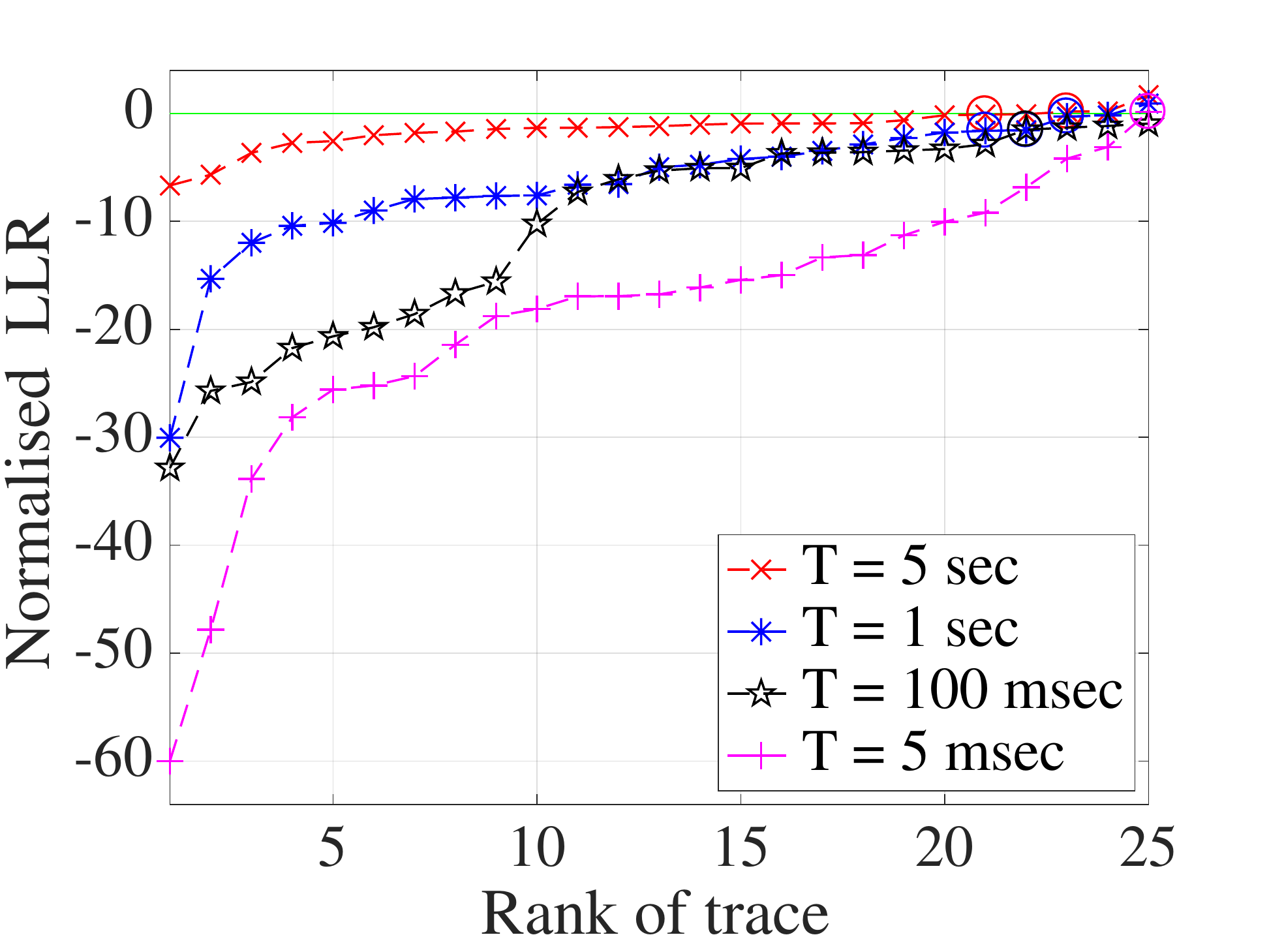}}
	\bigskip
	\subcaptionbox{Twente traces}[.32\linewidth][c]{%
		\includegraphics[width=.32\linewidth]{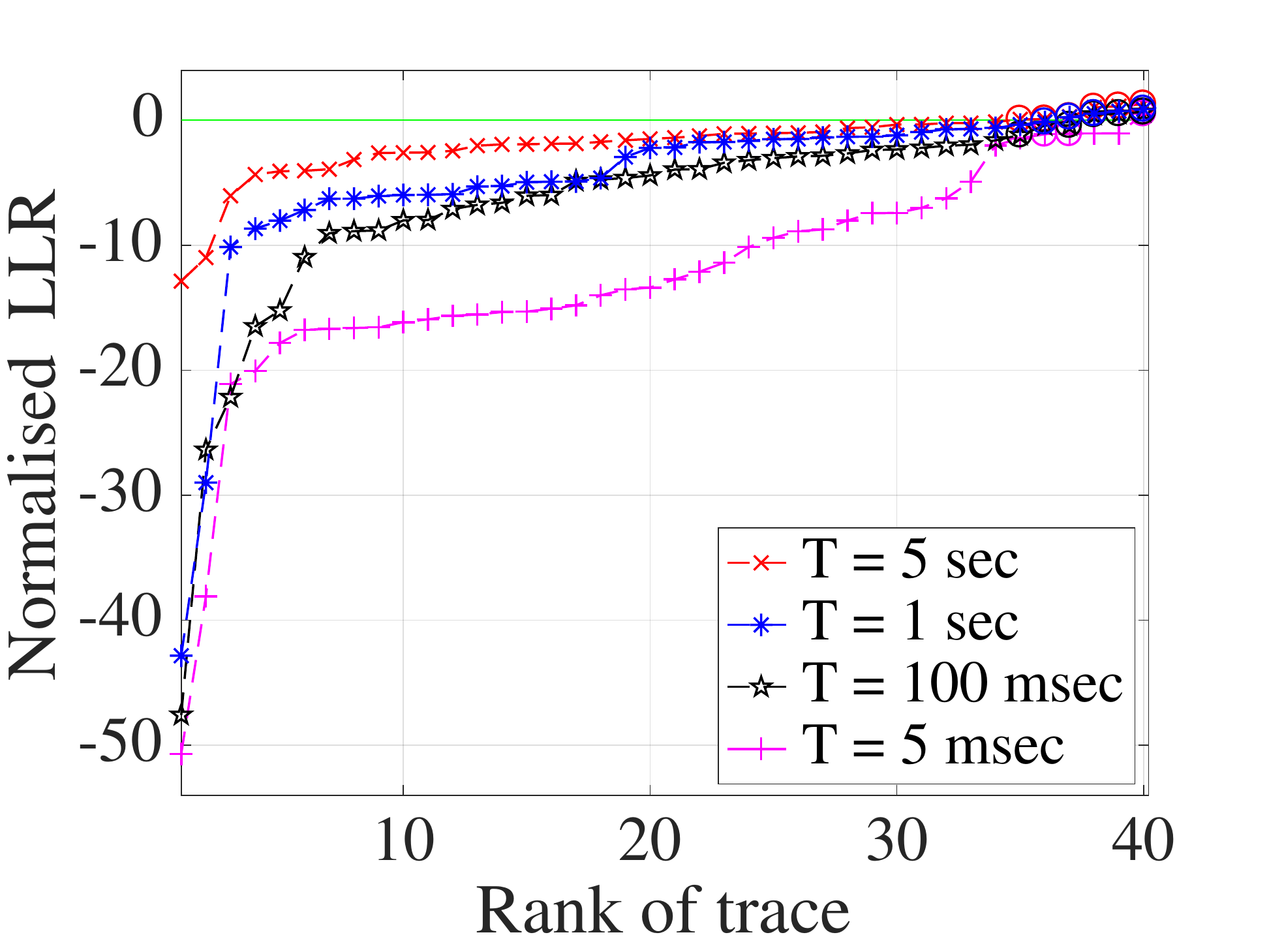}}\quad
	\subcaptionbox{MAWI traces}[.32\linewidth][c]{%
		\includegraphics[width=.32\linewidth]{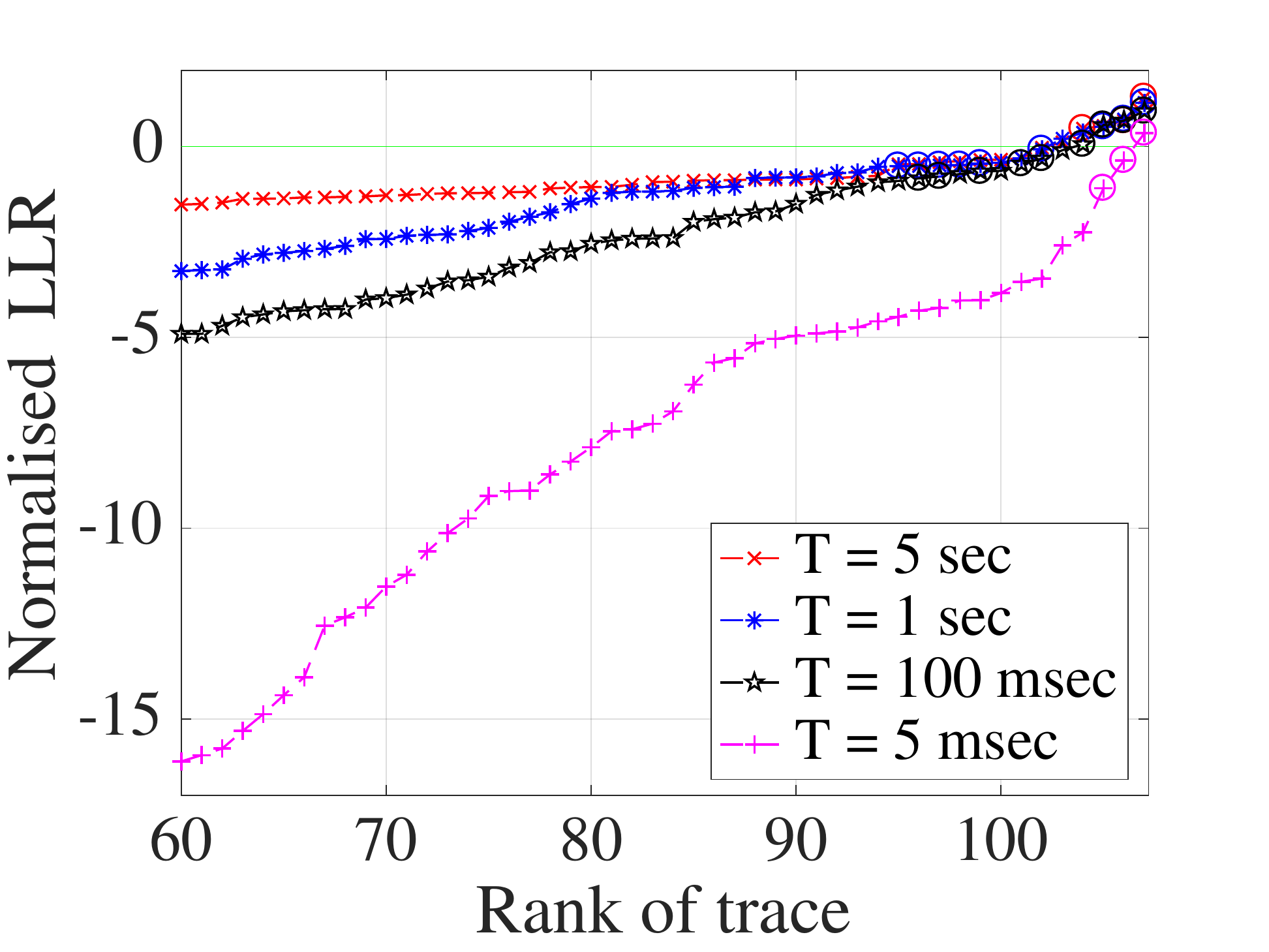}}\quad
	
	\caption{Normalised Log-Likelihood Ratio (LLR) test results for all studied traces and log-normal distribution. Aggregation timescales are 5 sec, 1 sec, 100 msec and 5 msec. Circled points in the plot are the ones with $p$-value greater than $0.1$, i.e. likelihood test is inconclusive with respect to fitting the log-normal distribution to the traffic data.}
	\label{LRResultsDiffTValues}
\end{figure*} 

\subsection{Fitting the log-normal distribution to Internet traffic data}
\label{fit-log-normal}

Figure~\ref{LRtestResults} shows the results of the LLR test for all $229$ traces with log-normal, exponential and Weibull distribution as the alternative to power-law. For this test we have aggregated traffic at a timescale $T=100$ msec. The points marked with a circle are the ones with $p>0.1$. It is clear that the log-normal distribution (black line in Figure~\ref{LRtestResults}) is the best fit for the studied traces; i.e. $\Re<0$ and $p<0.1$ for most traces when the alternative distribution (to the power-law which is almost always rejected) is the log-normal one\footnote{For clarity, in Figures~\ref{LRtestResults}(e) and~\ref{LRResultsDiffTValues}(e) we only plot traces 60 -- 107. For traces 1 -- 59, $\Re$ is less than $0$ and the respective $p$-value is less than $0.1$; i.e. the alternative distribution is the best fit for the respective trace}. The log-normal distribution is not the best fit for $1$ out of $27$ CAIDA  traces, $2$ out $30$ Waikato traces, $1$ out of $25$ Auckland traces, $5$ out of $40$ Twente traces and $9$ out of $107$ MAWI traces. We examined these traces in more detail and discuss them in Section~\ref{anomalous}.

For the vast majority of traces the power-law distribution is favoured over the exponential one (i.e. $\Re>0$), as shown in Figure~\ref{LRtestResults}. Thus, the exponential distribution cannot be considered as a good model for our traffic traces. On the other hand, the Weibull appears to be a better fit over the power-law distribution; however, when compared to the log-normal distribution, it still performs poorly (i.e. $\Re>0$ or $\Re<0$ but $p>0.1$) for a substantial amount of traces.  

Identifying the log-normal distribution as the best fit for the vast majority of traffic traces at $T=100$ msec is very encouraging. This specific traffic aggregation timescale has been commonly studied in the literature~\cite{ResDimension,transaction2015}.  
Next we investigate what the best model is for a range of aggregation timescales. The results are shown in Figure~\ref{LRResultsDiffTValues}. As reflected by the $\Re$ and $p$-values, the log-normal distribution is the best fit for the vast majority of captured traces at all examined timescales ($5$ msec to $5$ sec)\footnote{Note that it is possible that the network traffic may not follow a log-normal distribution at very fine or coarse aggregation granularities.}. This is a strong result suggesting the generality of our observations. The good log-normal fit at time scales as small as $5$ msec is important for practical applications of the log-normal model.  

We also examined Q-Q plots for a large number of traces\footnote{Due to lack of space, Q-Q plots are not included as we would have to present plots for each trace, separately.}. The log-normal distribution appeared to be a better fit than other tested distributions and no deviations from the expected pattern were observed in the body or tail of the distribution.

\subsection{Anomalous traces} 
\label{anomalous}

\begin{figure}[!h]
	\centering
	\subcaptionbox{Anomalous trace}[.48\linewidth][c]{%
		\includegraphics[width=.48\linewidth]{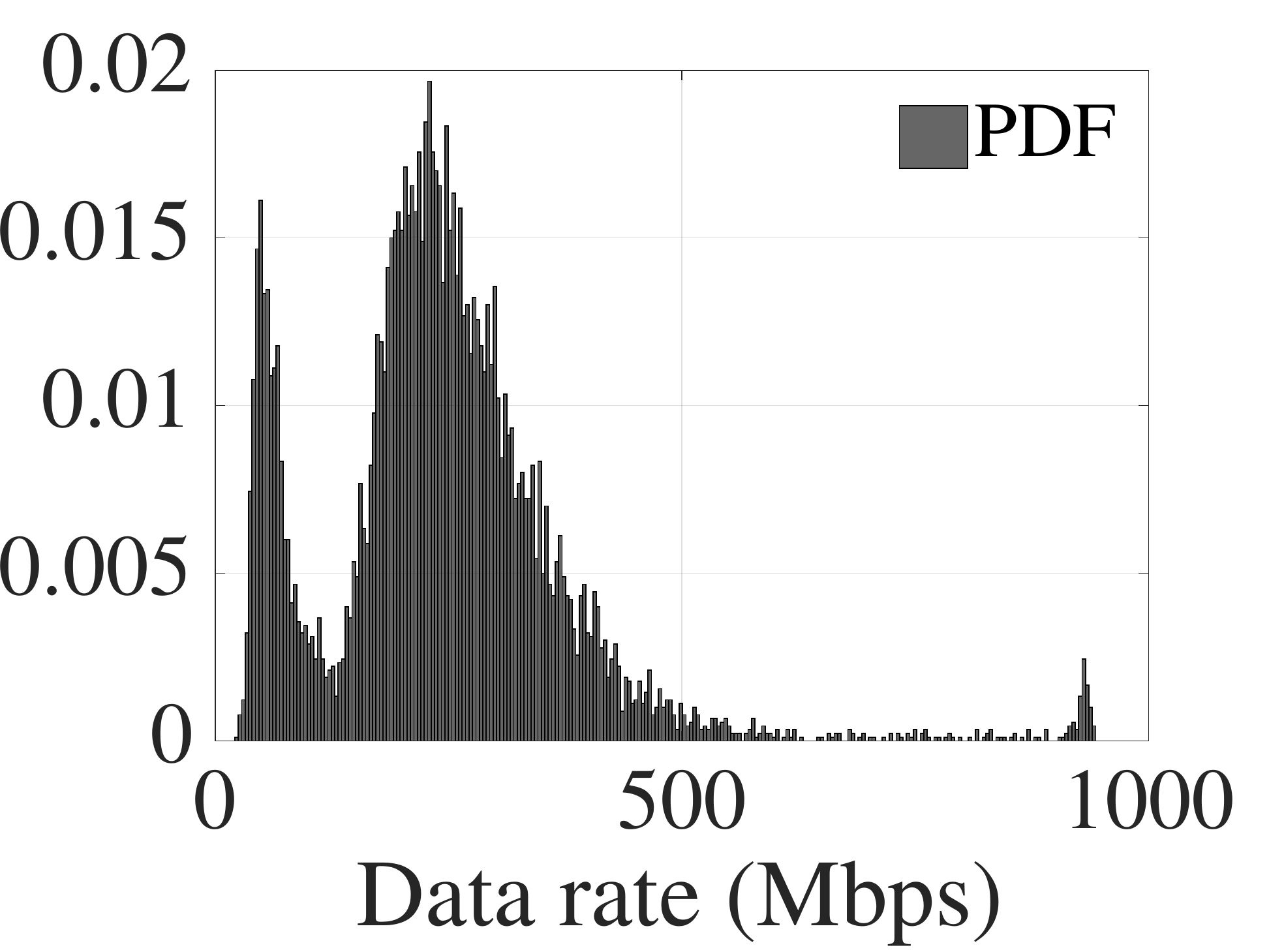}}\quad
	\subcaptionbox{Log-normal trace}[.48\linewidth][c]{%
		\includegraphics[width=.48\linewidth]{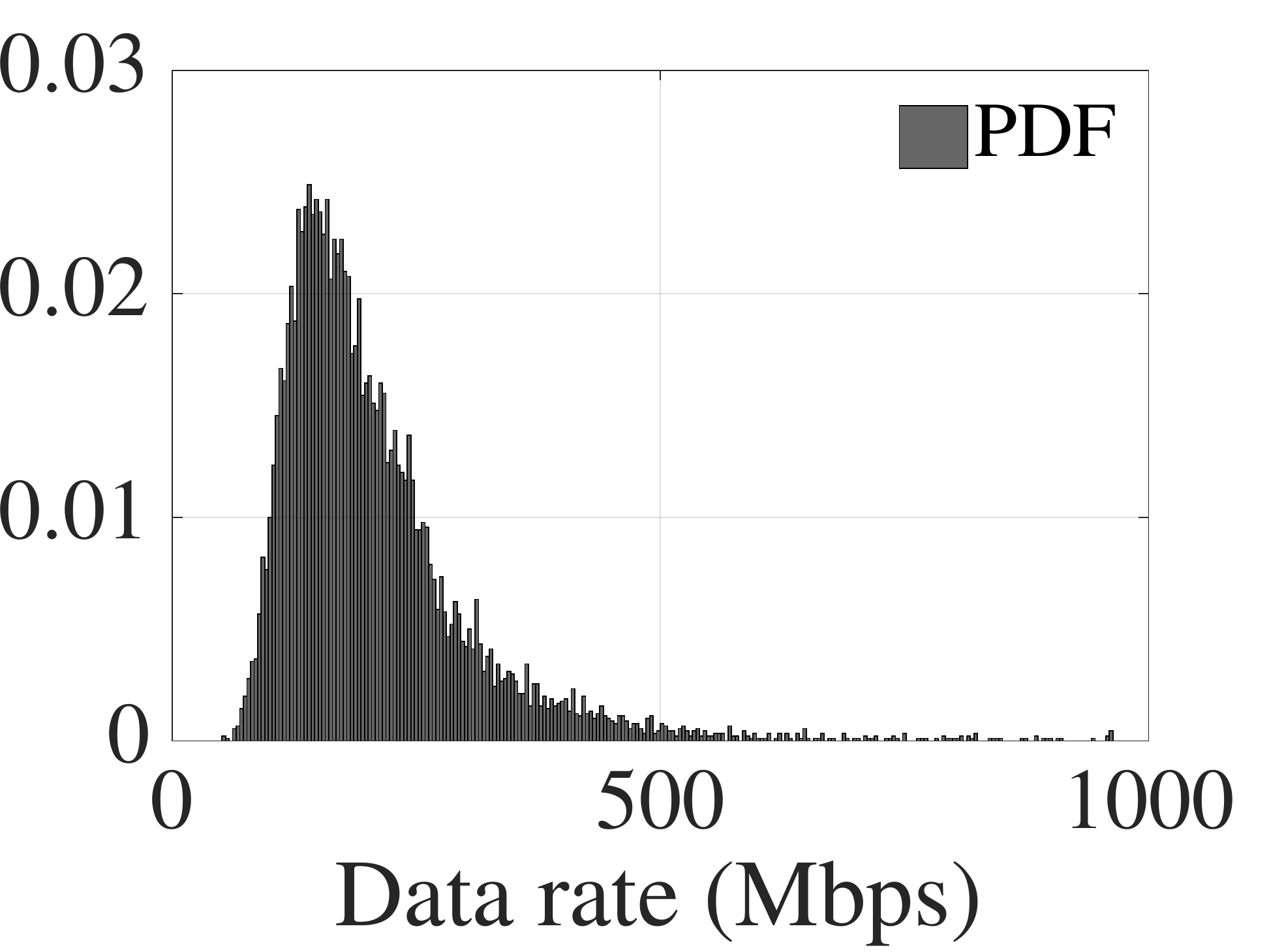}}\quad
	\caption{PDF of an anomalous and non-anomalous trace.}
	\label{anomalousTraces} 
\end{figure}

As mentioned in Section~\ref{fit-log-normal}, there are a small number of traces for which the log-normal distribution is not a good fit (none of the other examined distributions is, either). Figure~\ref{anomalousTraces}(a) shows the PDF plot for one of the $8$ anomalous MAWI traces. Figure~\ref{anomalousTraces}(b) shows the PDF for another MAWI trace for which the log-normal distribution is a good fit. It is obvious from Figure~\ref{anomalousTraces}(a) that the link was either severely underutilised (see large spike on the left part of the plot area) or fully utilised (see smaller spike at the right part of the plot area) for higher data rates. All traces for which the log-normal distribution was not a good fit exhibited similar behaviour and (aggregated) traffic patterns. On the contrary, we did not observe any such behaviour for the majority of traces for which the log-normal distribution was the best fit.  A likely explanation for the anomalous traces is that those traces contain either periods of over-capacity (traffic is at 100\% of link capacity) or periods where the link is broken (no traffic).
\subsection{Fitting the log-normal and Gaussian distributions using the correlation coefficient test}
\label{fit-correlation-coefficient-test}

\begin{figure*}[t]
	\centering
	\subcaptionbox{CAIDA traces}[0.188\linewidth][ctth]{%
		\includegraphics[height=3cm, width=3.88cm ]{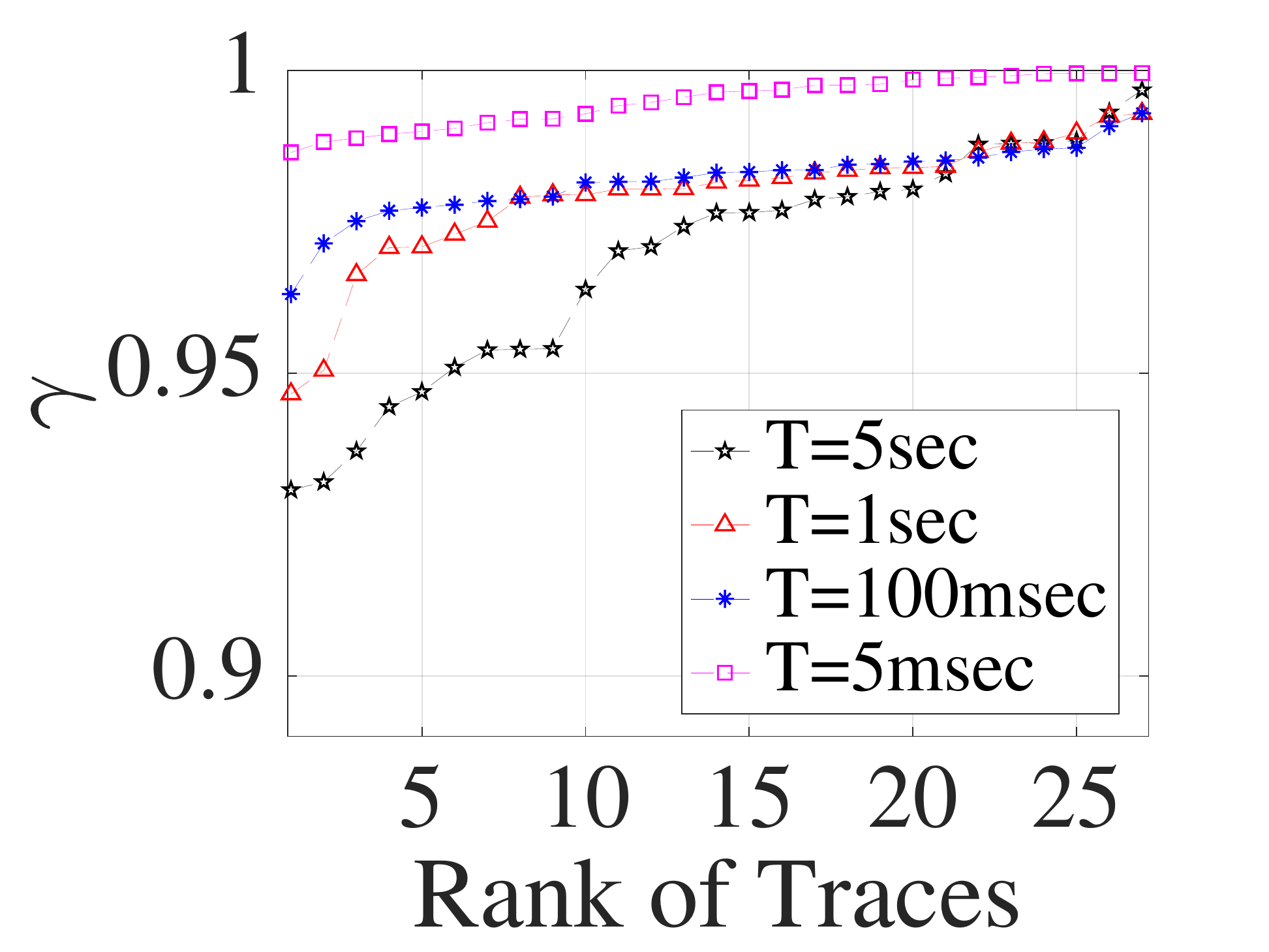}}\quad
	\subcaptionbox{Waikato traces}[0.188\linewidth][c]{%
		\includegraphics[height=3cm, width=3.88cm ]{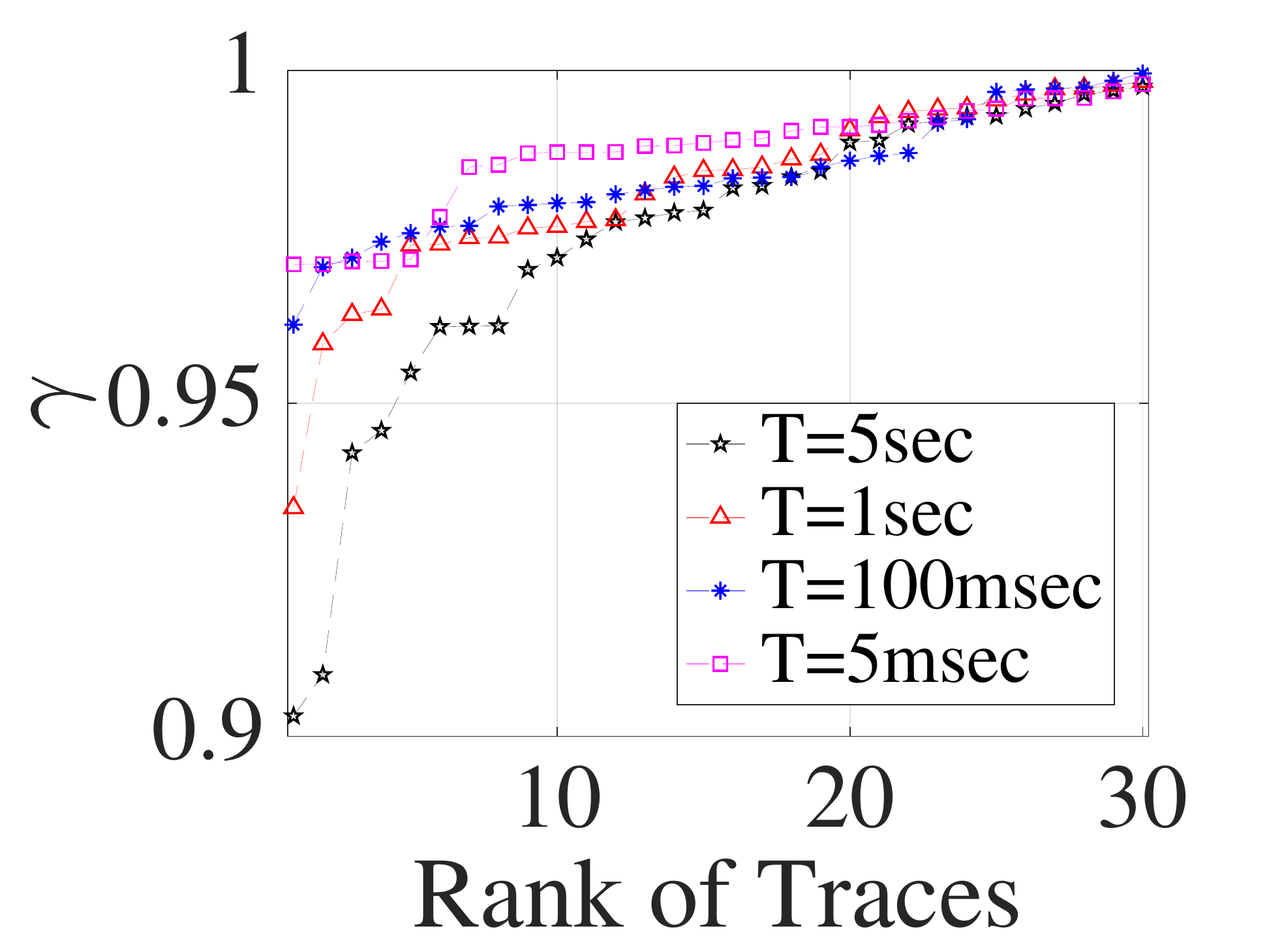}}\quad
	\subcaptionbox{Auckland traces}[0.188\linewidth][c]{%
		\includegraphics[height=3cm, width=3.88cm ]{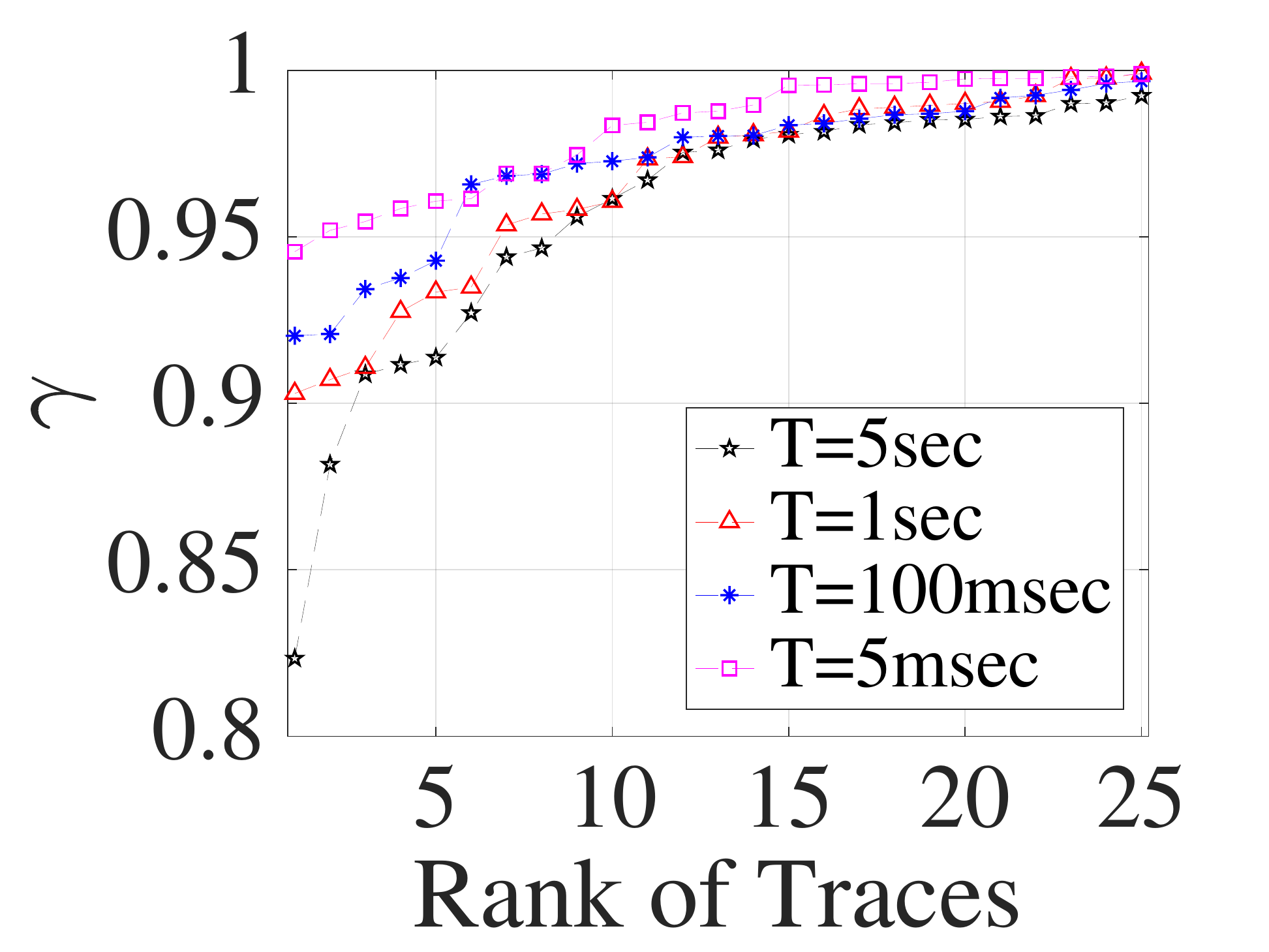}}
	\subcaptionbox{Twente traces}[0.188\linewidth][c]{%
		\includegraphics[height=3cm, width=3.88cm ]{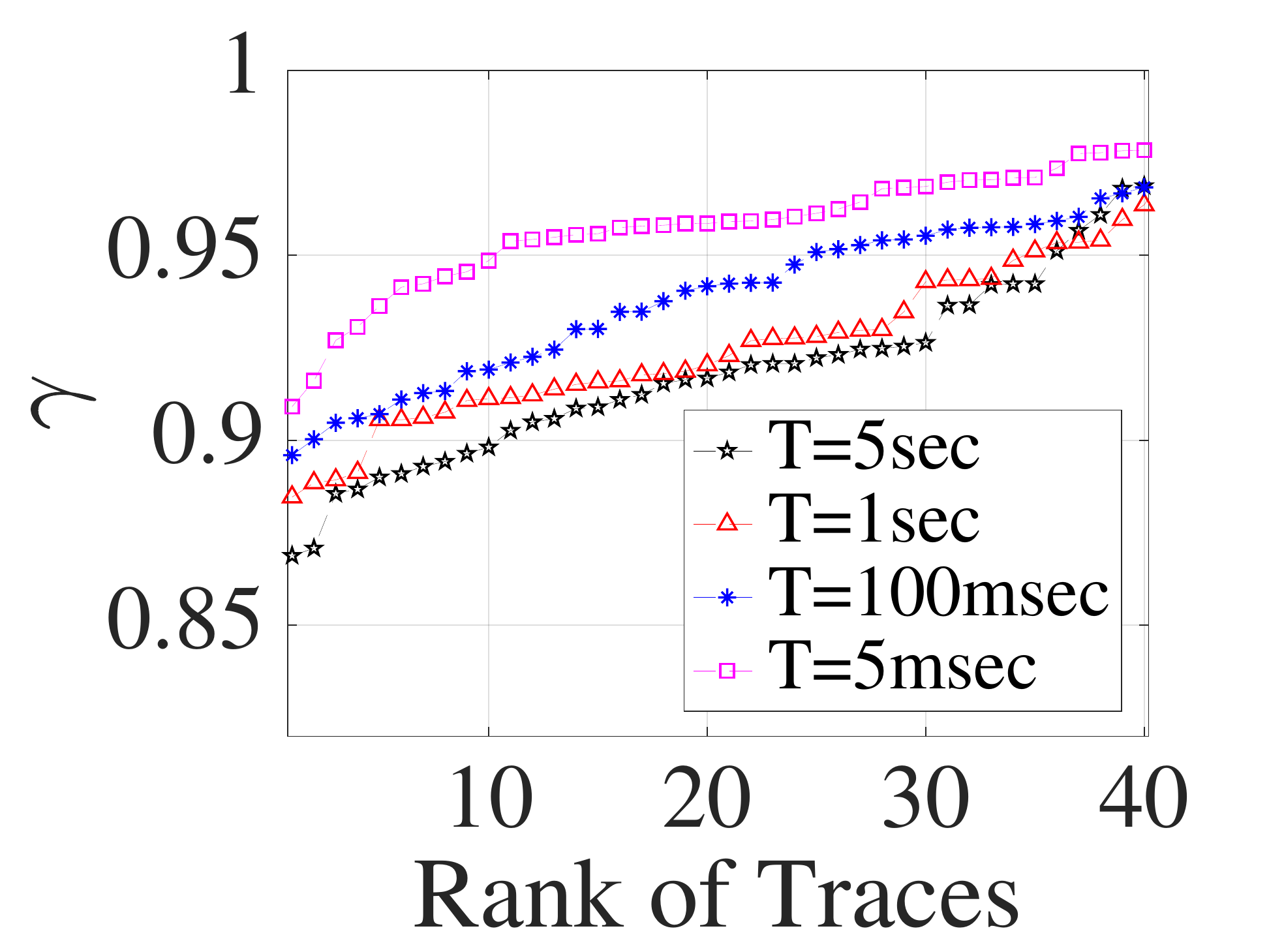}}
	\subcaptionbox{MAWI traces}[0.188\linewidth][c]{%
		\includegraphics[height=3cm, width=3.88cm ]{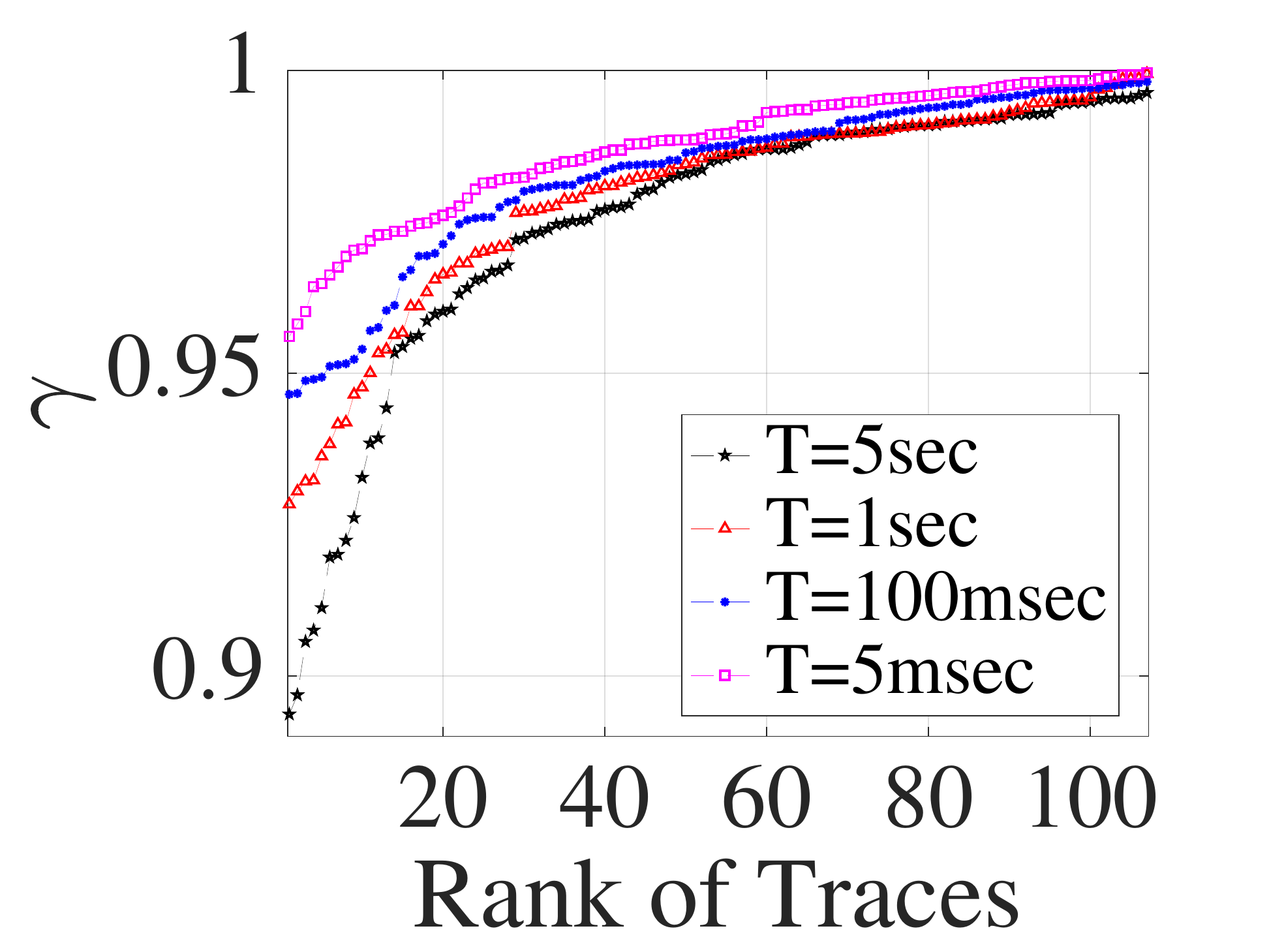}}	
	\bigskip 
	\subcaptionbox{CAIDA traces}[0.188\linewidth][ctth]{%
		\includegraphics[height=3cm, width=3.88cm ]{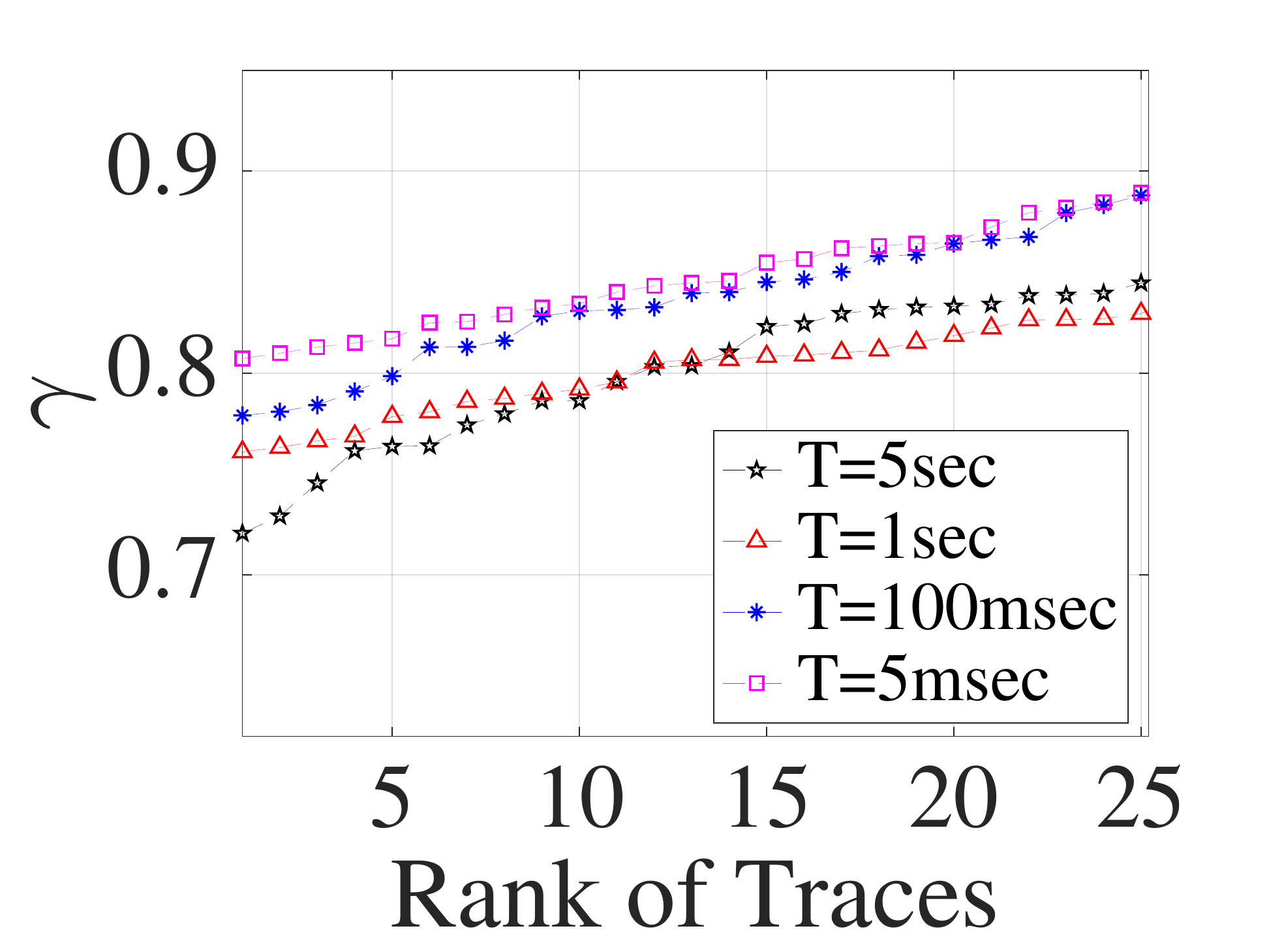}}\quad
	\subcaptionbox{Waikato traces}[0.188\linewidth][c]{%
		\includegraphics[height=3cm, width=3.88cm ]{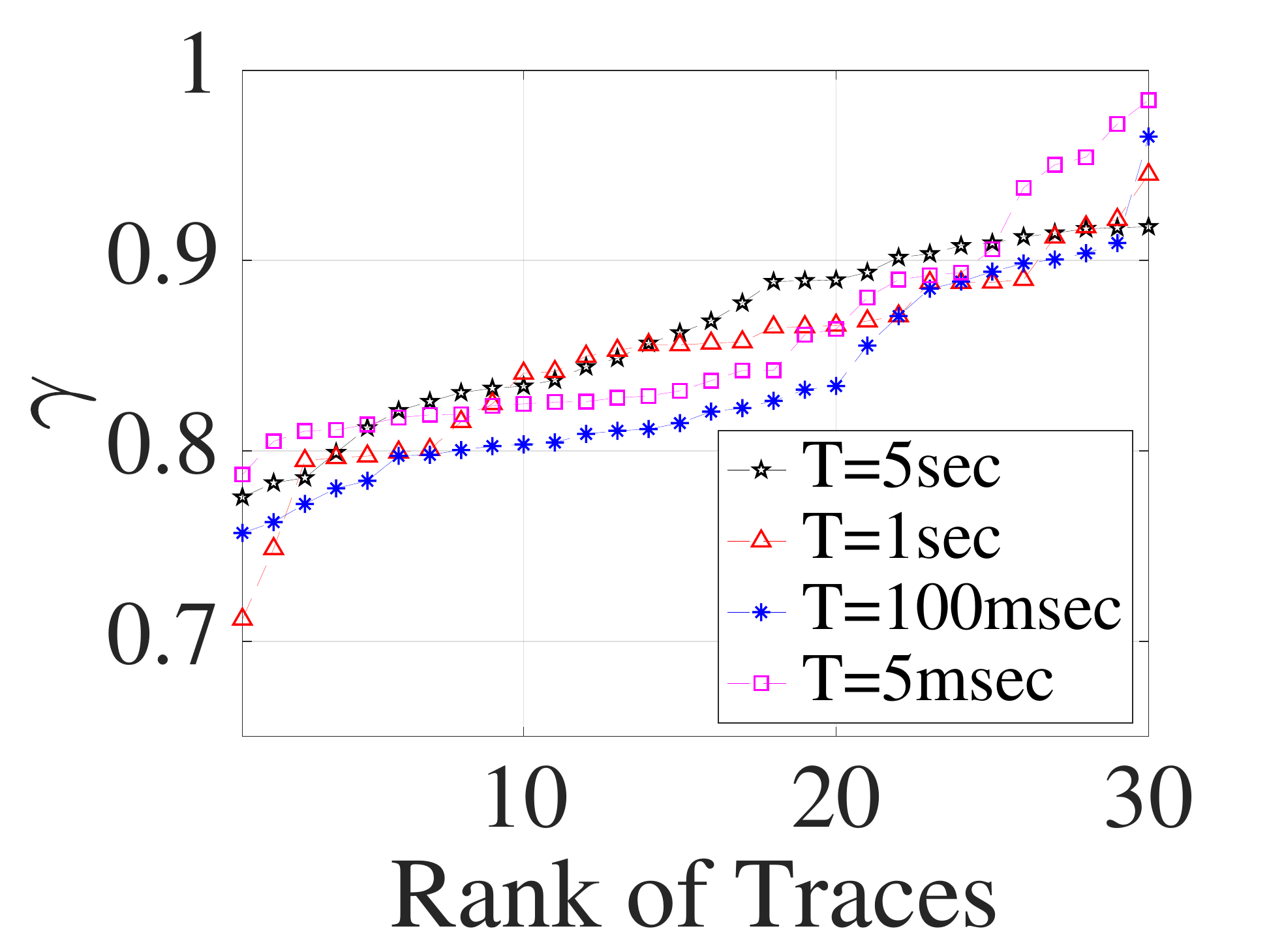}}\quad
	\subcaptionbox{Auckland traces}[0.188\linewidth][c]{%
		\includegraphics[height=3cm, width=3.88cm ]{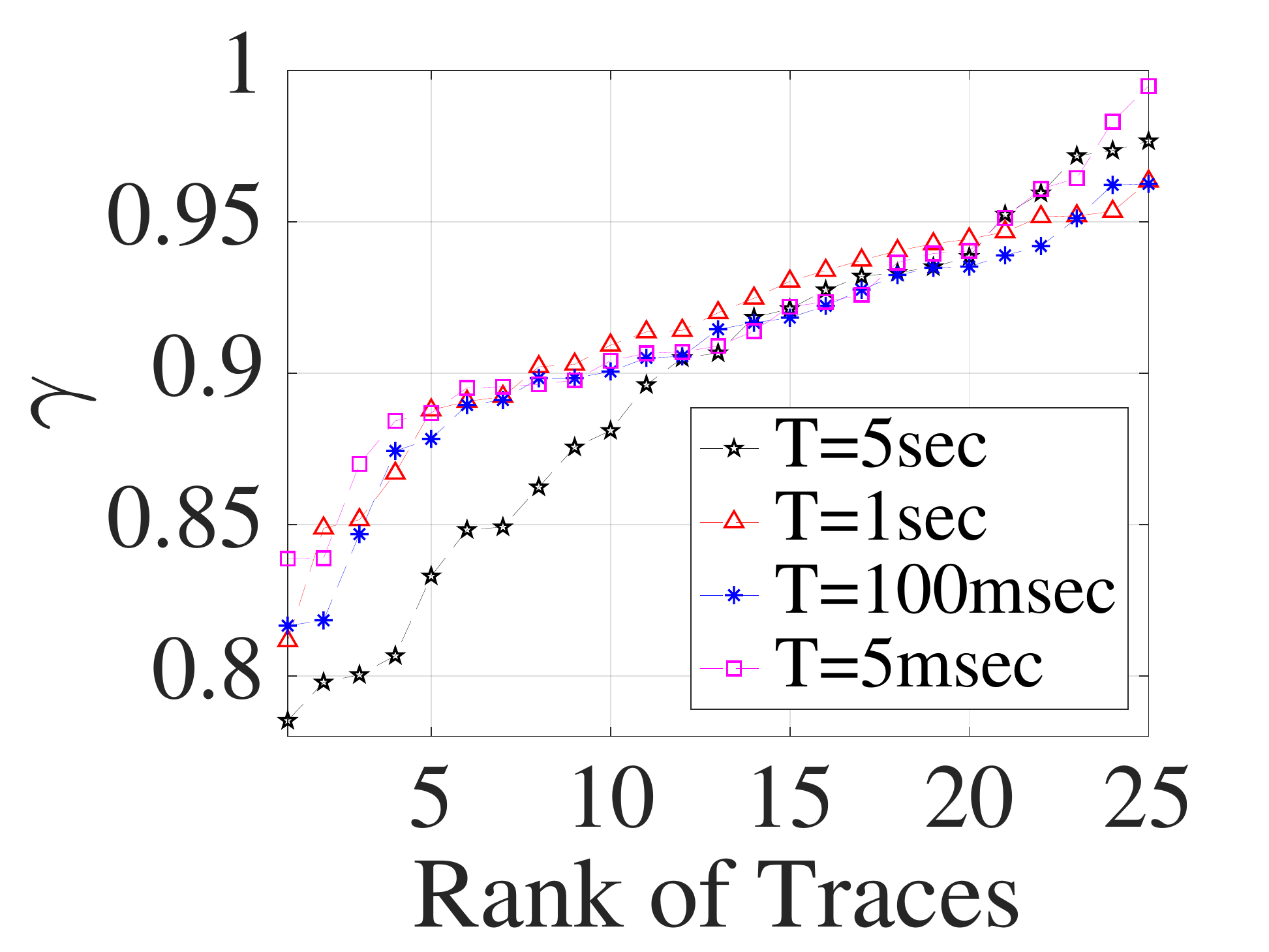}}
	\subcaptionbox{Twente traces}[0.188\linewidth][c]{%
		\includegraphics[height=3cm, width=3.88cm ]{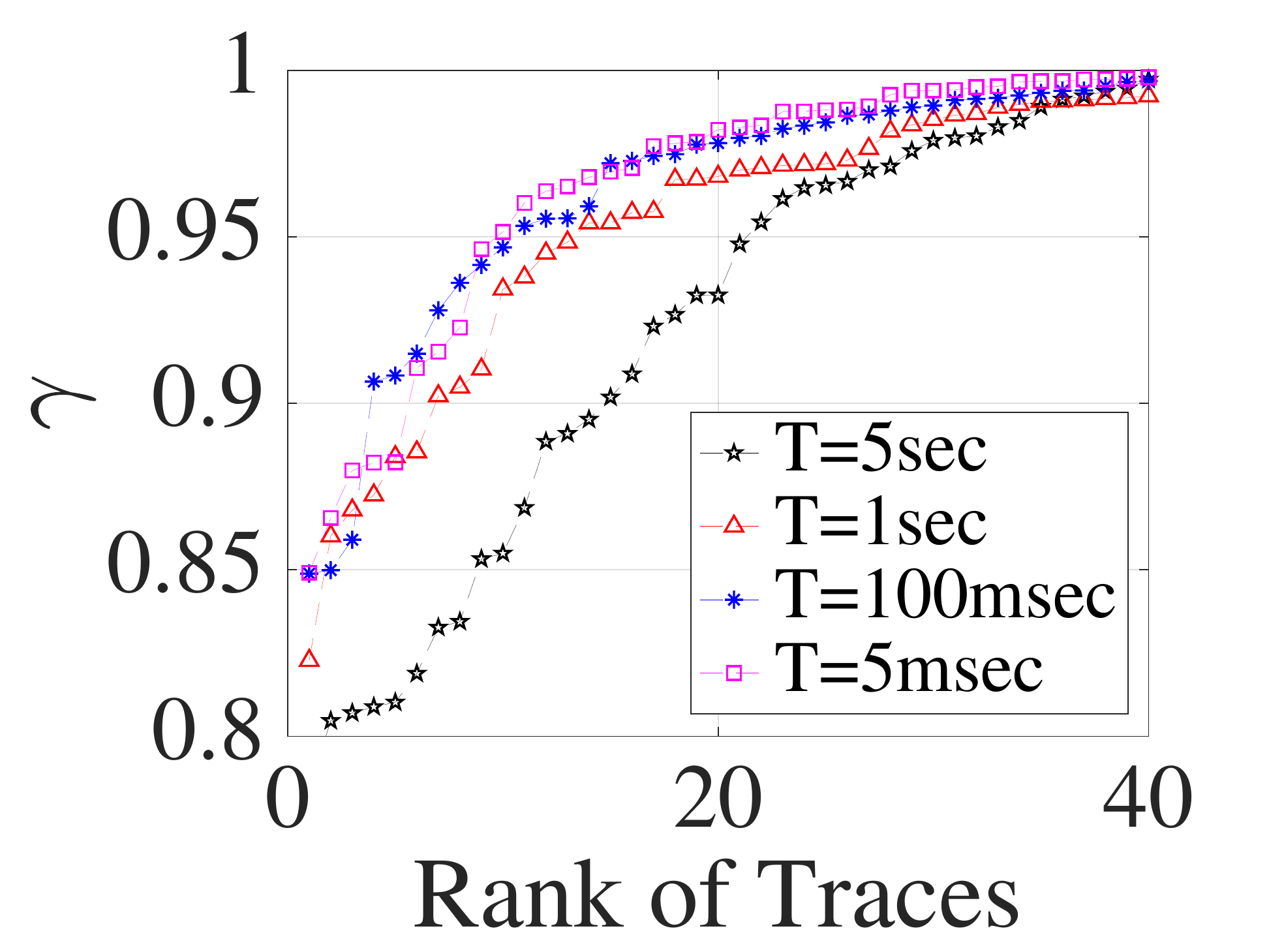}}
	\subcaptionbox{MAWI traces}[0.188\linewidth][c]{%
		\includegraphics[height=3cm, width=3.88cm ]{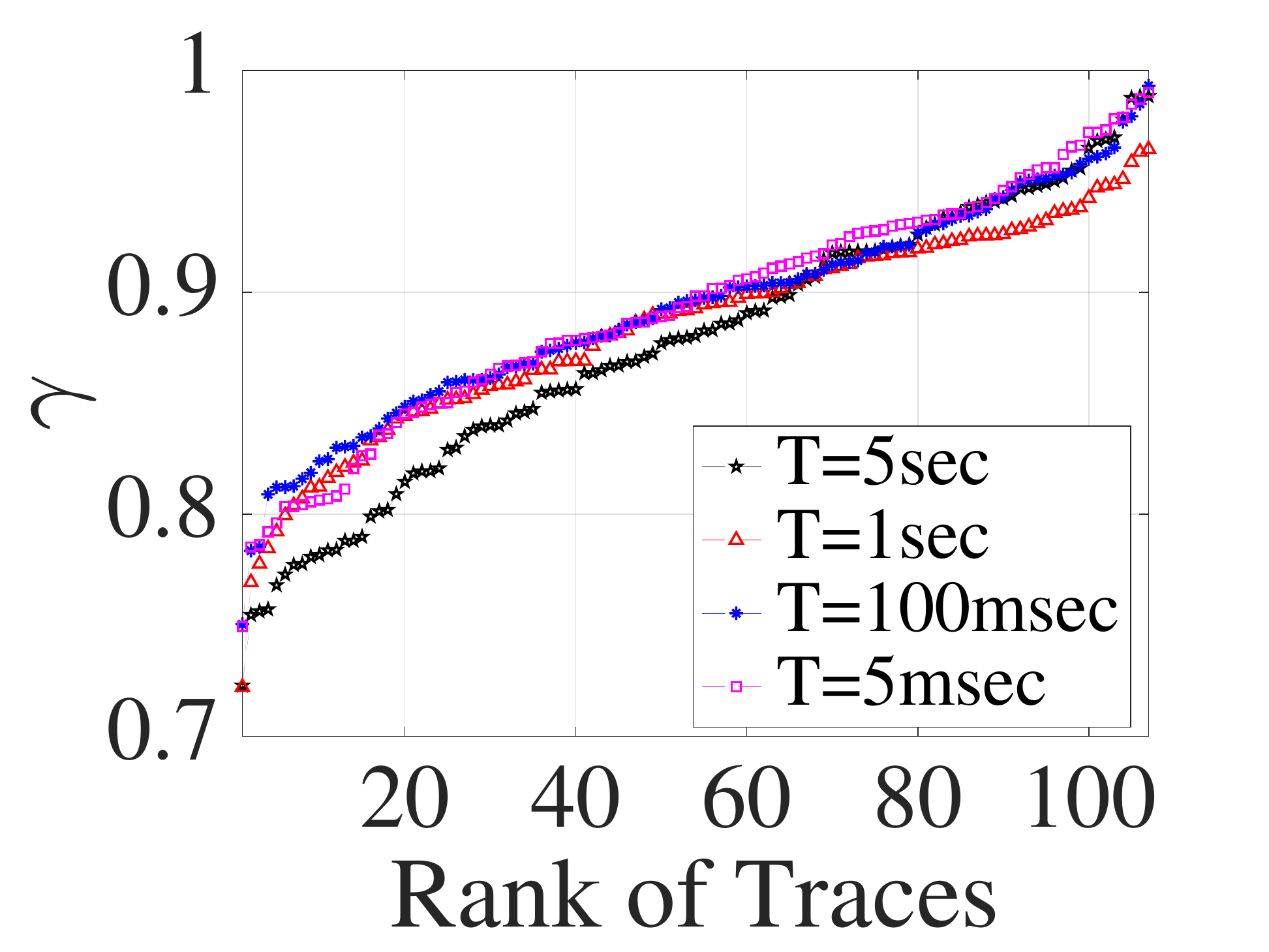}}	
	\bigskip 
	\subcaptionbox{CAIDA traces}[0.187\linewidth][c]{%
		\includegraphics[height=3cm, width=3.88cm ]{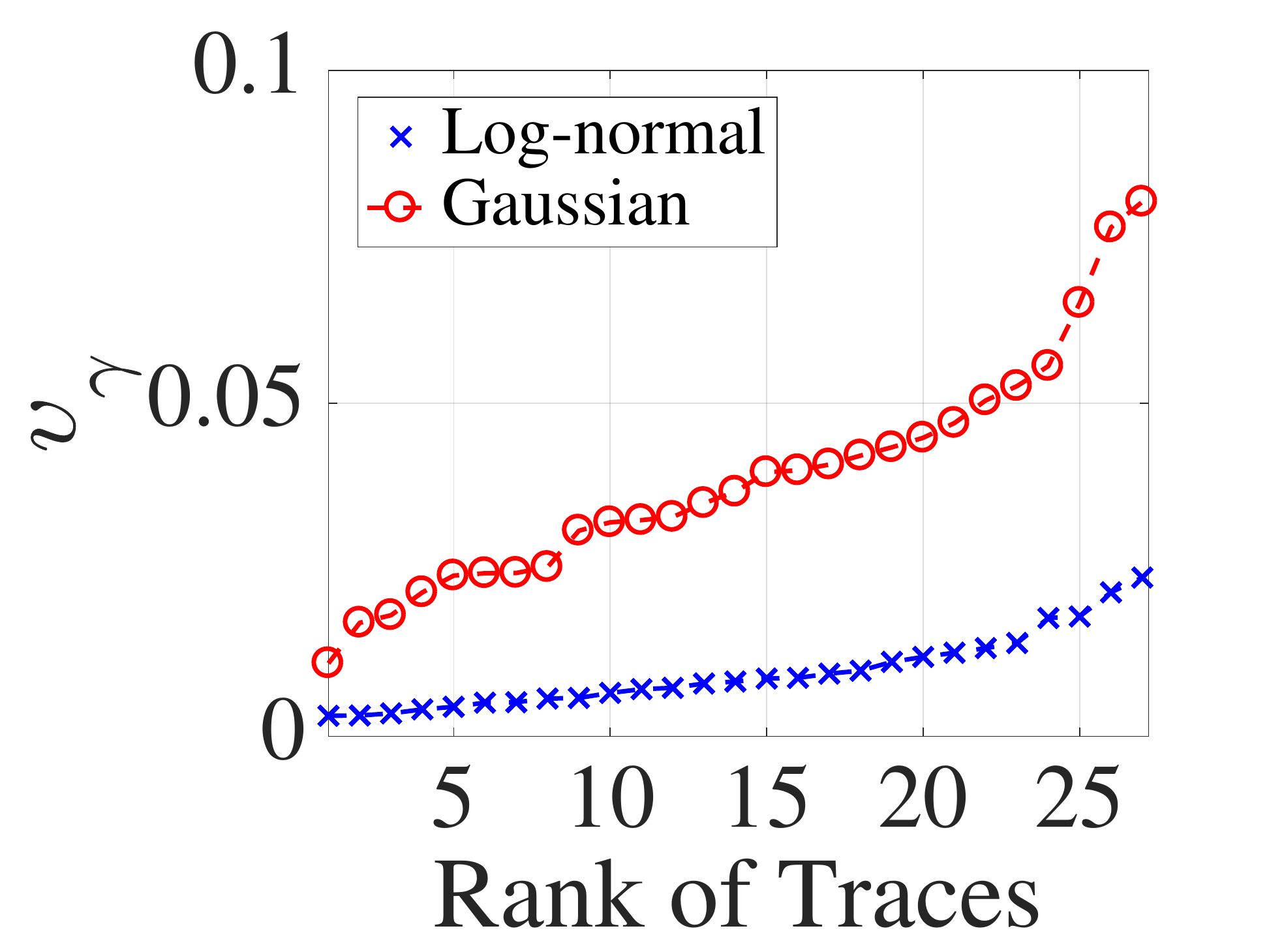}}\quad
	\subcaptionbox{Waikato traces}[0.187\linewidth][c]{%
		\includegraphics[height=3cm, width=3.88cm ]{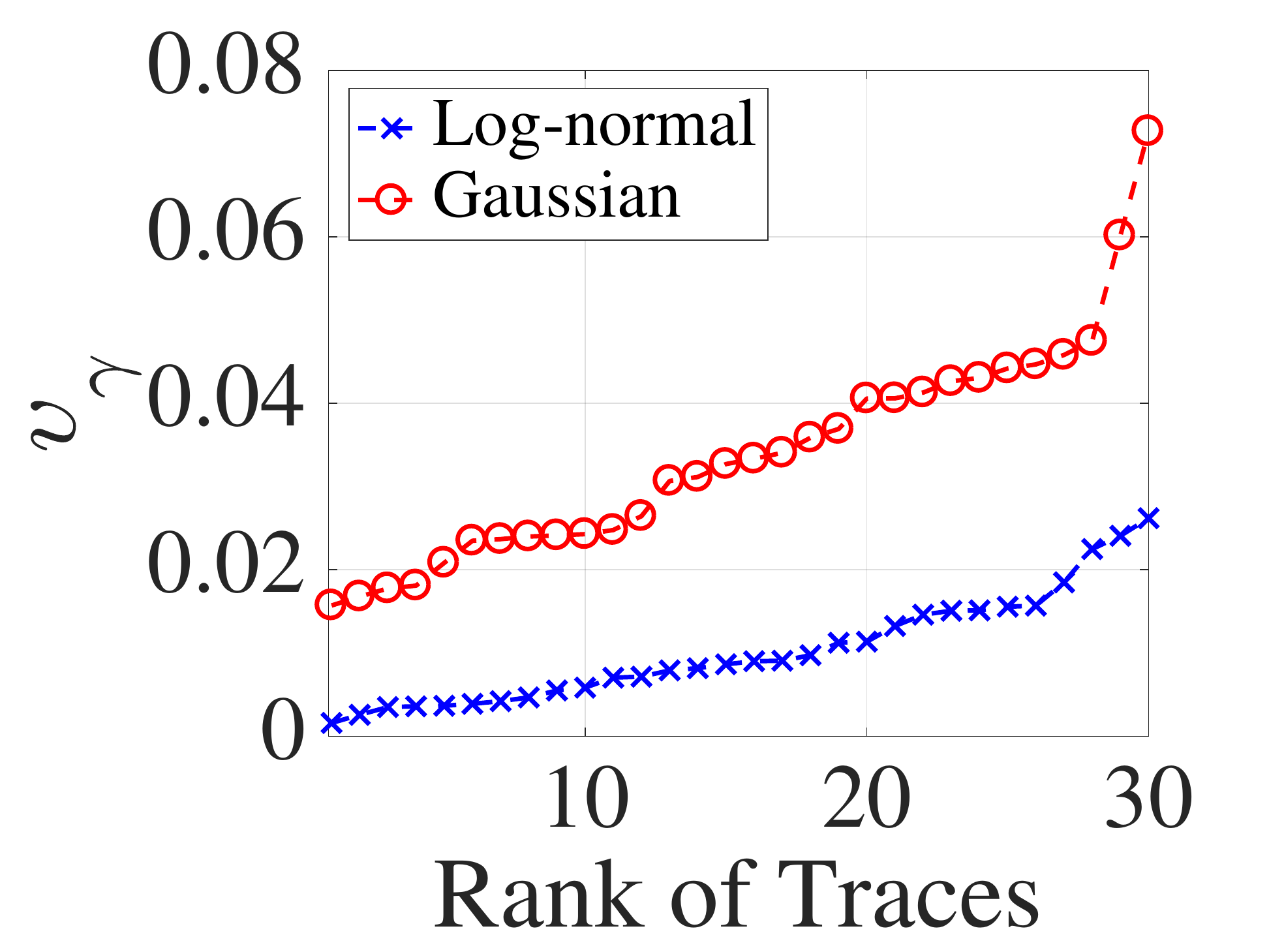}}\quad
	\subcaptionbox{Auckland traces}[0.187\linewidth][c]{%
		\includegraphics[height=3cm, width=3.88cm ]{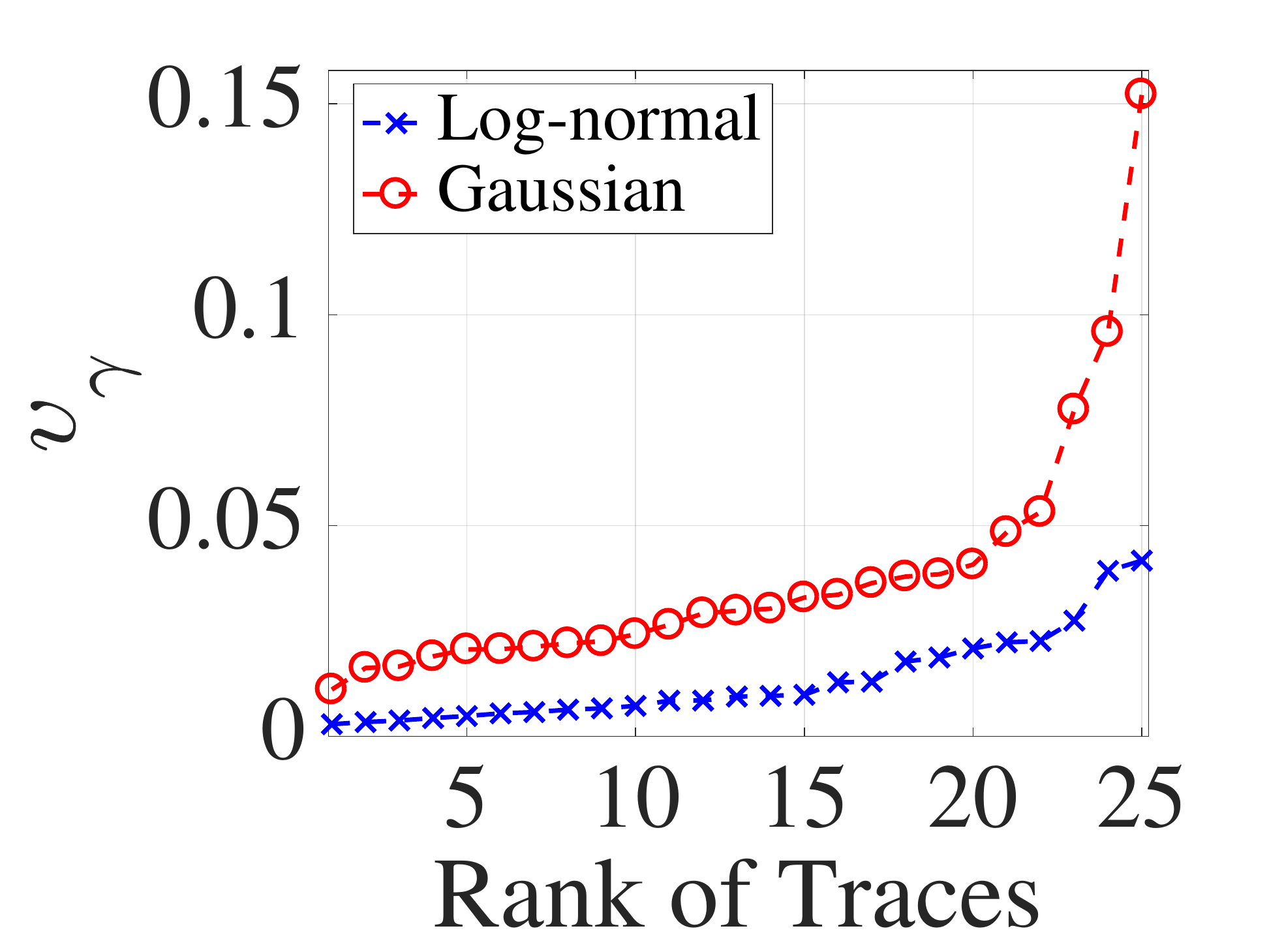}}
	\subcaptionbox{Twente traces}[0.187\linewidth][c]{%
		\includegraphics[height=3cm, width=3.88cm ]{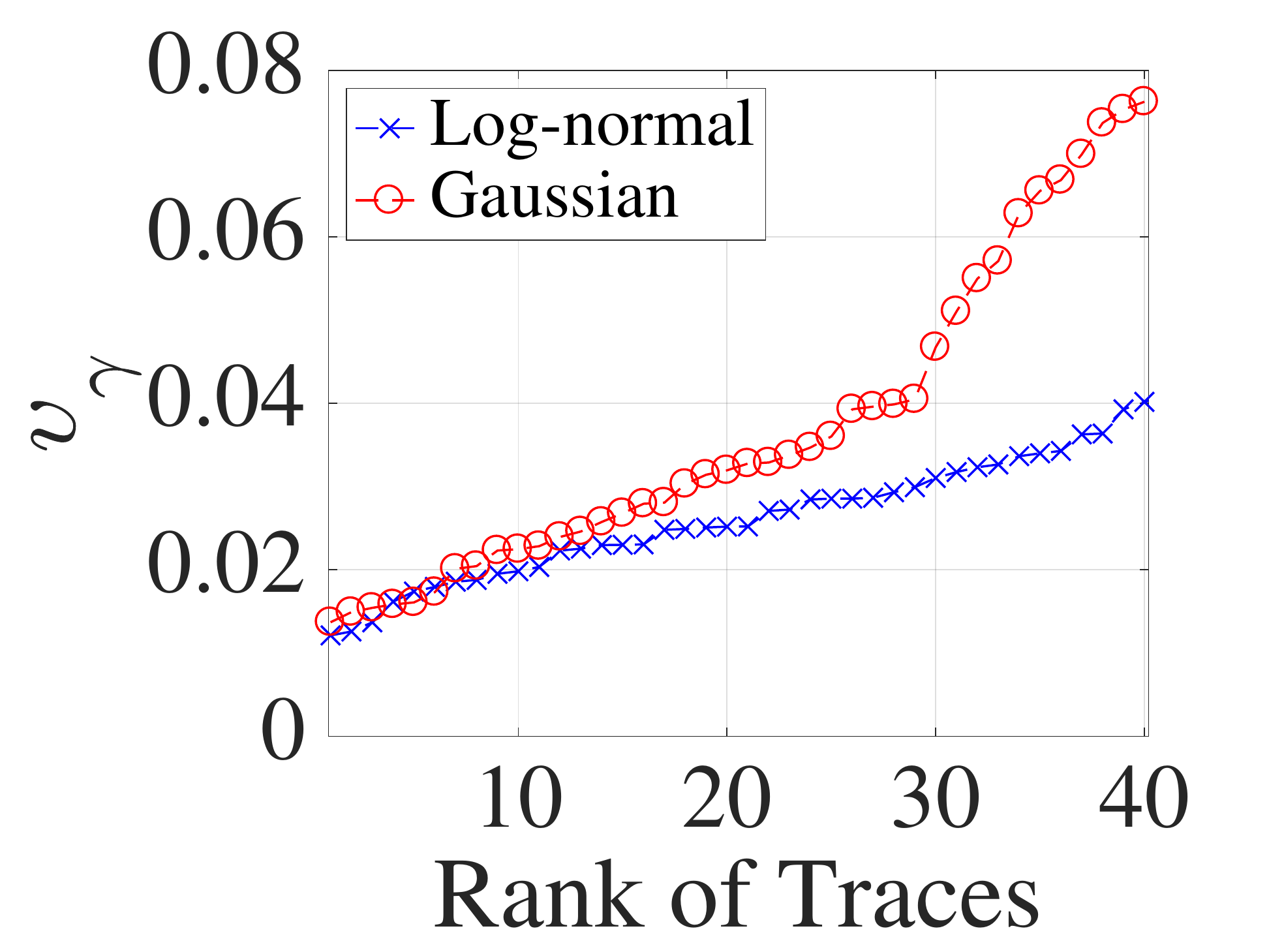}}\quad
	\subcaptionbox{MAWI traces}[0.187\linewidth][c]{%
		\includegraphics[height=3cm, width=3.88cm ]{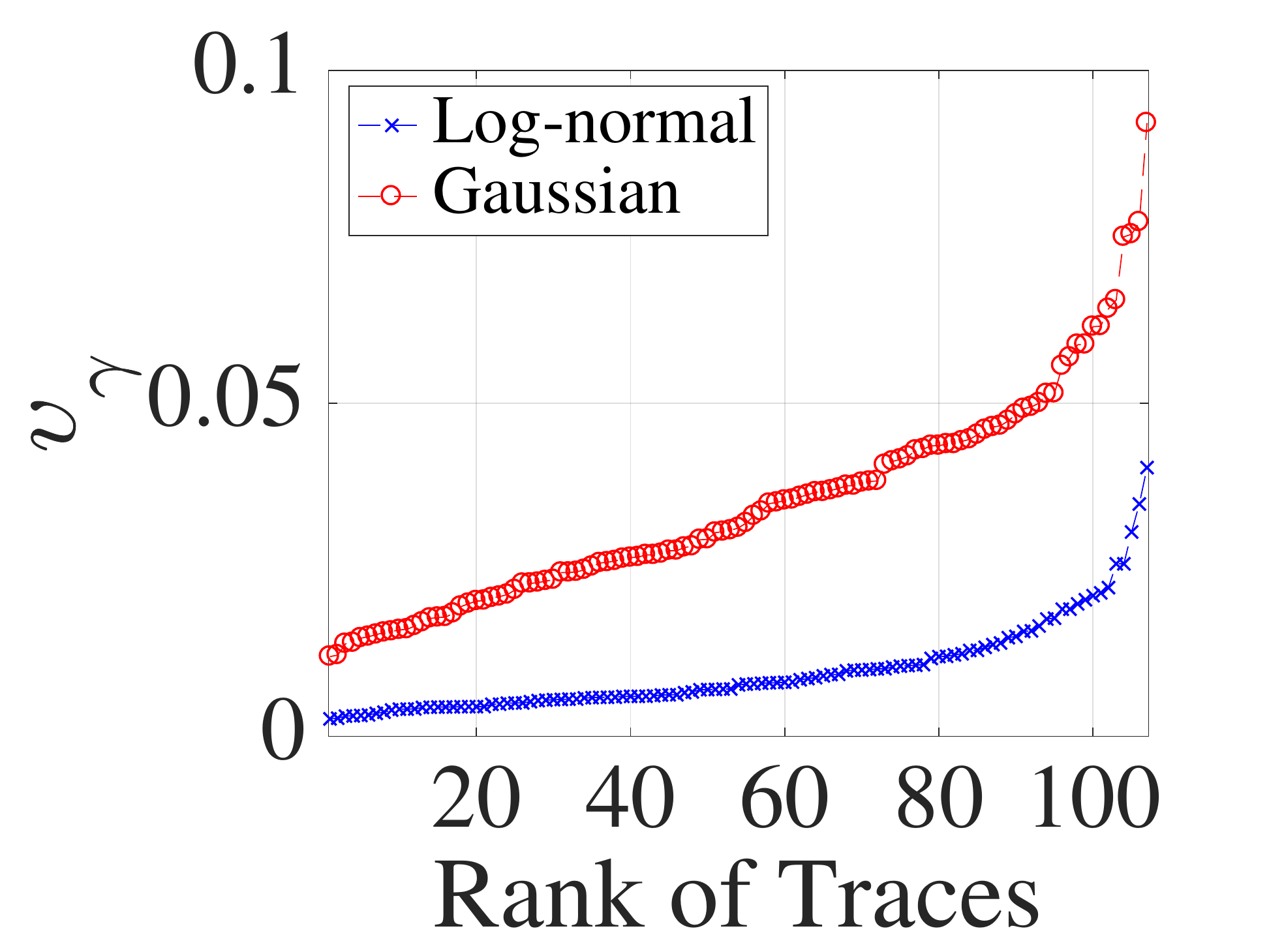}}\quad	
	\bigskip
	\subcaptionbox{CAIDA traces }[.186\linewidth][c]{%
		\includegraphics[height=3cm, width=3.86cm ]{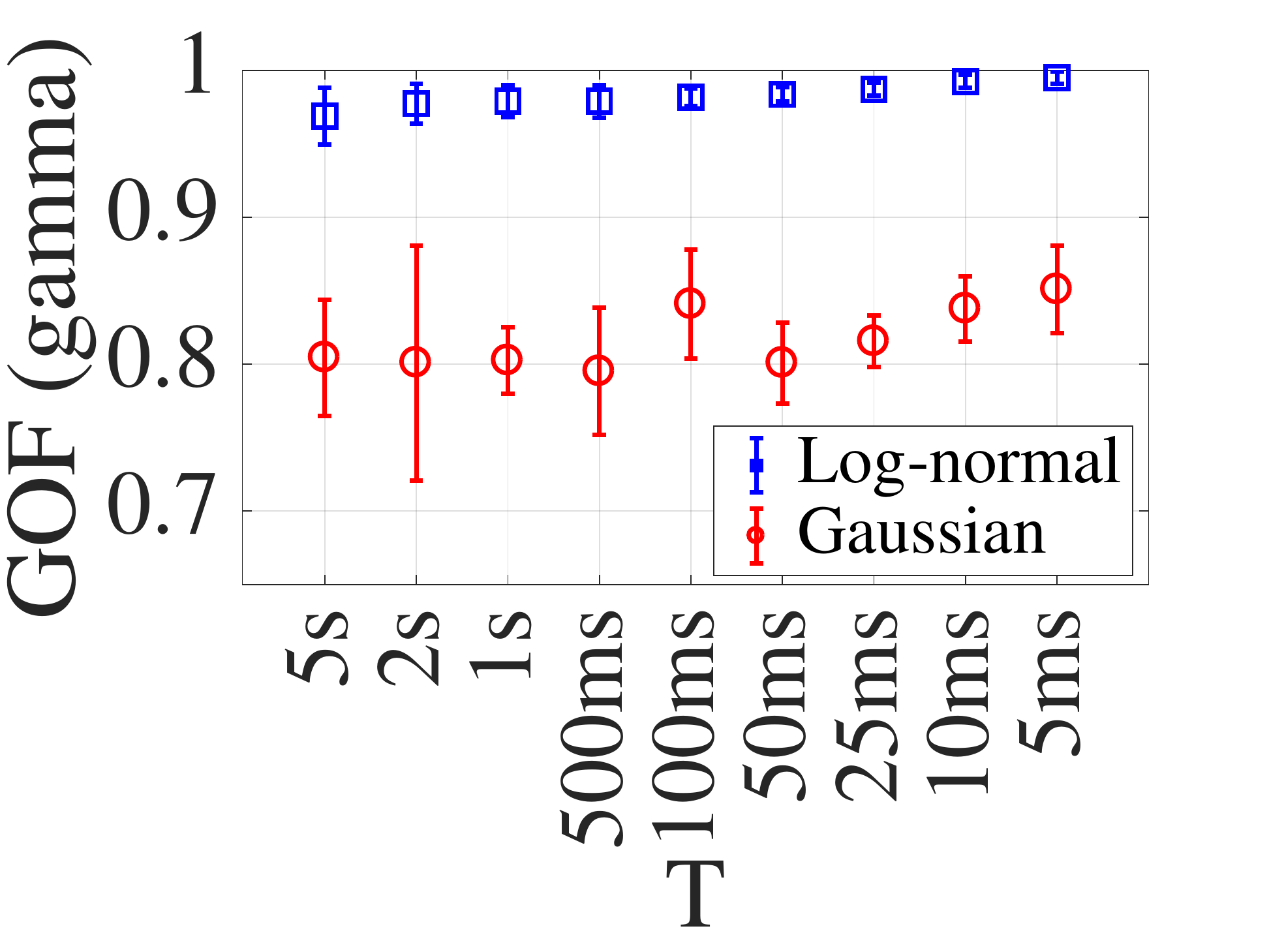}}\quad
	\subcaptionbox{Waikato traces}[.186\linewidth][c]{%
		\includegraphics[height=3cm, width=3.86cm ]{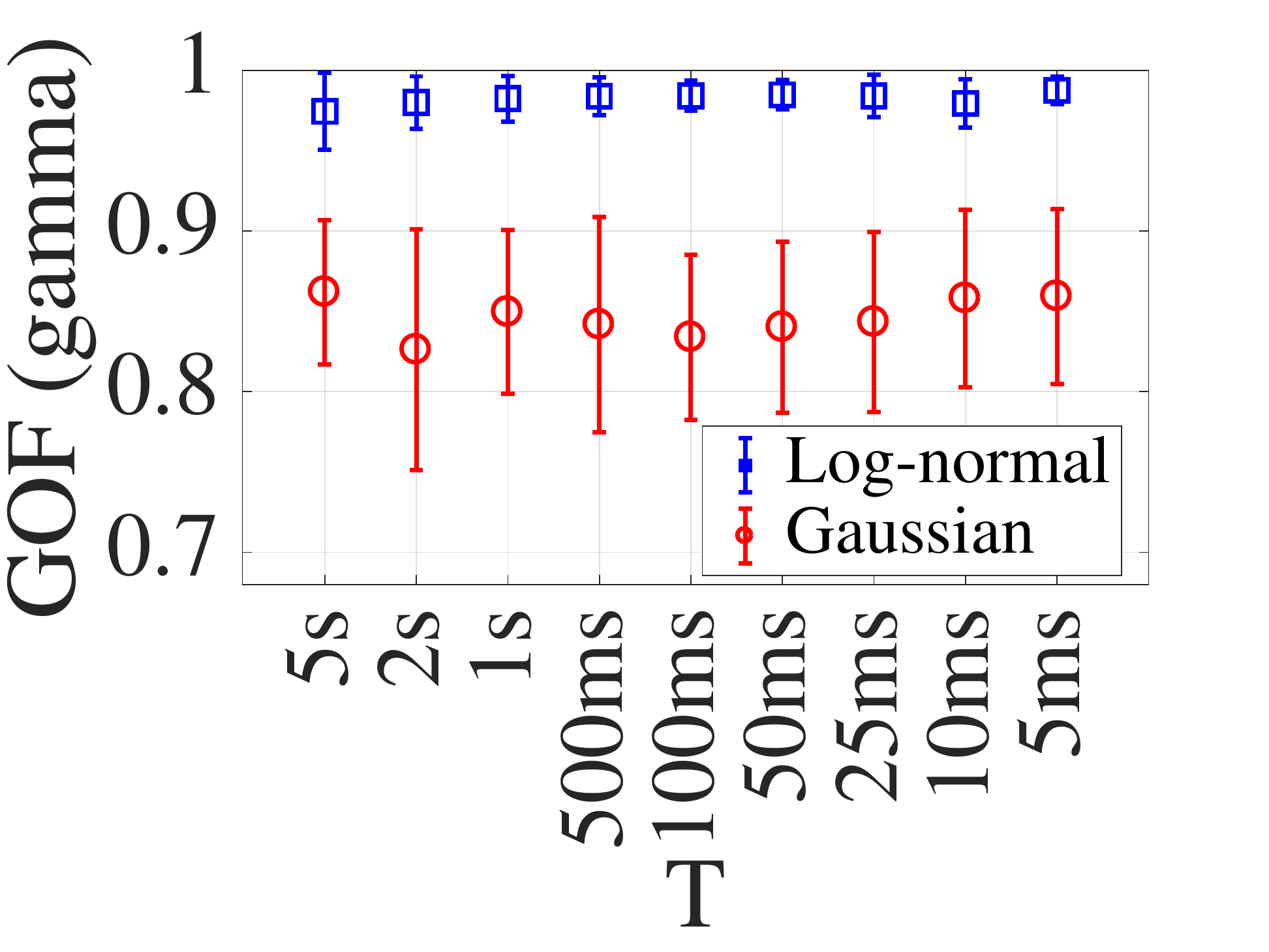}}\quad
	\subcaptionbox{Auckland traces}[.188\linewidth][c]{%
		\includegraphics[height=3cm, width=3.86cm ]{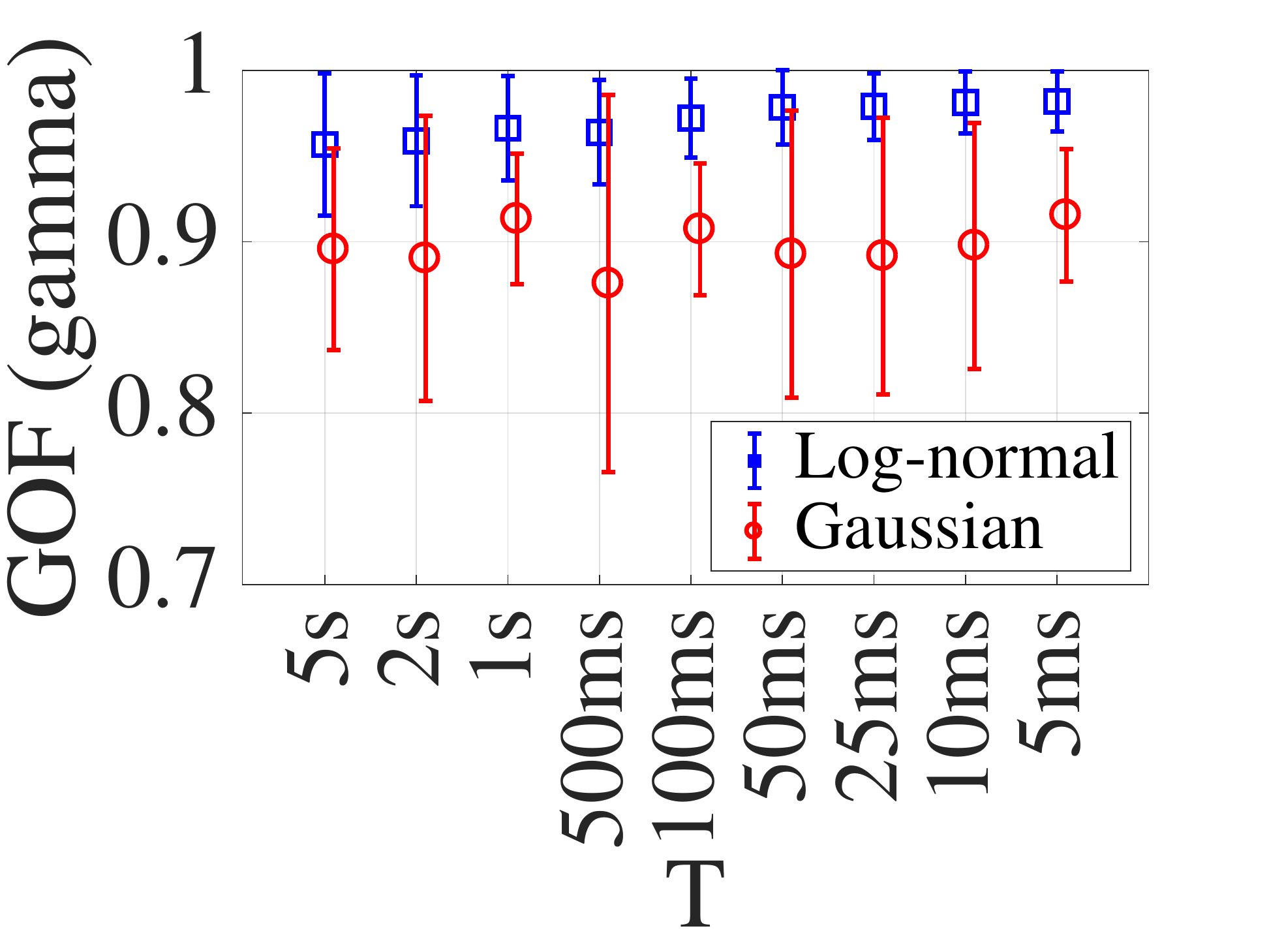}}
	\subcaptionbox{Twente traces}[.186\linewidth][c]{%
		\includegraphics[height=3cm, width=3.86cm ]{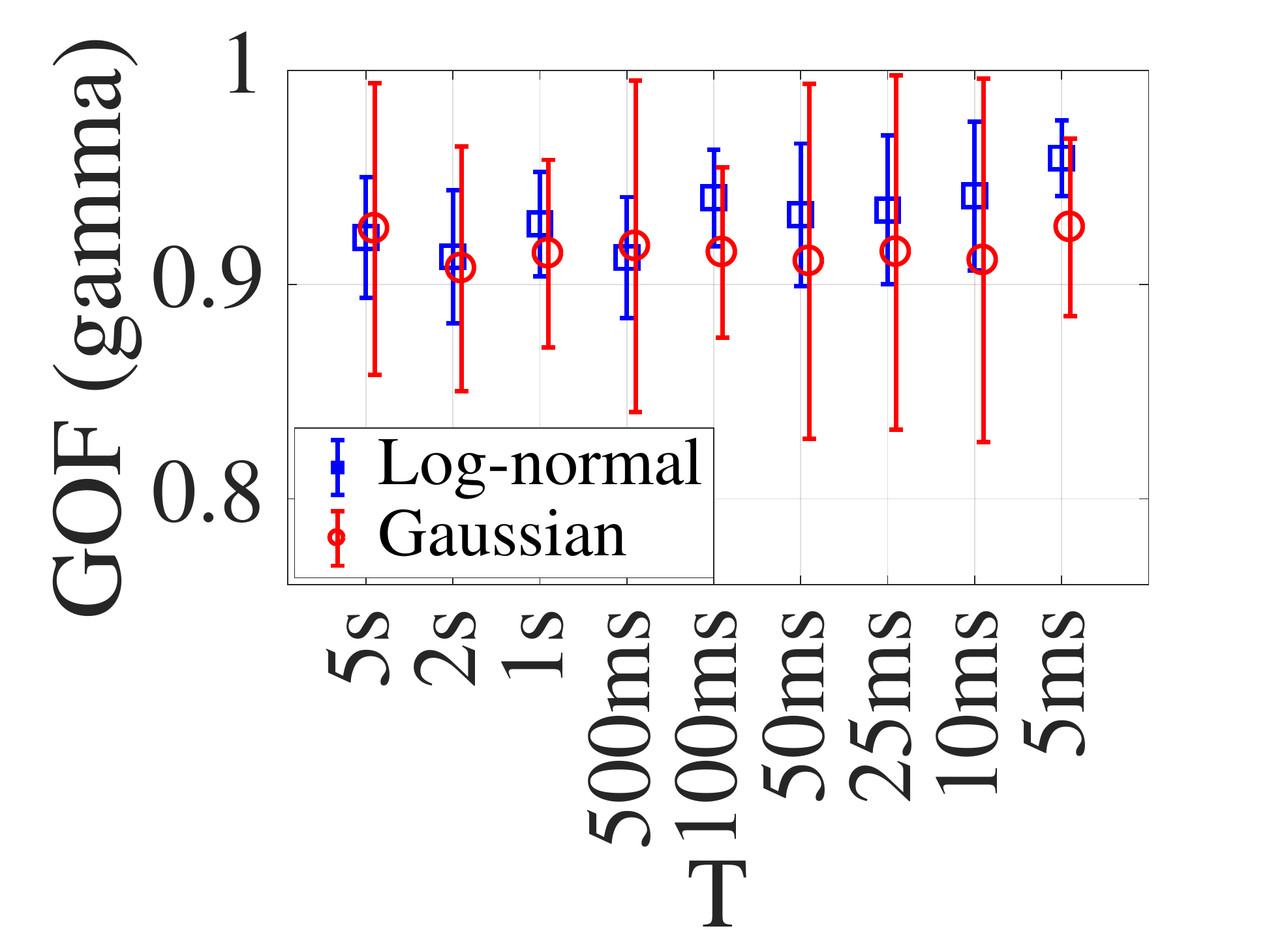}}\quad
	\subcaptionbox{MAWI traces}[.186\linewidth][c]{%
		\includegraphics[height=3cm, width=3.86cm ]{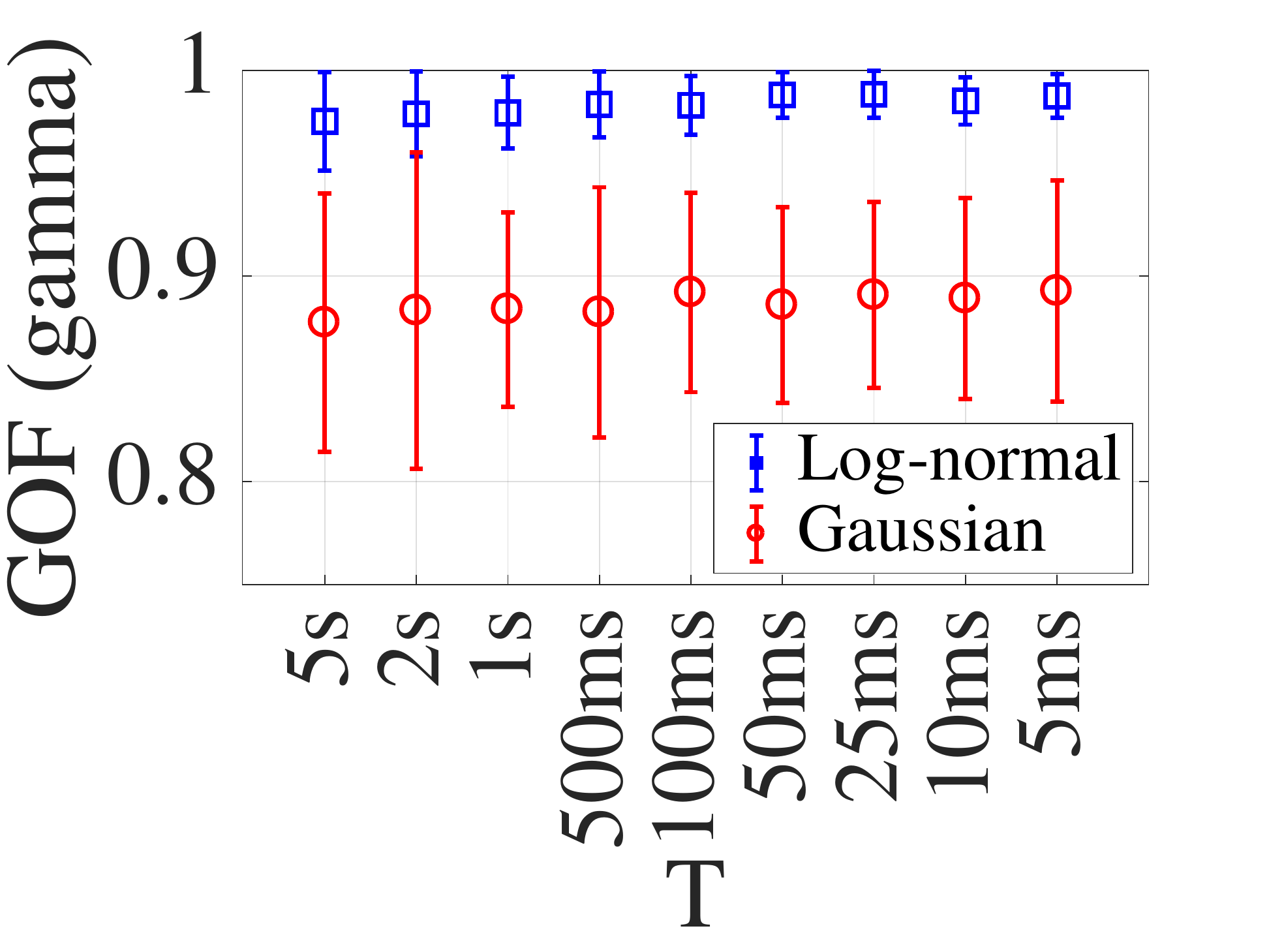}}\quad	
	\caption{Correlation coefficient test results for all studied traces and different timescales.}
	\label{gammaRes} 	
\end{figure*}

The linear correlation coefficient test has been widely used to assess the fit of a distribution to empirical data. To reinforce the results of Section~\ref{fit-log-normal}, we employ the linear correlation coefficient assuming that the log-normal distribution is the best fit (as we showed in Section \ref{sec:fitting}). We compare the results of this test for both the log-normal and Gaussian distributions. We use the linear correlation coefficient as defined in~\cite{sigcomm2002nonGaussian}:
\begin{equation}
 \gamma = \frac{\sum_{i=1}^{n} \left ( S_{(i)}- \hat{\mu} \right )\left ( x_{i}-\hat{x} \right )}{\sqrt{\sum_{i=1}^{n}\left ( S_{(i)}-\hat{\mu} \right )^{2}.\sum_{i=1}^{n}\left ( x_{i}-\hat{x} \right )^{2}}} 
\end{equation}
where $S_{(i)}$ is the observed sample $i$, and $\hat{\mu}=\frac{1}{n}\sum_{i=1}^{n}S_{(i)}$ is the samples' mean value. $x_{i}$ is sample $i$ from the reference distribution (log-normal in our case), which can be calculated from the inverse CDF of the reference random variable $x_{i} =F^{-1}\left ( \frac{i}{n+1} \right )$ and $\hat{x}=\frac{1}{n}\sum_{i=1}^{n}x_{i}$ is the respective mean value. The value of the correlation coefficient can vary between $-1\leq \gamma \leq 1$, with a $1$, $0$ and $-1$ indicating perfect correlation, no correlation and perfect anti-correlation, respectively. Strong goodness-of-fit (GOF) is assumed to exist when the value of $\gamma$ is greater than $0.95$~\cite{ResDimension}.
 
We measure the linear correlation coefficient for all datasets at four different aggregation timescales (ranging from 5 msec to 5 sec) and plot the results in Figures~\ref{gammaRes}(a) to~\ref{gammaRes}(e) for the log-normal distribution and Figures~\ref{gammaRes}(f) to~\ref{gammaRes}(j) for the Gaussian distribution. Traces are ordered by the value of $\gamma$ for the given timescale. It can be clearly seen that $\gamma>0.95$ for most traces when employing the test for the log-normal distribution, but this is not the case for the Gaussian distribution. $\gamma$ is larger for smaller aggregation timescales indicating that the log-normal distribution is an even better fit as the aggregation gets finer. For very small values of $T$, i.e. lower than 1 msec, data samples exhibit binary behaviour, where either a packet is transmitted or not during each examined time frame~\cite{transaction2015}. We have examined $\gamma$ for very short (and large) aggregation timescales, and can confirm the absence of a model describing the data (for brevity, we have omitted the relevant figures).

Next, we calculate $\upsilon_{\gamma}$ (the variation of $\gamma$) for each dataset. $\upsilon_{\gamma}$ gives an indication of the stability of $\gamma$ for each dataset, for all timescales tested. This metric is defined as:
\begin{equation}
\upsilon_{\gamma} = \sqrt{  var(\gamma _{T_{1}},\gamma _{T_{2}}, \gamma _{T_{3}},\gamma _{T_{4}} )}  
\end{equation}
where $ {T_{1}}= 5$ sec, $ {T_{2}}= 1$ sec, $ {T_{3}}= 100$ msec and $ {T_{4}}= 5$ msec. Figures~\ref{gammaRes}(k) to~\ref{gammaRes}(o) show the results for each dataset with the traces ranked by $\upsilon_{\gamma}$. For log-normal model, $\upsilon_{\gamma}$ is very small (below $0.045$) for all traces, therefore we can conclude that $\gamma$ is almost constant for all studied aggregation timescales. While $\upsilon_{\gamma}$ is higher for Gaussian model. Furthermore, the error bars in Figures~\ref{gammaRes}(p) to~\ref{gammaRes}(t) represent the standard deviation of the correlation coefficient at different timescales (see x-axis). This again shows that for log-normal model $\gamma$ is larger than $0.95$ (at different T values) for most CAIDA and MAWI traces, while it is larger than $0.9$ for all other datasets. This is not the case with the Gaussian model, where most ${\gamma}$ values are less than $0.9$. 
 
Overall, the correlation coefficient test reinforces the results extracted in Section~\ref{fit-log-normal}, providing strong evidence that the log-normal distribution is the best fit for all studied traces. Superior
performance of our model can also be seen from comparison of our
results for correlation coefficient with those in \cite{ieee-network-2009} where the Gaussian
model was used.

\section{Bandwidth Provisioning}
\label{sec:provision}

It has been previously suggested that network link provisioning could be based on fitted traffic models instead of relying on straightforward empirical rules~\cite{ieee-network-2009}. In this way, over- or under-provisioning can be mitigated or eliminated even in the presence of strong traffic fluctuations. Such approaches rely on having a statistical model that accurately describes the network traffic. This is therefore an excellent area for applying our findings on fitting the log-normal distribution to Internet traffic data. In the literature, the following inequality (the authors call it the ``link transparency formula") has been used for bandwidth provisioning~\cite{transaction2015}:
\begin{equation}
  P\left ( A(T)\geq CT \right ) \leq \varepsilon.    \label{link-tran}  
\end{equation}
  
In words, this inequality states that the probability that the captured traffic $A(T)$ over a specific aggregation timescale  $T$ is larger than the link capacity has to be smaller than the value of a performance criterion $\varepsilon$. The value of $\varepsilon$ is chosen carefully by the network provider in order to meet a specific SLA~\cite{ieee-network-2009}. Likewise, the value of the aggregation time $T$ should be sufficiently small so that the fluctuations in the traffic can be modelled as well, taking into account the buffering capabilities of network switching devices\footnote{Large traffic fluctuations at very short aggregation timescales are smoothed by the presence of buffers at network routers and switches.}.

We compare bandwidth provisioning  using Meent's approximation formula~\cite{ieee-network-2009} (assuming Gaussian) and using a log-normal traffic model.

\begin{figure}[!ht]
	\centering
	\includegraphics[scale=0.28]{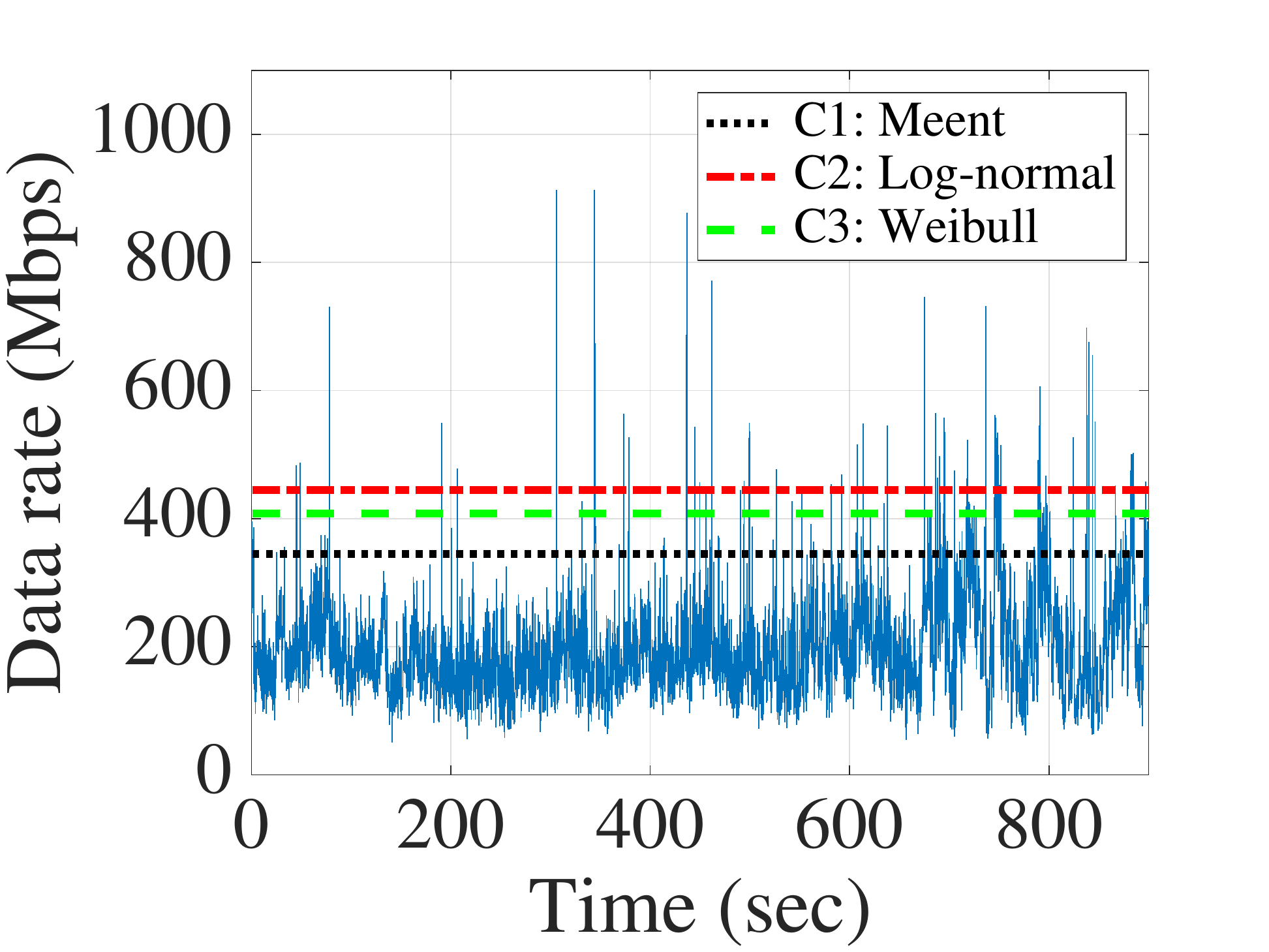}\quad
	\caption{Data rate of a MAWI trace ($T=100$ msec and $\varepsilon=0.01$). The horizontal lines represent the calculated link capacity based on different models.}
	\label{time-domain} 	
\end{figure}

\subsection{Bandwidth provisioning using Meent's formula} 
To find the minimum required link capacity, Meent et al.~\cite{ieee-network-2009} proposed a bandwidth provisioning approach that is based on the assumption that the traffic follows a Gaussian distribution. Meent's dimensioning formula is defined as follows~\cite{ieee-network-2009}:
\begin{equation}
\  C1=\mu +\frac{1}{T} \sqrt{-2log(\varepsilon) .\upsilon (T)}   \label{meents-euq}
\end{equation}
where $\mu$ is the average value of the traffic, $\upsilon (T)$ is the variance at timescale $T$ and $\varepsilon$ is the performance criterion. The link capacity is obtained by adding a safety margin value 
\[
\textrm{Safety margin = }\sqrt{-2log(\varepsilon)}  \textrm{ . } \sqrt{\frac{\upsilon (T)}{T^{2}} }
\]
to the average of the captured traffic (see Equation~\ref{meents-euq}). This safety margin value depends on $\varepsilon$ and the ratio $\sqrt{\upsilon (T)/T^{2}}$. As the value of  $\varepsilon$ decreases the safety margin increases. For example, when the value of  $\varepsilon$ decreases from $10^{-2}$ to $10^{-4}$, then  value of the safety margin increases by $40\%$. This is different from conventional link dimensioning methods, where the safety margin is fixed to be 30\% above the average of the presented traffic~\cite{cisco,ieee-network-2009}. Traffic tails are represented using the Chernoff bound, as follows:
\begin{equation}
P  \left  ( A(T)\geq CT  \right )\leq  e^{-SCT}E\left [e^{SA(T)}  \right].
\end{equation}
Here \noindent${E\left [e^{SA(T)} \right ]}$ is the moment generation function (MGF) of the captured traffic $A(T)$.
\begin{figure*}[t]
	\centering
	\subcaptionbox{target $\varepsilon = 0.5$}[.23\linewidth][c]{%
		\includegraphics[width=.22\linewidth]{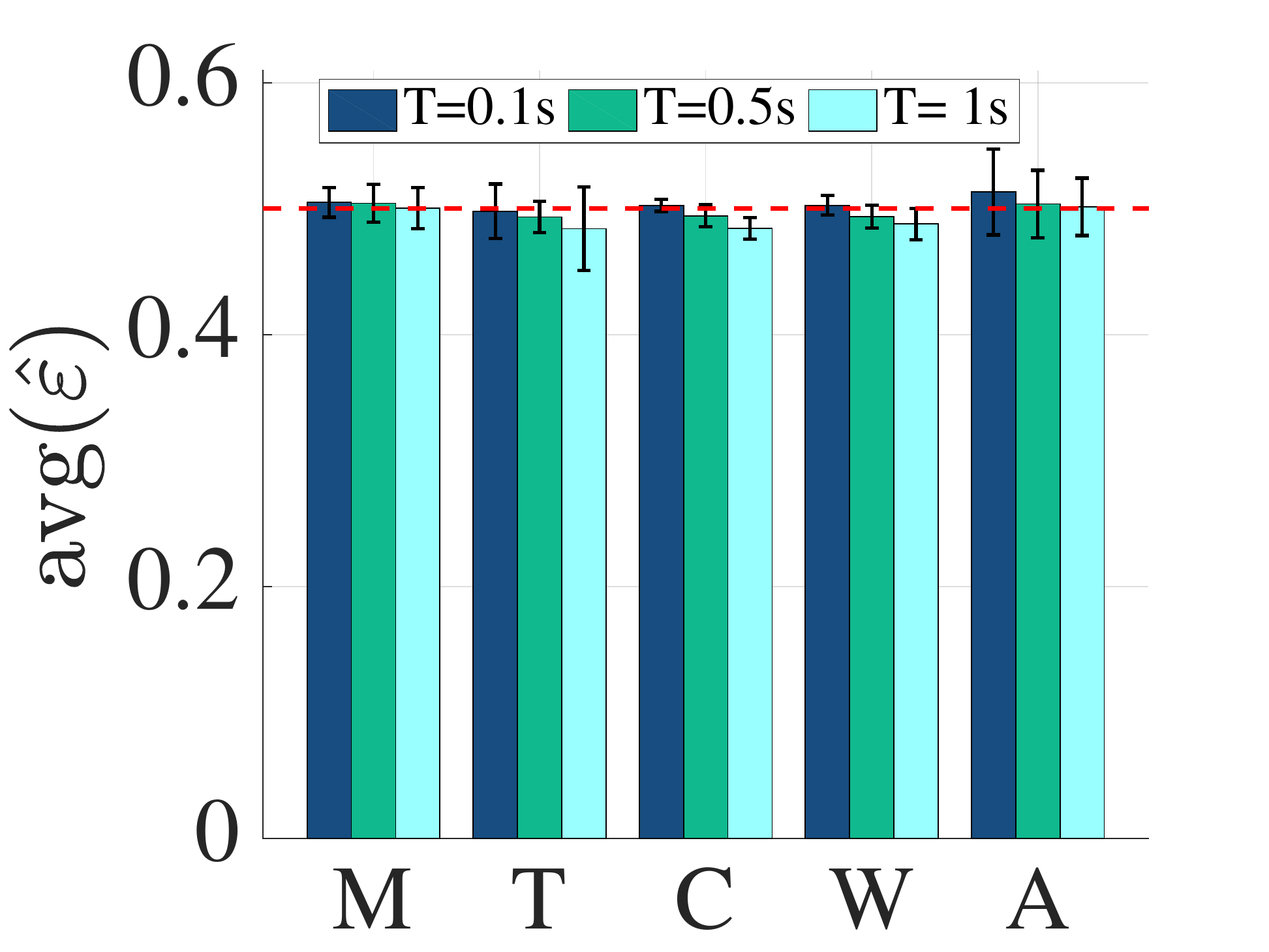}}\quad
	\subcaptionbox{target $\varepsilon = 0.1$}[.23\linewidth][c]{%
		\includegraphics[width=.22\linewidth]{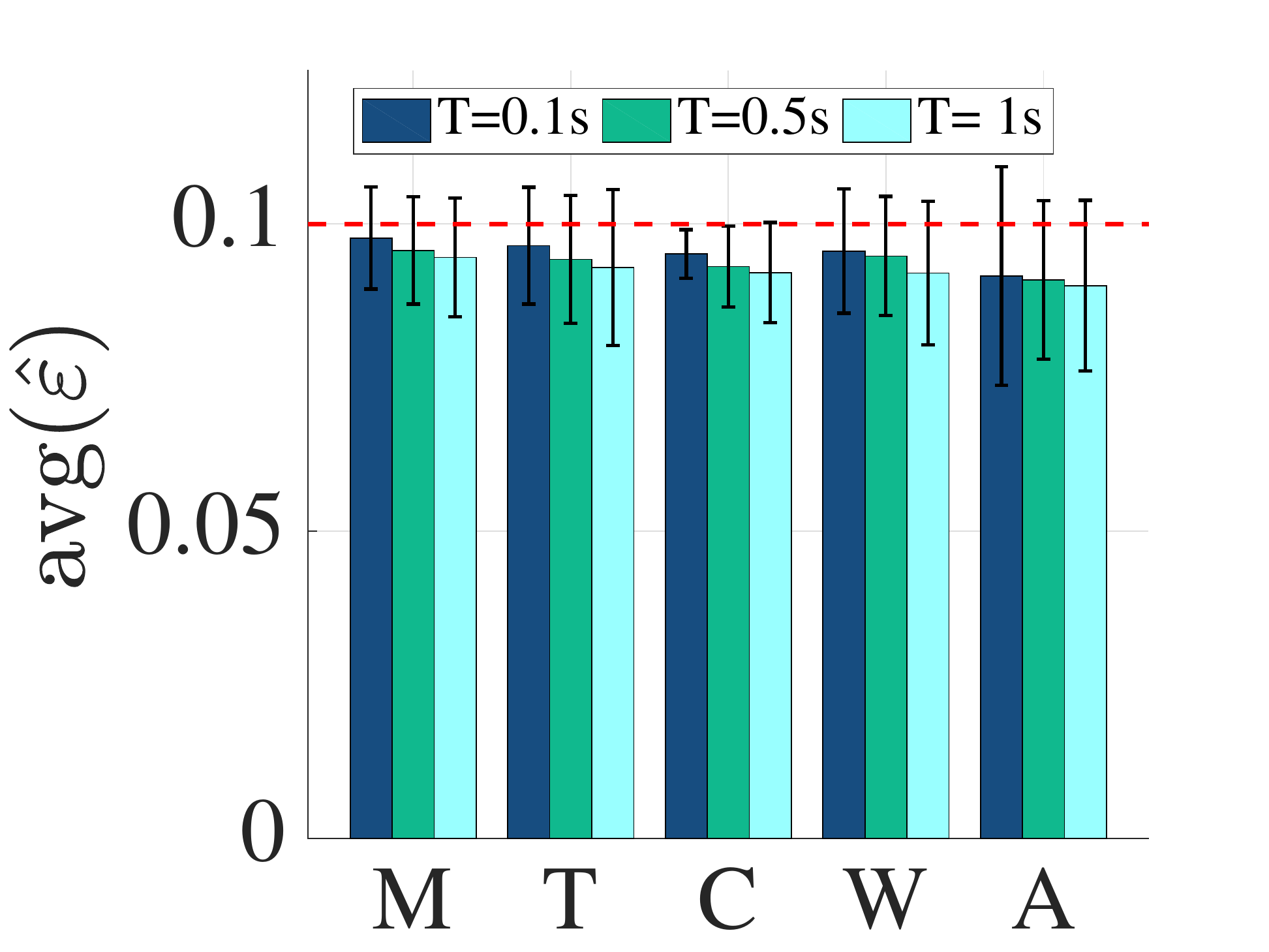}}\quad
	\subcaptionbox{target $\varepsilon = 0.05$}[.23\linewidth][c]{%
		\includegraphics[width=.22\linewidth]{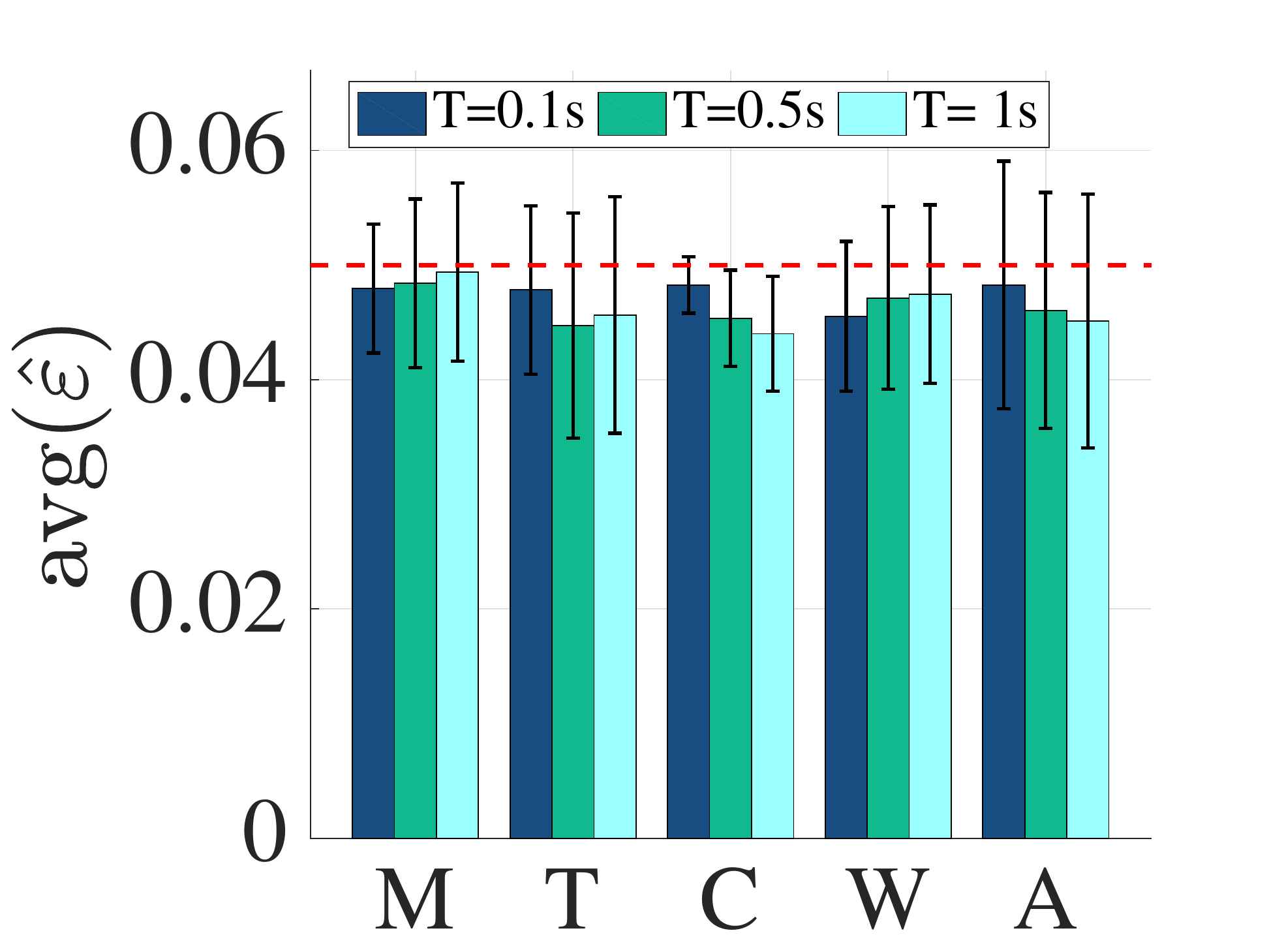}}
	\subcaptionbox{target $\varepsilon = 0.01$}[.23\linewidth][c]{%
		\includegraphics[width=.22\linewidth]{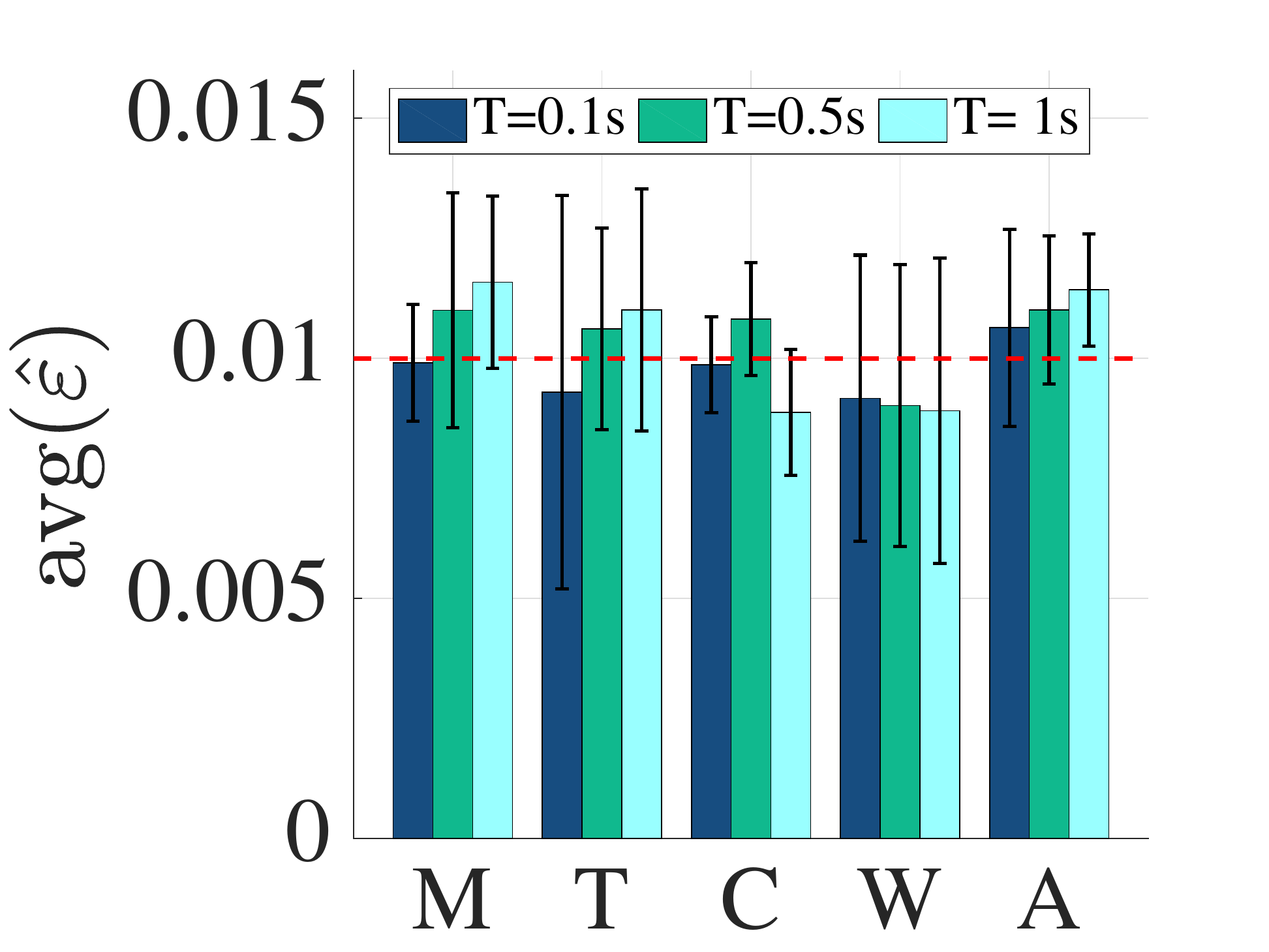}}
 
 
	\centering
	\subcaptionbox{target $\varepsilon = 0.5$}[.23\linewidth][c]{%
		\includegraphics[width=.22\linewidth]{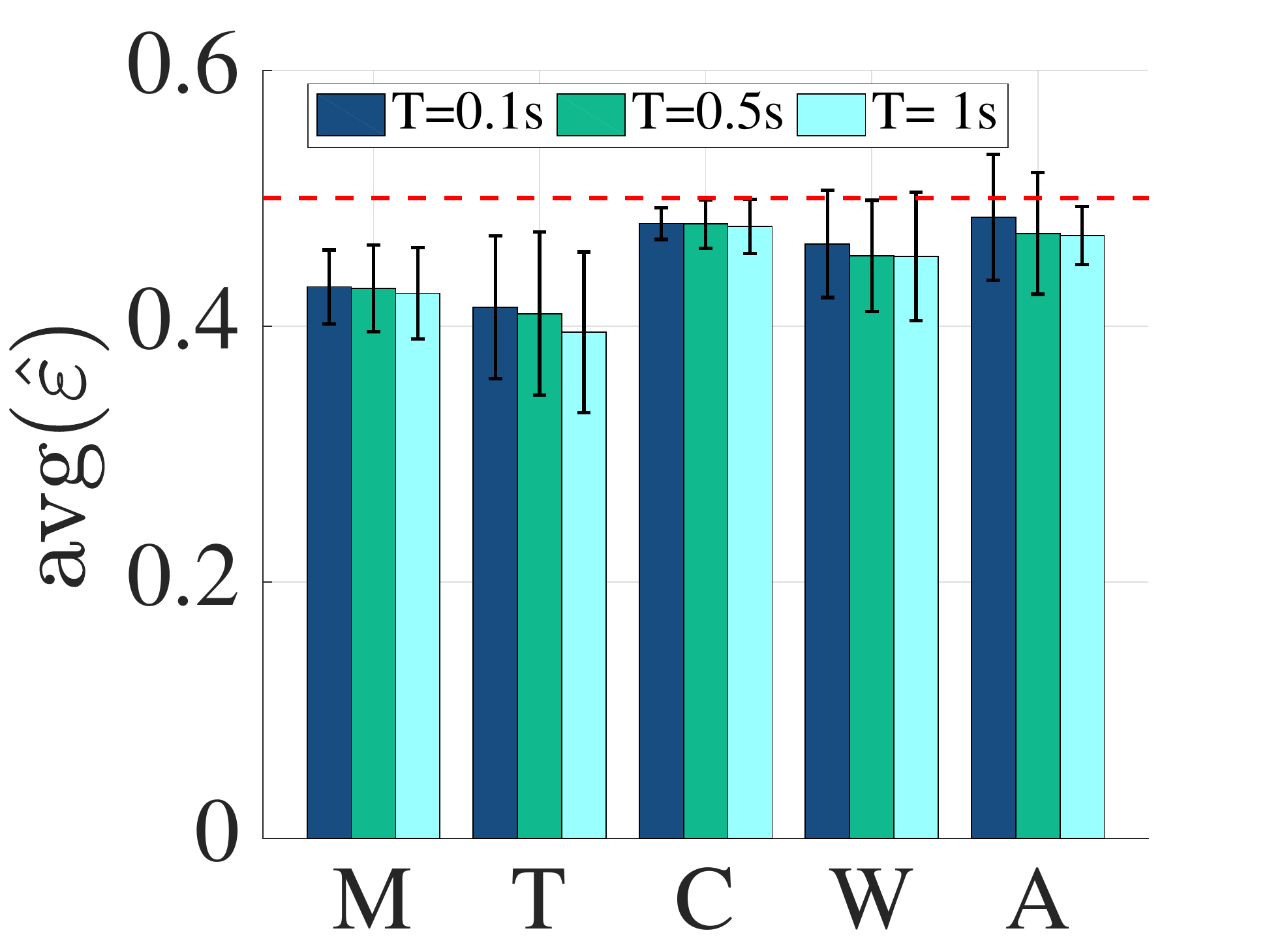}}\quad
	\subcaptionbox{target $\varepsilon = 0.1$}[.23\linewidth][c]{%
		\includegraphics[width=.22\linewidth]{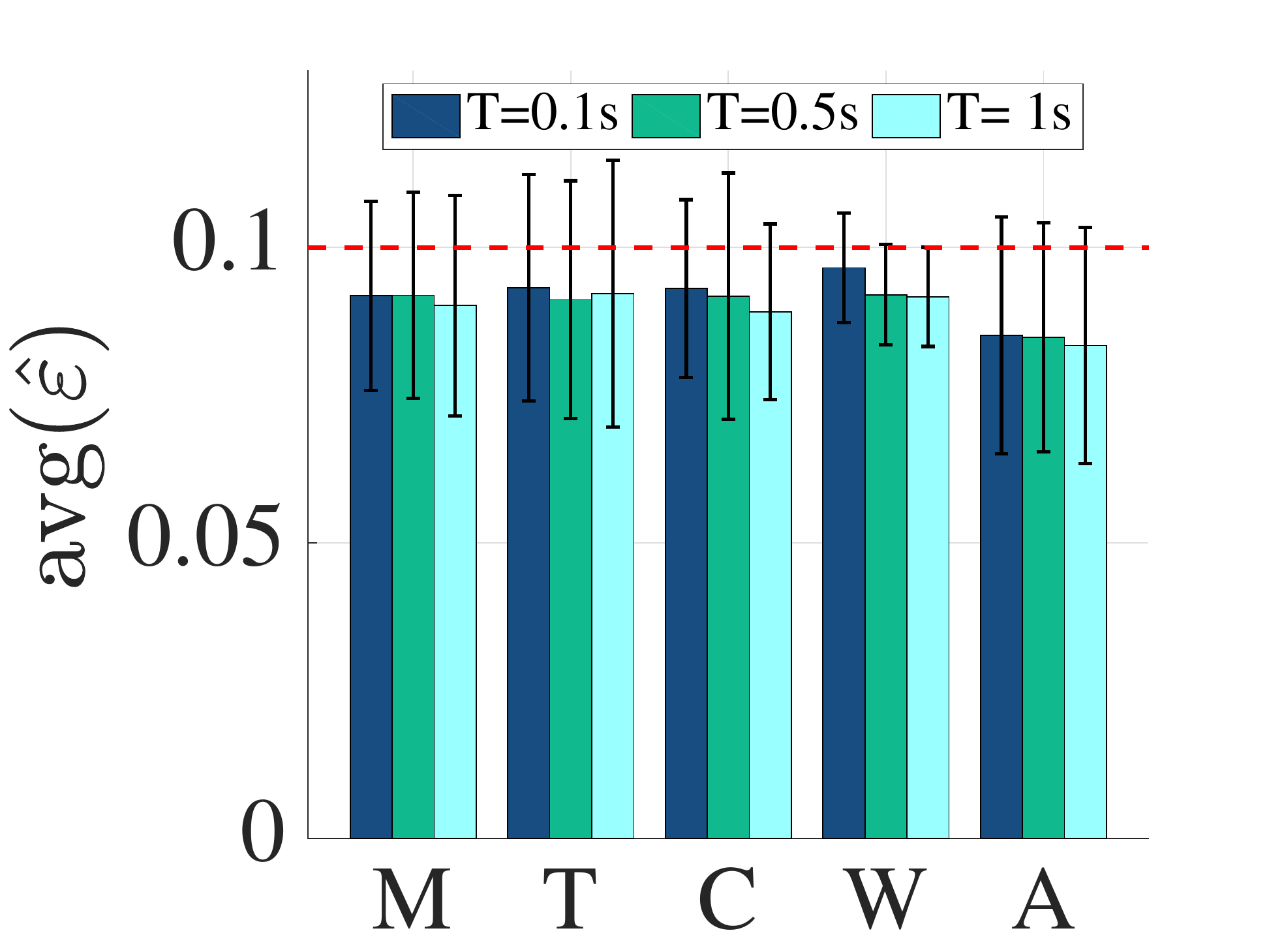}}\quad
	\subcaptionbox{target $\varepsilon = 0.05$}[.23\linewidth][c]{%
		\includegraphics[width=.22\linewidth]{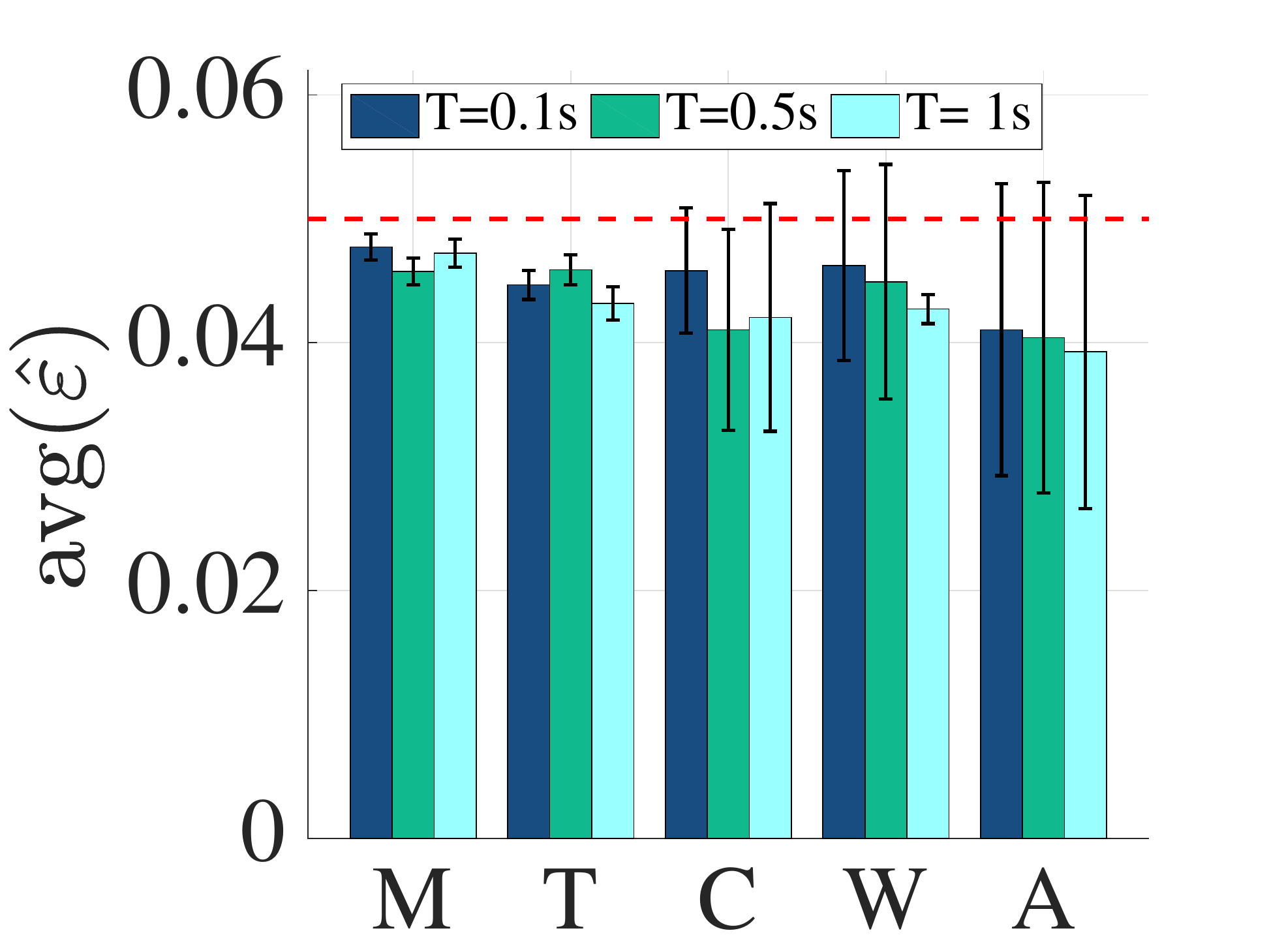}}
	\subcaptionbox{target $\varepsilon = 0.01$}[.23\linewidth][c]{%
		\includegraphics[width=.22\linewidth]{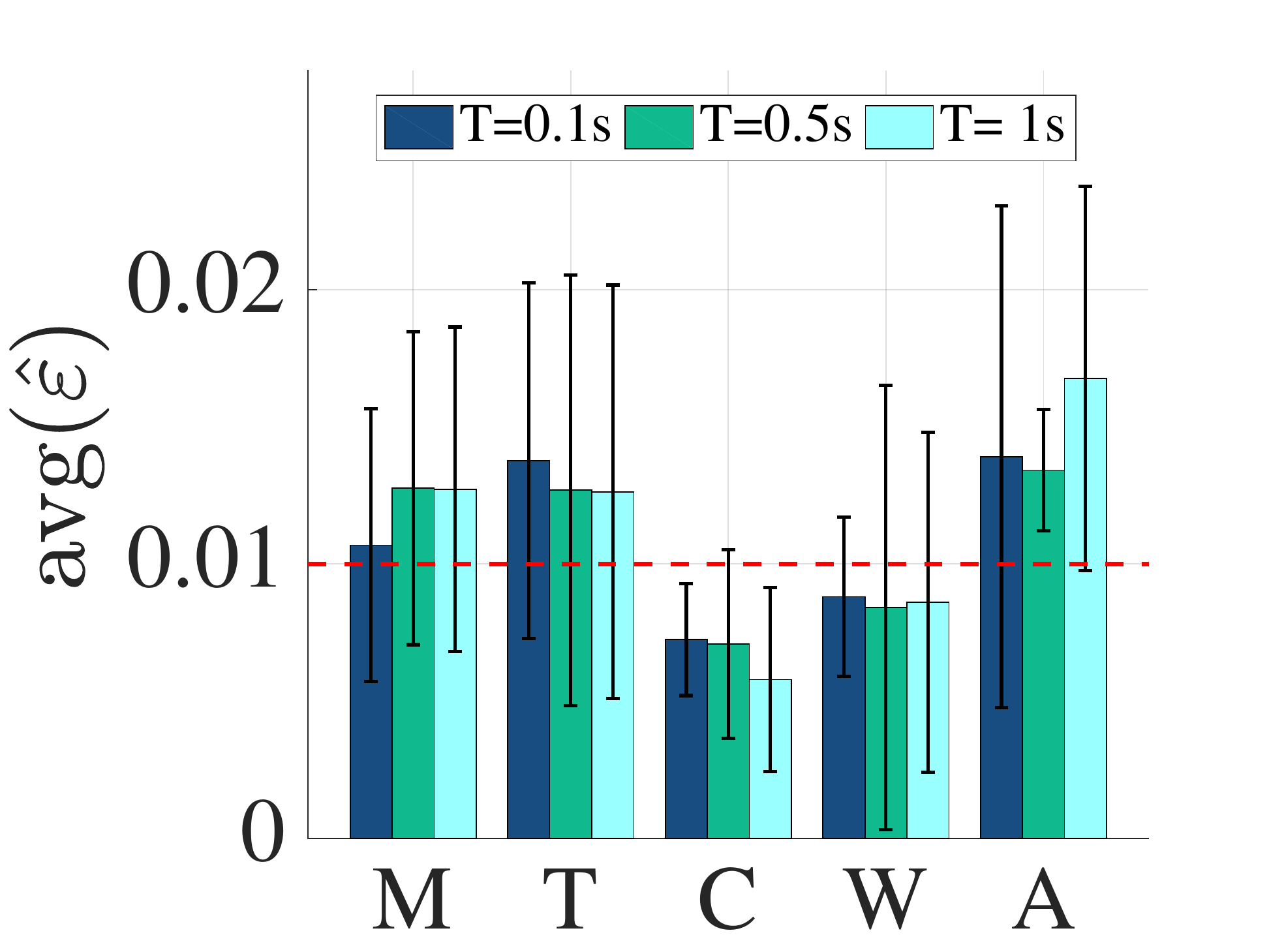}}
 
 
	\centering
	\subcaptionbox{target $\varepsilon = 0.5$}[.23\linewidth][c]{%
		\includegraphics[width=.22\linewidth]{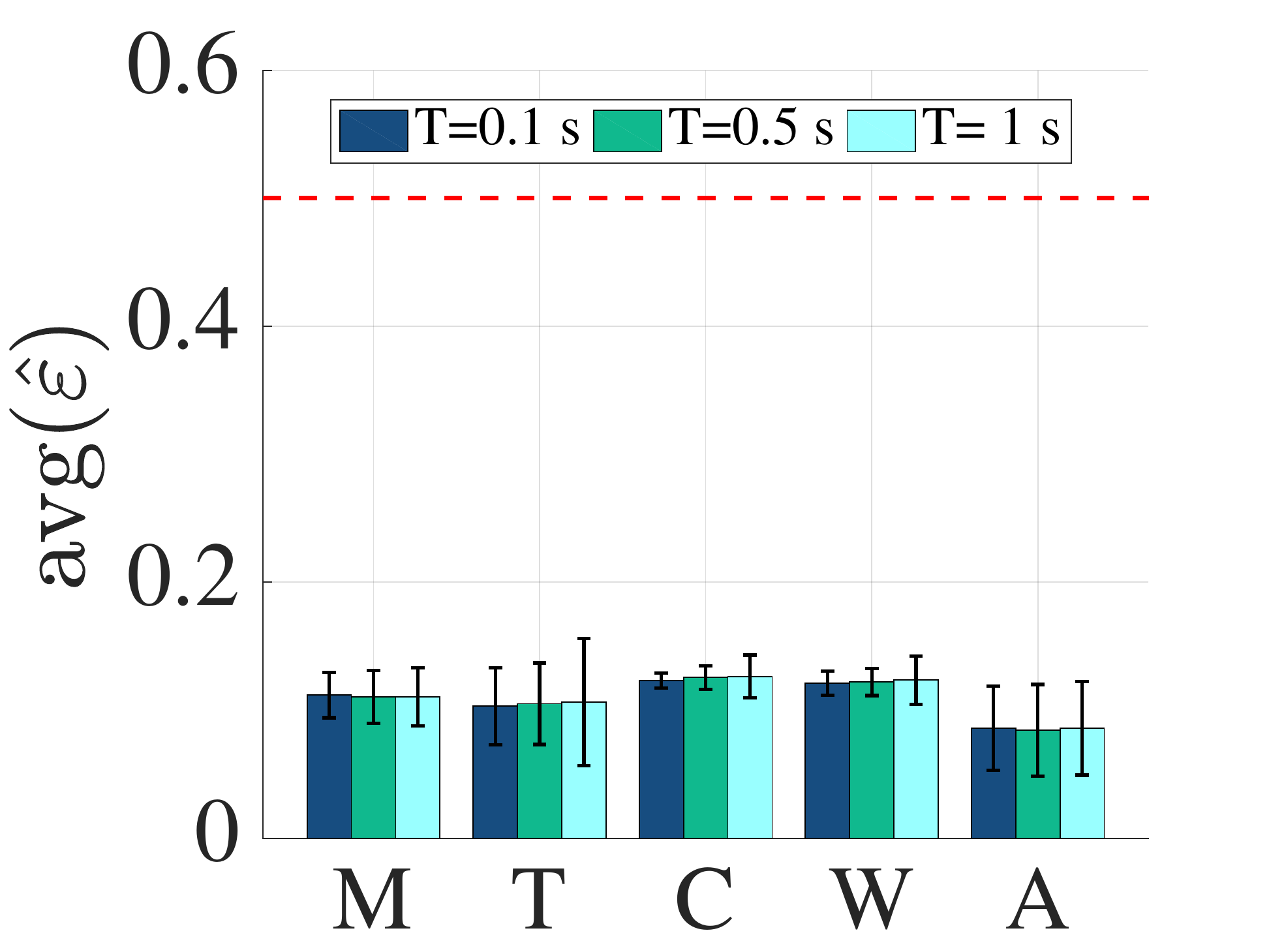}}\quad
	\subcaptionbox{target $\varepsilon = 0.1$}[.23\linewidth][c]{%
		\includegraphics[width=.22\linewidth]{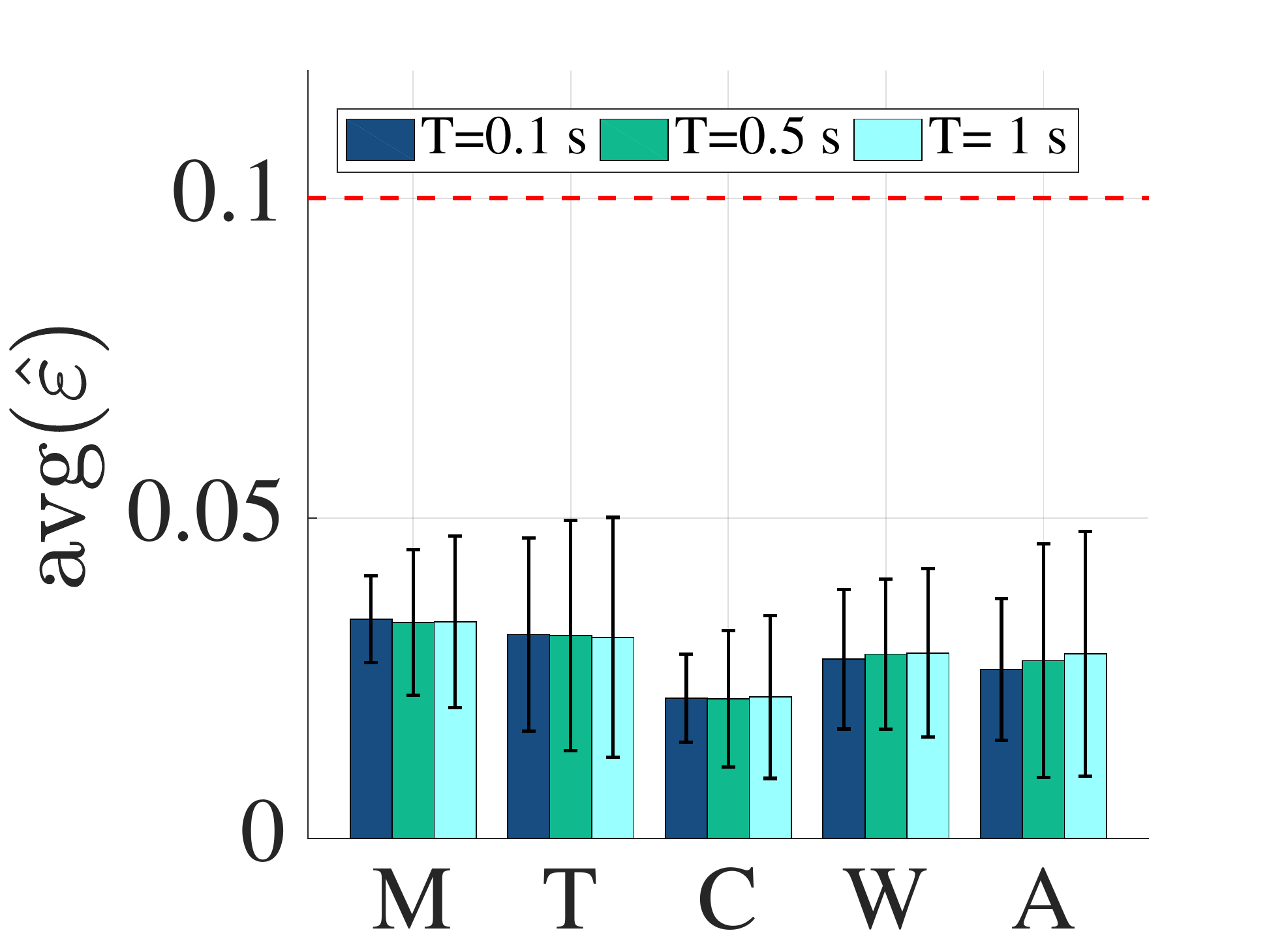}}\quad
	\subcaptionbox{target $\varepsilon = 0.05$}[.23\linewidth][c]{%
		\includegraphics[width=.22\linewidth]{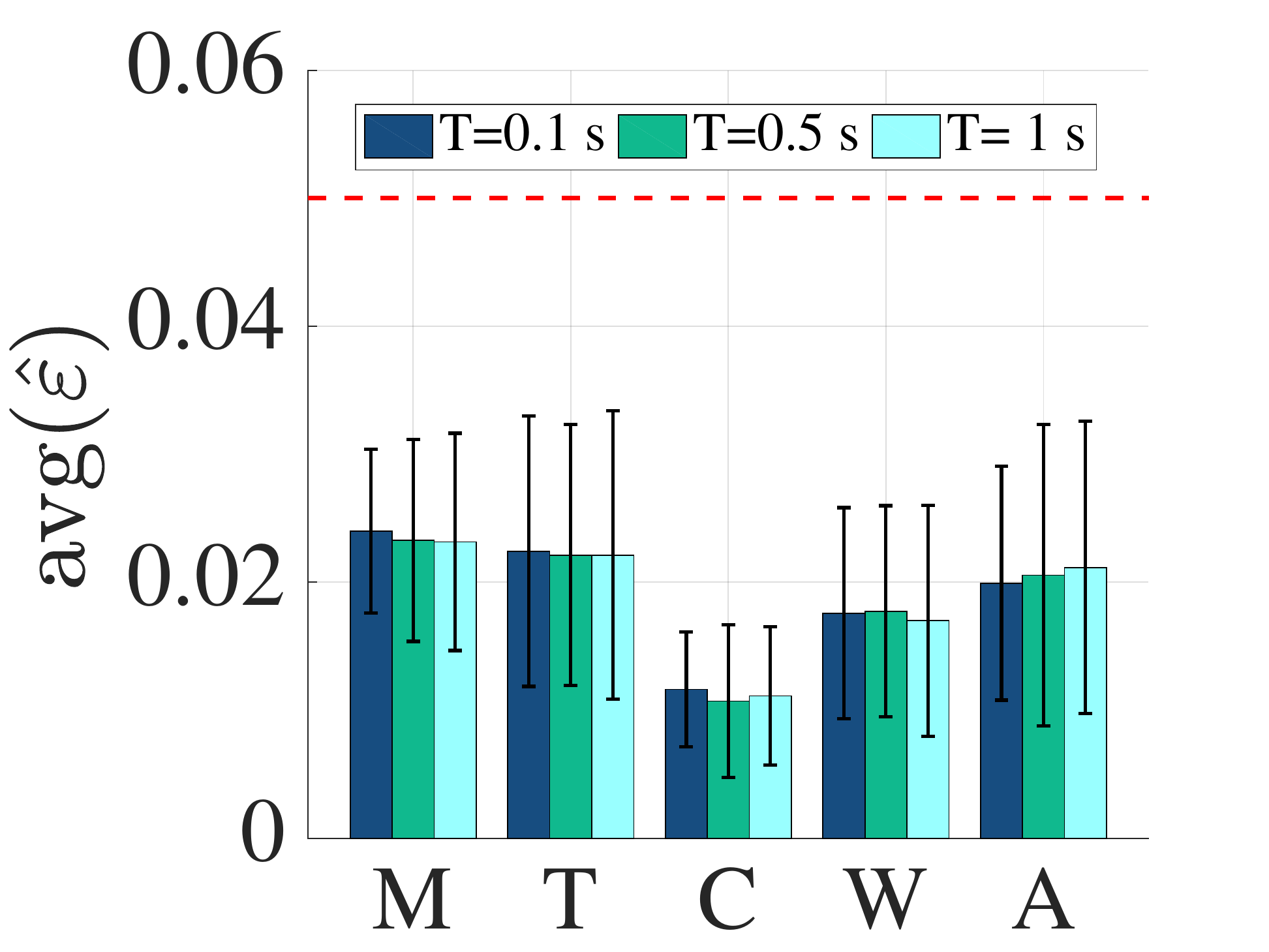}}
	\subcaptionbox{target $\varepsilon = 0.01$}[.23\linewidth][c]{%
		\includegraphics[width=.22\linewidth]{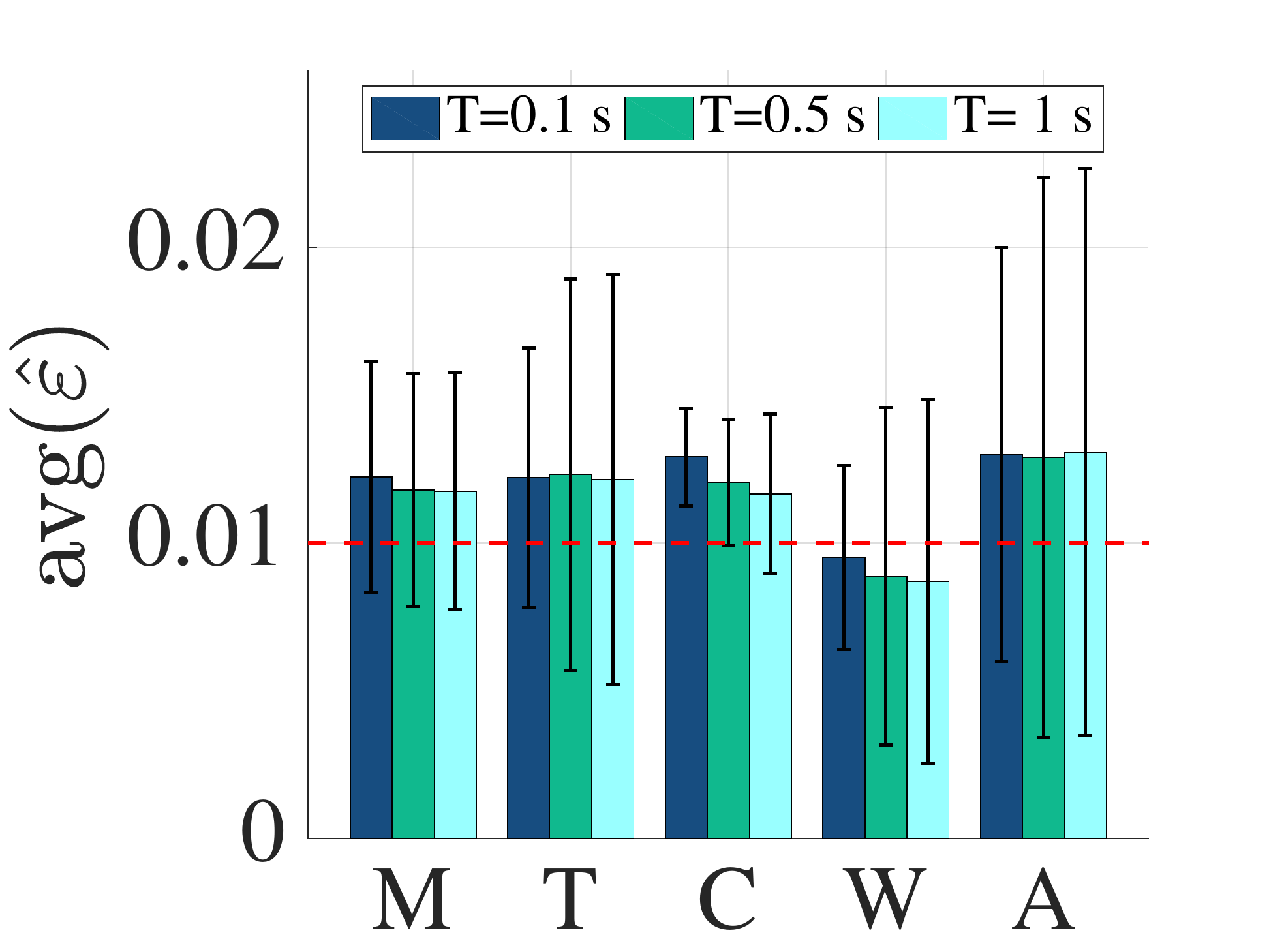}}
 	\caption{Link dimensioning based on (a-d) log-normal model, (e-h) Weibull model and (i-l) Meent's formula: avg$(\hat{ \varepsilon})$ for different datasets (M: MAWI, T: Twente, C: CAIDA, W: Waikato, A: Auckland), aggregation timescales ($100$ msec, $500$ msec and $1$ s), and target values of $\varepsilon$ (0.5, 0.1, 0.05 and 0.01). Error bars represent stderr $|\varepsilon -\hat{\varepsilon}|$.}
    \label{empEpsiRes} 	
\end{figure*}

\subsection{Bandwidth provisioning based on the log-normal model}

Here we investigate whether we could achieve more reliable bandwidth provisioning by adopting the log-normal traffic model. We calculate the mean and variance from the captured trace and generate the respective log-normal model. Then, we use the CDF function ($F$) to solve the link transparency formula shown in Equation~\ref{link-tran}. Hence, $F$ is defined as $F(C) = P(A(T)/T<C) $, which can be solved to find $C$, as follows: 
\begin{equation}
C2=F^{-1}\left (1-\varepsilon \right).
\label{cdf-lognormal}
\end{equation}

\subsection{Comparison of bandwidth provisioning approaches}
\label{comparison}

In this section, we compare the bandwidth provisioning approaches described above. The performance indicator is the empirical value of the performance criterion, which is denoted by $\hat{ \varepsilon}$ and defined as follows:
\begin{equation} 
\hat{ \varepsilon}= \frac{\# \left \{ A_{i}| A_{i} \geq CT\right \}}{n}  \textrm{ , } i\in 1\ldots n.  
\label{empepsilon}  
\end{equation}

In words, this empirical value is the percentage of all the data samples of the captured traffic which are measured larger than the estimated link capacity. Ideally, $\hat{ \varepsilon}$ would be equal to the target value of the performance criterion $\varepsilon$. The difference between $\hat{ \varepsilon}$ and $\varepsilon$ is due to the fact that the chosen traffic model is not accurately describing the real network traffic. A simple example of the described comparison approach is illustrated in Figure~\ref{time-domain}, in which we plot the captured data rate for a MAWI trace ($T=100$ msec)\footnote{Note that in all subsequent figures we have also included results for a Weibull model to get insights about bandwidth provisioning using a heavy-tailed distribution.}. The calculated capacity values from each approach when the target $\varepsilon$ is $0.01$ are $C1=344.8$ Mbps and $C2=444.3$ Mbps (represented by the horizontal lines in Figure~\ref{time-domain}). The empirical value can be calculated by using Equation~\ref{empepsilon}, which gives $\hat{ \varepsilon}_{1}=0.042$ and $\hat{ \varepsilon}_{2}=0.012$. Obviously, with the first approach the network operator would not be able to meet the target $\varepsilon=0.01$, while with the second approach the empirical value is close to the target.

We next compare results of bandwidth provisioning calculations based on the (a) Meent's formula, (b) Weibull model and (c) proposed log-normal model. Figure~\ref{empEpsiRes}(a)-(d) shows the average of the empirical value (avg$(\hat{ \varepsilon})$) for all traces in each dataset at $T= 0.1$ sec, $T= 0.5$ sec and $T=1$ sec. The value of $T$ is chosen to be sufficiently small so that the fluctuations in the traffic can be modelled as well. Each model is tested for four different values of the performance criterion: $\varepsilon = 0.5$, $\varepsilon = 0.1$, $\varepsilon = 0.05$ and $\varepsilon = 0.01$. In Figure~\ref{empEpsiRes}(a)-(d) we clearly see that the log-normal model is able to satisfy the required performance criterion $\varepsilon$ at different aggregation time-scales for all datasets. In contrast, Meent's formula failed to allocate sufficient bandwidth, which results in missing the target performance criterion $\varepsilon$ for all datasets and target performance values, as depicted in Figure~\ref{empEpsiRes}(i)-(l) (see horizontal red line). The Weibull distribution performs better comparing to Meent's formula, but bandwidth provisioning using the log-normal model is far superior, as can be seen from Figures~\ref{empEpsiRes}(a)-(d) and~\ref{empEpsiRes}(e)-(h).

\section{95th percentile pricing scheme based on log-normal model}
\label{sec:pricing}

Traffic billing is typically based on the 95th percentile method \cite{billPAM}. Traffic volume is measured at border network devices (typically aggregated at time intervals of 5 minutes) and bills are calculated according to the 95-percentile of the distribution of measured volumes; i.e. network operators calculate bills by disregarding occasional traffic spikes. Forecasting future bills, which is important for ISPs and clients, can be done using a model of the traffic calculated through previously sampled traffic. In this section, we apply our findings on Internet traffic modelling in predicting the cost of traffic according to the 95th percentile method.
\begin{figure*}[t]
	\setlength{\belowcaptionskip}{-4pt}
	\centering
	\subcaptionbox{CAIDA}[.28\linewidth][c]{%
		\includegraphics[width=.28\linewidth]{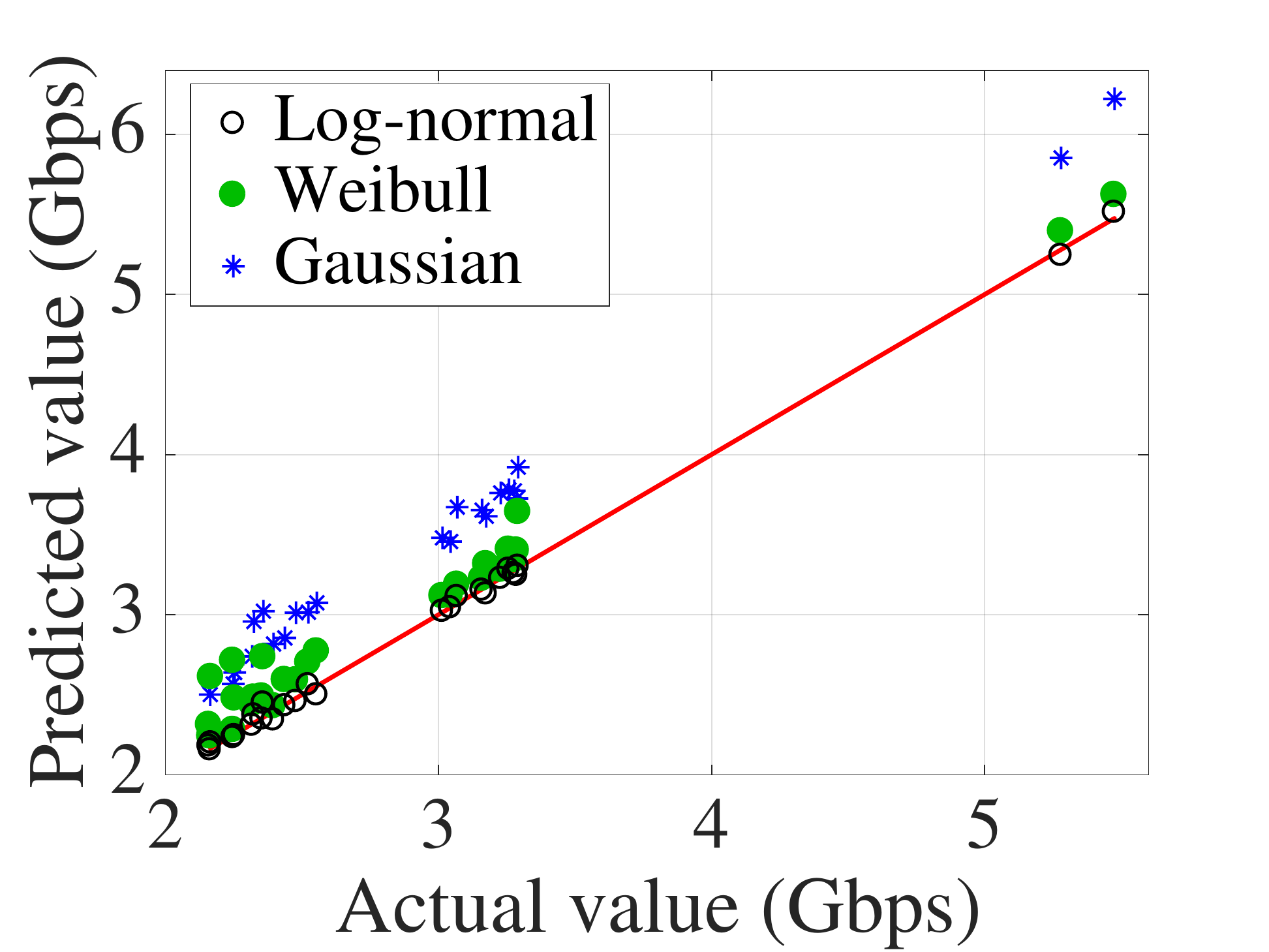}}\quad
	\subcaptionbox{Waikato}[.28\linewidth][c]{%
		\includegraphics[width=.28\linewidth]{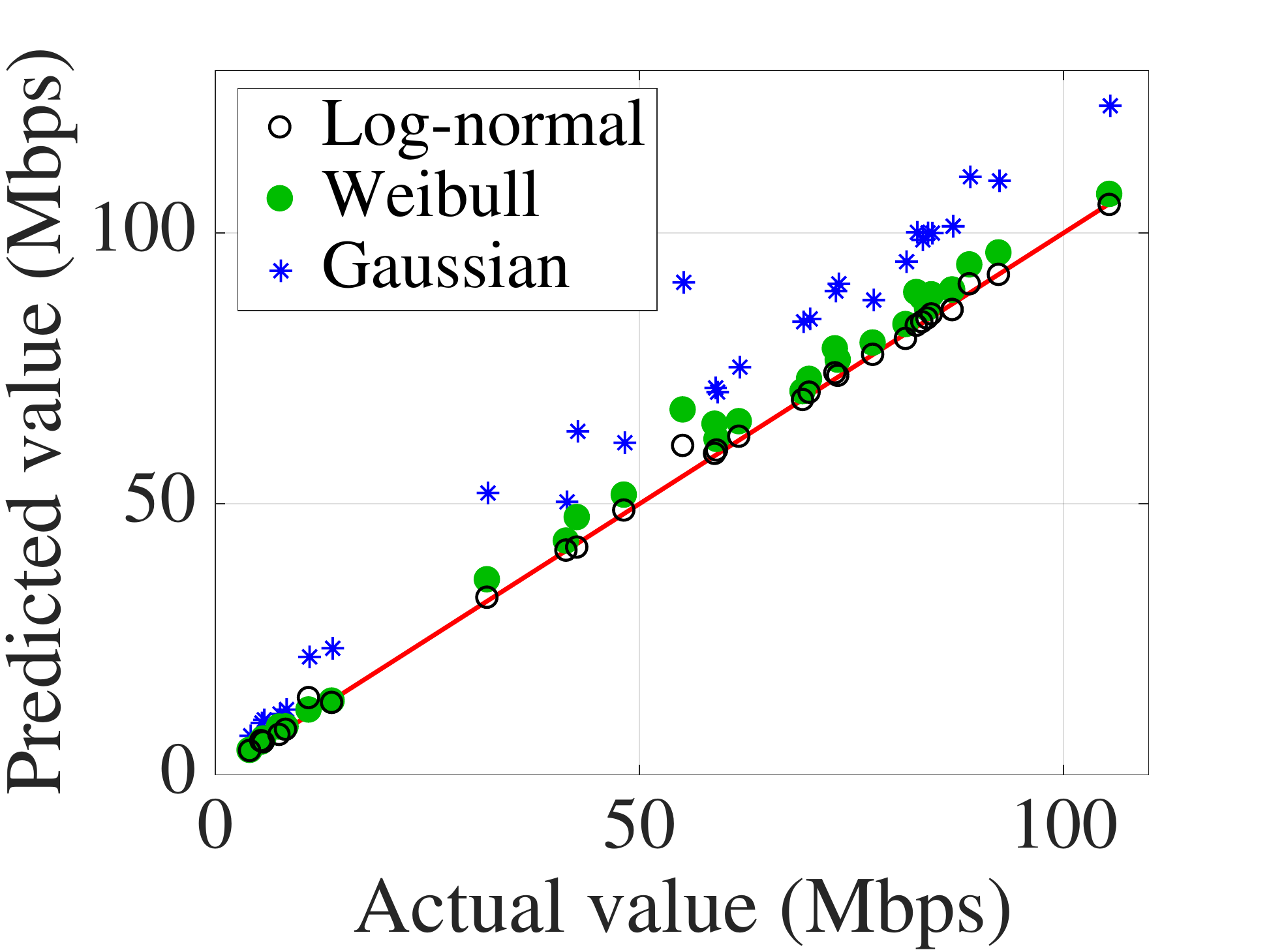}}\quad
	\subcaptionbox{Auckland}[.28\linewidth][c]{%
		\includegraphics[width=.28\linewidth]{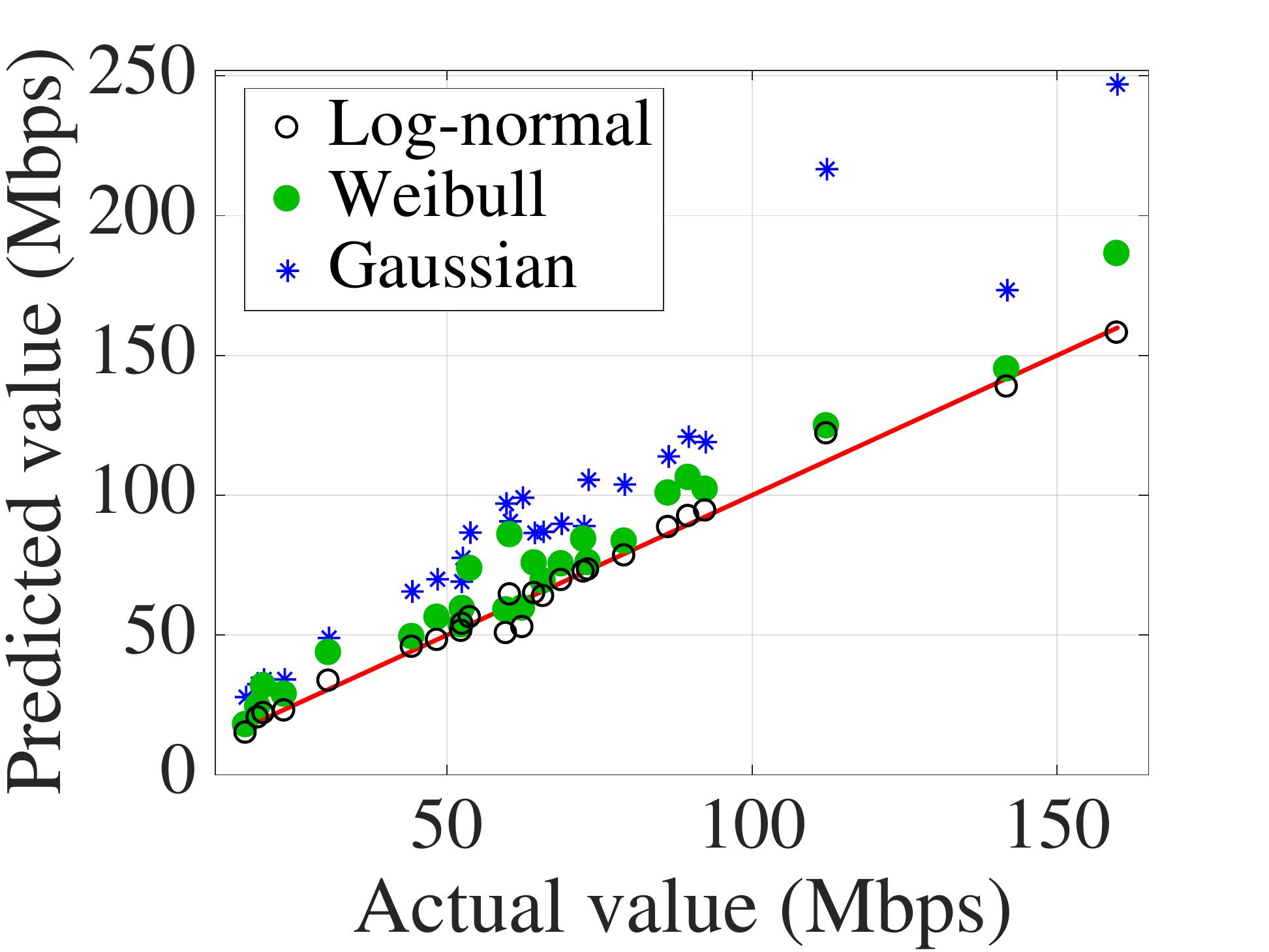}}
	\bigskip
	\\
	\subcaptionbox{Twente}[.28\linewidth][c]{%
		\includegraphics[width=.28\linewidth]{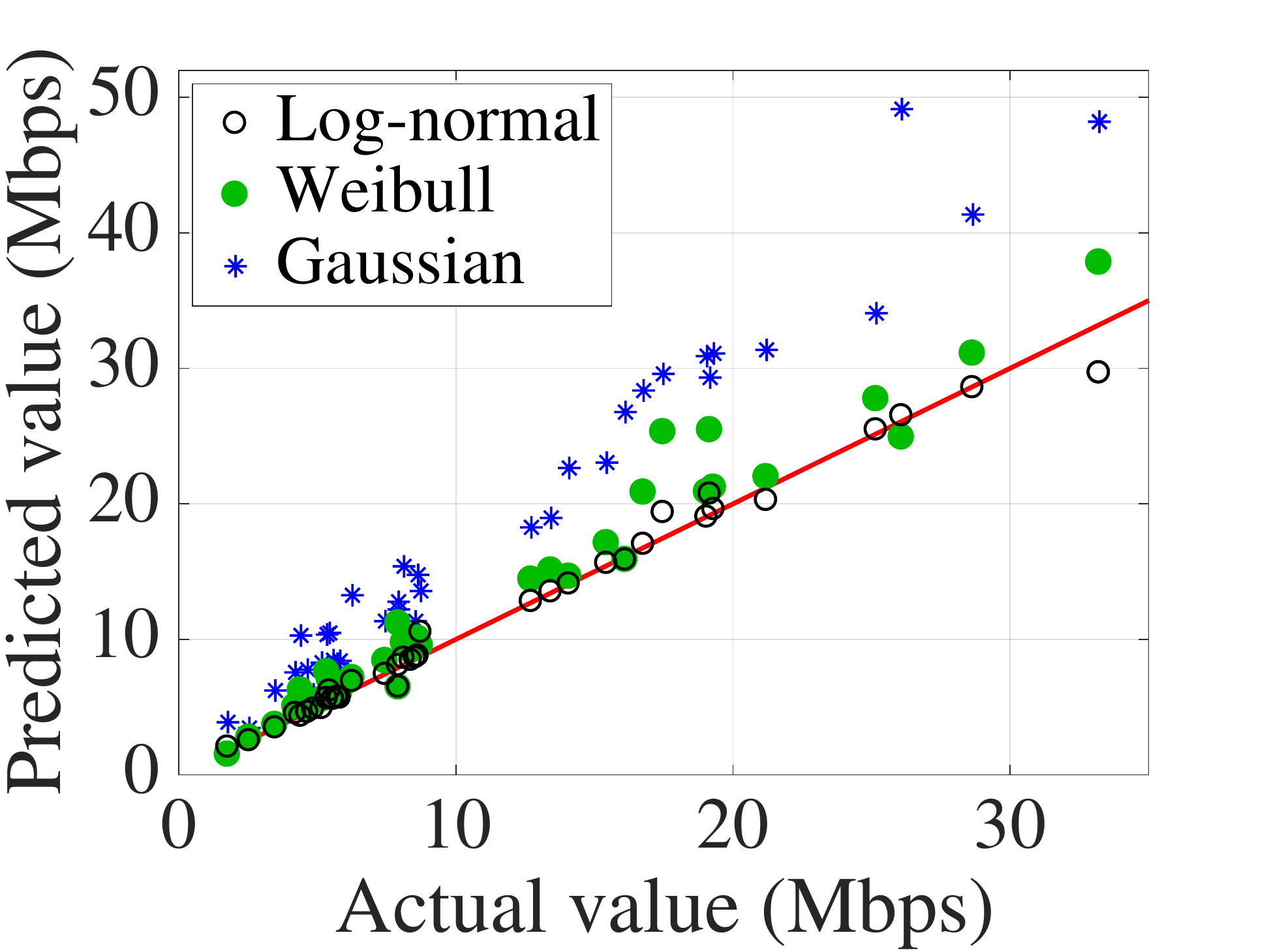}}\quad
	\subcaptionbox{MAWI}[.28\linewidth][c]{%
		\includegraphics[width=.28\linewidth]{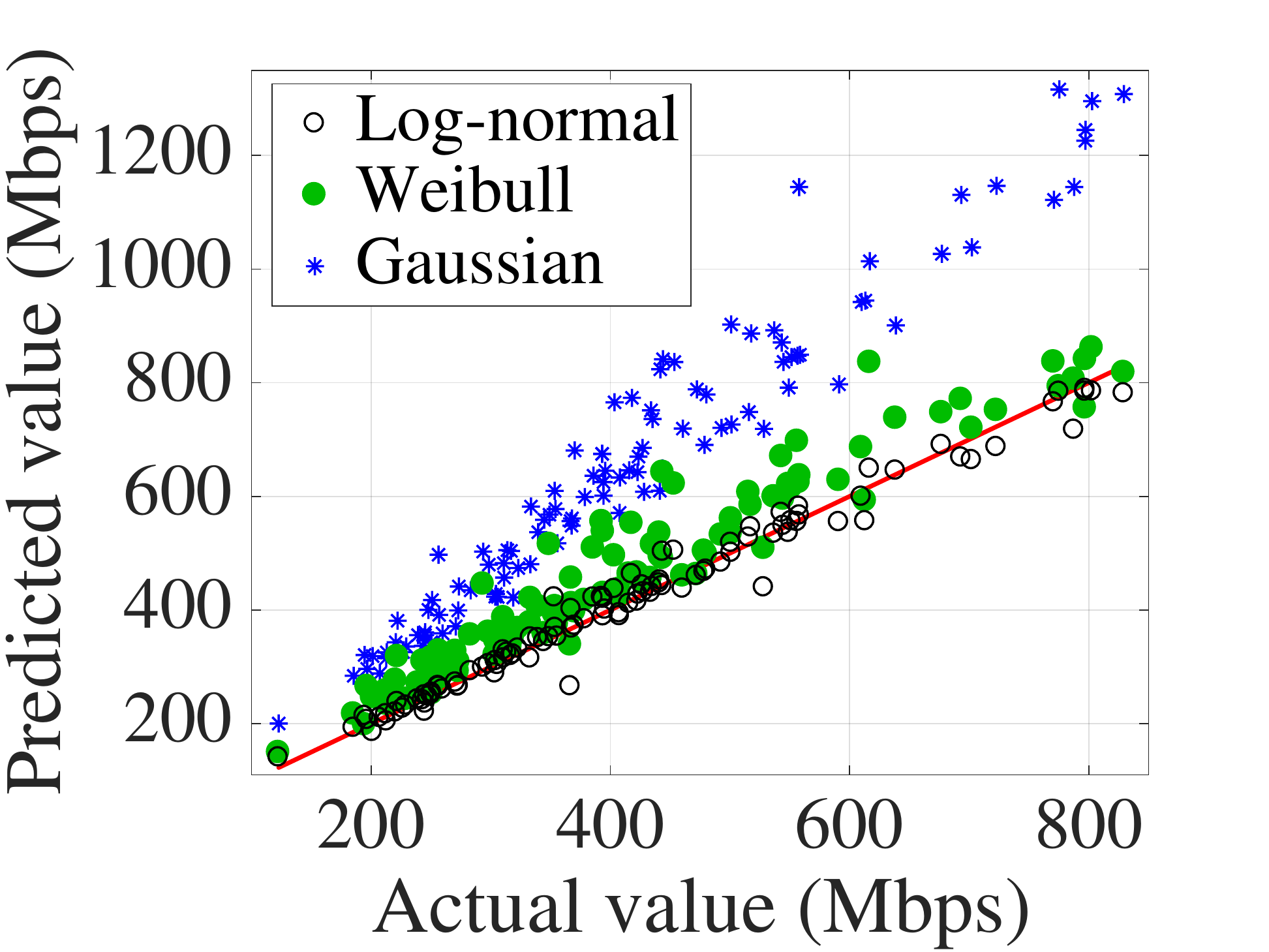}}\quad
	\caption{95th percentile values (actual vs predicted rates) based on log-normal, Weibull and Gaussian models. An ideal model would result in points in the plot area that fall exactly on the red line.}
	\label{percentile} 	
\end{figure*}
For each network trace we calculate the actual 95th percentile of the traffic volume. The majority of the studied traffic traces were 15-minute long but operators typically use measurements traffic volumes for much longer periods, therefore we scale down the calculation of the 95th percentile by dividing each trace (900 seconds) into 90 groups (10 seconds length each). The authors appreciate that by using 15-minute rather than day long traces we omit any study of diurnal effects in the distribution. We note that the sum of several log-normal distributions is itself very accurately represented by a log-normal distribution~\cite{mitchell68}. Hypothetically, therefore, if 96 consecutive 15-minute traces fit a log-normal distribution (with different parameters for each) then the resulting 24 hour trace is also likely to be a good fit to a log-normal. We also note that the distributions tested were on a level playing field in that they would all be affected equally by the shorter duration of the data sets.

We calculate the 95th percentile for the observed traffic. We then fit a Gaussian, Weibull and log-normal distribution to each trace (for $T=100$ msec) and calculate the 95th percentile of the fitted distribution. We plot the actual 95th percentile against the three predictions in Figure~\ref{percentile} with a red reference line to show where perfect predictions would be located. It is clear that the log-normal model provides much more accurate predictions of the 95th percentile than the Gaussian model. As with the bandwidth dimensioning case discussed in Section \ref{sec:provision}, the Weibull is better than the Gaussian model but worse than the proposed log-normal model.
   
We employ the normalised root mean squared error (NRMSE) as a goodness of fit to the results in Figure~\ref{percentile}. NRMSE measures the differences between values predicted by a hypothetical model and the actual values. In other words, it measures the quality of the fit between the actual data and the predicted model. Table~\ref{percentileGoF} shows the NRMSE for all datasets and the three considered models. It is clear that the lowest NRMSE value is for the log-normal model, which is the best model compared to the Gaussian and Weibull ones.  
\begin{table}[]
		\caption{Goodness of fit (GOF) using normalised root mean squared error (NRMSE)}
	\begin{tabular}{|c|c|c|c|l|l|}
		\hline
		Model/Dataset & CAIDA  & Waikato & Auckland & Twente & MAWI   \\ \hline
		Log-normal & 0.0399 & 0.0401  & 0.1058   & 0.0979 & 0.1528 \\ \hline
		Weibull    & 0.2410 & 0.1148  & 0.2984   & 0.2123 & 0.4145 \\ \hline
		Gaussian   & 0.5544 & 0.4193  & 0.6866   & 0.5741 & 0.9828 \\ \hline
	\end{tabular}

	\label{percentileGoF} 	
\end{table}
\section{Related work}
\label{sec:related}

Reliable traffic modelling is important for network planning, deployment and management; e.g. for traffic billing and network dimensioning. Historically, network traffic has been widely assumed to follow a Gaussian distribution. In~\cite{Gaussian-everywhere,Gaussian-revisited}, the authors studied network traces and verified that the Gaussianity assumption was valid (according to simple goodness-of-fit tests they used) at two different timescales. In~\cite{BusyhourTraff}, the authors studied traffic traces during busy hours over a relatively long period of time and also found that the Gaussian distribution is a good fit for the captured traffic. Schmidt et al.~\cite{2014-ifip-conf} found that the degree of Gaussianity is affected by short and intensive activities of single network hosts that create sudden traffic bursts. All the above mentioned works agreed on the Gaussian or `fairly Gaussian' traffic at different levels of aggregations in terms of timescale and number of users. The authors in~\cite{sigcomm2002nonGaussian,heavy-sigcomm-2001} examined the levels of aggregation required to observe Gaussianity in the modelled traffic, and concluded that this can be disturbed by traffic bursts. The work in~\cite{iccTrafficcharact,12-GLOBECOM2002} reinforces the argument above, by showing existence of large traffic spikes at short timescales which result in high values in the tail. Compared to existing literature, our findings are based on a modern, principled statistical methodology, and traffic traces that are spatially and temporally diverse. We have tested several hypothesised distributions and not just Gaussianity.

An early work drawing attention to the presence of heavy tails in Internet file sizes (not traffic) is that of Crovella and Bestavros~\cite{self-sim97}. Deciding whether Internet flows could be heavy-tailed became important as this implies significant departures from Gaussianity. The authors in~\cite{heavytails2017} provided robust evidence for the presence of various kinds of scaling, and in particular, heavy-tailed sources and long range dependence in a large dataset of traffic spanning a duration of 14 years.

Understanding the traffic characteristics and how these evolve is crucial for ISPs for network planning and link dimensioning. Operators typically over-provision their networks. A common approach to do so is to calculate the average bandwidth utilisation~\cite{proveGaussianIFIP} and add a safety margin. As a rule of thumb, this margin is defined as a percentage of the calculated bandwidth utilisation~\cite{cisco}. Meent et al.~\cite{ieee-network-2009} proposed  a new bandwidth provisioning formula, which calculates the minimum bandwidth that guarantees the required performance, according to an underlying SLA. This approach relies on the statistical parameters of the captured traffic and a performance parameter. The underlying fundamental assumption for this to work is that the traffic the network operator sees follows a Gaussian distribution. Same approach has been used in~\cite{transaction2015}.

The 95th percentile method is used widely for network traffic billing. Dimitropoulos et al.~\cite{billPAM} have found that the computed 95th percentile is significantly affected by traffic aggregation parameters. However, in their approach they do not assume any underlying model of the traffic; instead, they base their study on specific captured traces. Stanojevic et al.~\cite{95percentileIMC} proposed the use of Shapley value for computing the contribution of each flow to the 95th percentile price of interconnect links. Works ~\cite{Tosendornot,bulkTransfers,Tuangou,requestmapping} propose calculating the 95th percentile using experimental approaches. Xu et al.~\cite{gaussianPercentile} assume that network traffic follows a Gaussian distribution``through reasonable aggregation'' and propose a cost efficient data centre selection approach based on the 95th percentile.

\section{Conclusion}
\label{sec:conclusion}

The distribution of traffic on Internet links is an important problem that has received relatively little attention.  We use a well-known, state-of-the-art statistical framework to investigate the problem using a large corpus of traces.  The traces cover several network settings including home user access links, tier 1 backbone links and campus to Internet links.  The traces are from times from  2002 to 2018  and are from a number of different countries.  We investigated the distribution of the amount of traffic observed on a link in a given (small) aggregation period which we varied from $5$ msec to $5$ sec.  The hypotheses compared were that the traffic volume was heavy-tailed, that the traffic was log-normal and that the traffic was normal (Gaussian).  The vast majority of traces fitted the log-normal assumption best and this remained true all timescales tried.  Where no distribution tested was a good fit this could be attributed either to the link being saturated (at capacity) for a large part of the observation or exhibiting signs of link-failure (no or very low traffic for part of the observation).

We investigate the impact of the distribution on two sample traffic engineering problems.  Firstly, we looked at predicting the proportion of time a link will exceed a given capacity.  This could be useful for provisioning links or for predicting when SLA violation is likely to occur.  Secondly, we looked at predicting the 95th percentile transit bill that ISP might be given.  For both of these problems the log-normal distribution gave a more accurate result than a heavy-tailed distribution or a Gaussian  distribution.  We conclude that the log-normal distribution is a good (best) fit for traffic volume on a normally functioning internet links in a variety of settings and over a variety of timescales, and further argue  that this assumption can make a large difference to  statistically predicted outcomes for applied network engineering problems.

In future work, we plan to test the stationarity of the traffic traces.

\bibliographystyle{IEEEtran}
\bibliography{bibRef}

\end{document}